%% file: draft6.tex
\documentstyle[12pt]{article}\textheight
230mm\textwidth 165mm 
\newcommand{\bi}{\bibitem}
\newcommand{\be}{\begin{eqnarray}}
\newcommand{\ee}{\end{eqnarray}}
\newcommand{\nn}{\nonumber}
\catcode`\@=11
\def\lsim{\mathrel{\mathpalette\@versim<}}
\def\gsim{\mathrel{\mathpalette\@versim>}}
\def\@versim#1#2{\vcenter{\offinterlineskip
\ialign{$\m@th#1\hfil##\hfil$\crcr#2\crcr\sim\crcr } }}
\catcode`\@=12

\begin{document}

\pagestyle{empty}

\noindent
\hspace*{10.7cm} \vspace{-3mm}  HIP-1998-82/TH\\
\hspace*{10.7cm} \vspace{-3mm}  KANAZAWA-98-20\\

\noindent
\hspace*{10.7cm} November 1998

\begin{center}
{\Large\bf  Running of Soft Parameters \\in\\
Extra Space-Time Dimensions}
\end{center}

\begin{center}
{\sc Tatsuo Kobayashi}$\ ^{(1),\dag}$, 
{\sc Jisuke Kubo}$\ ^{(2)}$,\\ 
{\sc Myriam Mondrag\' on}$\ ^{(3),*}$ and
{\sc George Zoupanos}$\ ^{(4),**}$  
\end{center}
\begin{center}
{\em $\ ^{(1)}$ 
Department of Physics,  High Energy Physics Division, 
University of Helsinki \vspace{-2mm}\\ and \vspace{-2mm}\\
Helsinki Institute of Physics, 
FIN-00014 Helsinki, Finland} \vspace{-2mm}\\
{\em $\ ^{(2)}$ 
Institute for Theoretical Physics, 
Kanazawa  University, 
Kanazawa 920-1192, Japan}\vspace{-2mm}\\
{\em $\ ^{(3)}$ Instituto de F\' isica,  UNAM,
Apdo. Postal 20-364,
M\' exico 01000 D.F., M\' exico}  \vspace{-2mm}\\
{\em $\ ^{(4)}$
Physics Department, Nat. Technical University, GR-157 80 
\vspace{-3mm} Zografou,
Athens, Greece.}
\end{center}

\vspace{0.5cm}
\begin{center}
{\sc\large Abstract}
\end{center}

\noindent
The evolution of the parameters including \vspace{-2mm}
those in the soft
supersymmetry-breaking (SSB) sector is studied in the 
minimal supersymmetric standard \vspace{-2mm} model (MSSM)  with 
a certain set of Kaluza-Klein towers which 
has been recently \vspace{-2mm} considered by
Dienes  {\em et al}.
We use the continuous Wilson renormalizaion group 
technique \vspace{-2mm}
to derive the matching condition between the effective, 
renormalizable and original, unrenormalizable 
 theories. \vspace{-2mm}
We investigate whether the assumption on a large 
compactification radius in the model \vspace{-2mm}
 is consistent with the gauge coupling unification,
the $b-\tau$  unification and the radiative breaking \vspace{-2mm}
of the electroweak gauge symmetry 
with the universal SSB terms.\vspace{-2mm} We calculate the
superpartner spectrum under the assumption of  \vspace{-2mm} 
the  universal SSB parameters 
to find differences between the model and the MSSM.


\vspace{2cm}
\footnoterule
\vspace{0.1cm}
\noindent
$^{\dag}$Partially supported  by the Academy of Finland \vspace{-3mm} 
(no. 37599).\\
\noindent
$^{*}$Partially supported \vspace{-2mm}
 by the UNAM Papiit project
IN110296. \\ \noindent
$ ^{**}$Partially supported  by the E.C. projects, \vspace{-3mm}
FMBI-CT96-1212 and  ERBFMRXCT960090,
and the Greek projects, PENED95/1170; 1981.

\newpage
\pagestyle{plain}

\section{Introduction}
Recently, motivated by the works of
refs. \cite{witten1,witten2} in which
the strong coupling limit of  heterotic superstrings has
been considered,  there have been  renewed interests to
consider Kaluza-Klein theories with a large compactification
radius \cite{antoniadis1}--\cite{hewett}.
In an extreme case, the radius could be in the range
of submillimeter,
whereas the Kaluza-Klein excitations with 
masses $\lsim 10$ TeV will 
become observable in future collider experiments 
\cite{antoniadis2}--\cite{arkani4}.
These scenarios could be even embedded into various superstring 
models with an anisotropic compactification
\cite{antoniadis1}--\cite{kakushadze2}, 
especially into models  based on Type I superstring 
which should describe the strong coupling limit of the $SO(32)$
heterotic superstring \cite{witten2}.

It is clear that the qualitative nature of
the traditional unification scenario changes
\cite{dienes1,abel1,ross1}
if the fields of the standard model (SM) or 
of the minimal supersymmetric standard model 
(MSSM) feel the existence of  extra
dimensions  whose scale is significantly 
smaller than the ordinary GUT scale 
$M_{\rm GUT} \sim 10^{15-16}$ GeV.
We would like to emphasize that, 
strictly speaking,  
the unification in those scenarios takes place
not in
$4$  rather in $D=4+\delta$ dimensions in which the original theory
is formulated. On one hand, quantities near and above 
$M_{\rm GUT}$,
including $M_{\rm GUT}$ itself, are  $D=4+\delta$ dimensional quantities.
At energies much below 
the compactification scale, on the other hand,
 all the massive Kaluza-Klein
states  decouple so
that we have a four dimensional effective theory.
Therefore,  there must be a
certain  matching condition between the four dimensional effective and
$D=4+\delta$ dimensional theories. Clearly, 
the four dimensional theory
does not know anything about the matching condition, but 
they can be derived from
the original  $D=4+\delta$ dimensional theory. In fact, in the treatment 
of Dienes {\em et al.} \cite{dienes1} 
on   the massive Kaluza-Klein
excitations,
one needs an outside information about the infrared 
and ultraviolet cut off to define a finite, four dimensional, low-energy
effective theory. As we will see in sect. 2, 
 we will derive the matching condition,
 in contrast to ref. \cite{dienes1}, from the requirement that
the evolution equations of couplings in the effective theory 
smoothly go over
in the large compactification-radius limit to what 
one finds in uncompactified, original, $D=4+\delta$
dimensional theory.
Specifically, we will employ 
the continuous Wilson renormalization 
group (RG) approach \cite{wilson1}
 which can be formulated in any space-time dimensions.
Of various existing  formulations \cite{wegner,polchinski1} 
of the continuous Wilson RG in literature, we will
follow the formulation of ref.  \cite{wegner}.
It turns out that the small discrepancy in the matching condition
compared with that of ref. \cite{dienes1}
 has no significant effect in the
application to the model of ref. \cite{dienes1}
which we also will consider in this paper.

Given the matching condition,
we will be staying in four dimensions to
extend the method of ref.  \cite{dienes1}
so as to include 
the soft supersymmetry-breaking (SSB) sector.
To this end, we will make use of the recent  progress on the
renormalization properties of SSB parameters 
\cite{yamada1}--\cite{avdeev1}. 
It has been shown
that the ultraviolet divergences of
 the SSB parameters are simply related to  those of 
the corresponding supersymmetric parameters \cite{yamada1}
so that 
the $\beta$-functions of the SSB parameters 
can be easily  obtained by applying certain 
differential operators on
the anomalous dimensions
 and the gauge coupling
$\beta$-function of the supersymmetric 
theory \cite{jack3,avdeev1}.
This method works only for four dimensional
theories
when using a mass-independent renormalization scheme
such as the dimensional reduction 
scheme \cite{yamada1,jack3,avdeev1}.
So it is not obvious that the method can be
applied  straightforwardly to 
softly broken supersymmetric  
theories with  Kaluza-Klein towers, because one defines the theory by
 cut off.
Fortunately, the (ultraviolet) divergent parts in one-loop order are
independent of renormalization scheme and have a simple structure so that
the massive Kaluza-Klein
excitations  do not disturb
the above mentioned relations among
the divergent parts of the SSB parameters and 
those of  the supersymmetric parameters.
In the second-half of sect 2, we will 
consider the model of ref. \cite{dienes1}, the MSSM with a certain
set of Kaluza-Klein towers, and calculate  the one-loop 
$\beta$-functions for the SSB parameters above the compactification scale
$\mu_0 =R^{-1}$.

Given the $\beta$-functions, we can discuss various aspects
of the model in a more concrete fashion, which will be
the subject of sect. 3.
We first consider the gauge coupling 
unification and find
 that the smaller the $\mu_0$ is, the larger is the predicted
value of the QCD coupling $\alpha_3(M_Z)$,
in accord with  the result of ref. \cite{ross1}.
We however need more detailed information on a possible  theory 
 above $\mu_0$ to control the corrections
such as the threshold effect at 
$M_{\rm GUT}$ in order to give a more precise
 prediction of
$\alpha_3(M_Z)$. Here we will consider the possibility that
the level of $U(1)_Y$ can be different from the usual
$SU(5)$-motivated value $5/3$.

We will calculate the mass of the bottom quark
under the assumption of the $b-\tau$ Yukawa 
coupling unification for the given top quark mass.
Our analysis points out that 
the $b-\tau$  unification can be consistent with a large 
 compactification radius.
In the final part of sect. 3, we investigate RG effects on 
the SSB parameters, where we assume that they are
universal at the GUT scale.
In particular,  we study the possibility of achieving
the  radiative electroweak symmetry breaking, 
the nature of
the lightest superparticle (LSP), and the constraint 
coming from the negative (mass)$^2$ of the stau.
As we will conclude in 
sect. 4, the basic low-energy
($\lsim O(1)$ TeV) feature of the MSSM 
 remains unchanged even if $\mu_0$ is as small as
$\sim O(10)$ TeV, but there exist a certain  chance
in the superpartner spectrum to experimentally
discriminate the model from the MSSM.

\section{$\beta$-functions in extra dimensions}
\subsection{Large radius limit and matching condition}
Suppose that we would like to
 study low-energy physics of 
a Kaluza-Klein 
theory which is defined in $D=4+\delta$ dimensions
with extra $\delta$ dimensions compactified.
 The theory is not renormalizable, and  presumably
trivial, but it can be well defined by introducing an
ultraviolet cut off $\Lambda_0$.
The natural framework to study the low-energy physics
is provided by the continuous Wilson renormalization 
group (RG) \cite{wilson1},
 which can be realized in terms of an 
integro-differential equation \cite{wegner,polchinski1}.
Here we would like to
follow the formulation of ref.  \cite{wegner}.
Since we expect that, in the limit that the radius
of the compactified dimensions approaches
infinity, the result goes over to 
what one finds in the uncompactified case,
we briefly sketch below the treatment of ref. \cite{wegner}
in the case of a scalar theory in 
uncompactified Euclidean $D$ dimensions.

The basic idea to the
non-perturbative RG approach \cite{wegner}
\footnote{See also ref. \cite{morris1}.
The continuous Wilson renormalization 
RG approach is called sometimes
the non-perturbative RG approach.}
is to divide the field $\phi(p)$
in the momentum space into
low and high energy modes,
\be
\phi (p) &=& \theta (\Lambda-|p|)\phi_{<}(p)+
\theta(|p|-\Lambda)\phi_{>}(p)~,
\ee
and integrate out the high energy modes
in the path integral to define the effective theory:
\be
S_{\rm eff} [~\phi_{<},\Lambda~]
&=&- \ln \{~\int {\cal D}\phi_{>} ~e^{-S[~\phi_{>},\phi_{<}~]}~\}~,
\ee
where $ S_{\rm eff}$ is the Wilson effective action.
It turns out that the difference
\be
\delta S_{\rm eff} &=&
S_{\rm eff} [~\phi_{<},\Lambda+\delta\Lambda~]
-S_{\rm eff} [~\phi_{<},\Lambda~]
\ee
for an infinitesimal  $\delta\Lambda$ becomes a Gaussian path
integral which can be in fact carried out.
That is, it is possible to calculate
$\Lambda (\partial S_{\rm eff}/\partial \Lambda)$
to write down a formal expression
for the RG flow equation  of the effective theory in the form
\be
\Lambda \frac{\partial S_{\rm eff}}{\partial \Lambda}
&=& {\cal O}(S_{\rm eff})~,
\label{nrg}
\ee
where ${\cal O}$ is a non-linear operator acting 
on the functional $S_{\rm eff}$. The explicit expression for 
${\cal O}$ was first obtained by Wegner and Houghton \cite{wegner},
but in practice, 
 the RG equation (\ref{nrg}) cannot be solved exactly.
There are various approaches to find approximate solutions
to the Wegner-Houghton (W-H) equation, and 
one of the
successful ones is the so-called local potential approximation
\cite{nicoll}--\cite{morris2}, which we would
like to adopt here. In this approximation, one makes an ansatz
for  the solution to the W-H equation
(\ref{nrg}) in the form
\be
S_{\rm eff} &=&\int d^D x \left(~
\frac{1}{2}\partial_M \phi \partial_M \phi +V(\phi^2)
\right)~,
\label{seff}
\ee
and finds that the W-H equation (\ref{nrg}) reduces 
to a partial differential equation 
for the potential $V$ \cite{hasen1,morris2},
\be
\Lambda \frac{\partial V}{\partial \Lambda}=
-\frac{A_D}{2}~
\ln(1+V'+2 \rho V'')-D V-(2-D)\rho V'~,
\label{pde1}
\ee
where we have defined:
\be
\rho &=& \frac{\phi^2}{2}~,~
V'=\frac{d V}{d
\rho}~,~A_D=\frac{2^{1-D}}{\pi^{D/2} \Gamma(D/2)} ~, 
\label{ad}
\ee
and all the quantities are made dimensionless by
multiplying them  with an appropriate power of $\Lambda$.
If we furthermore assume that the potential
$V$ is a power series in $\rho$, i.e.,
\be
V(\rho) &=&\sum_{n=0} \frac{\tilde{\lambda}_n(\Lambda)}{n!}
\rho^n~,
\ee
we find \cite{morris2}
\be
\Lambda\frac{d \lambda_0 }{d \Lambda}
&=&-\frac{A_D}{2} \ln (1+\Lambda^{-2}\lambda_1)~
\Lambda^{D}   ~,
\label{lambda0}\\
\Lambda\frac{d \lambda_1 }{d \Lambda}
&=&-A_D \frac{3 \lambda_2/2}{(1+\Lambda^{-2}\lambda_1)} ~
\Lambda^{D-2}  ~,\\
\Lambda\frac{d \lambda_2 }{d \Lambda}
&=& - A_D \left(-\frac{9 \lambda_2^2/2}{(1+\Lambda^{-2}\lambda_1)^2}~
\Lambda^{D-4}+
\frac{5 \lambda_3/2}{(1+\Lambda^{-2}\lambda_1)}~
\Lambda^{D-2}\right)~,
\label{lambda2}\\
\Lambda\frac{d \lambda_3 }{d \Lambda}
&=& - A_D \left(~
\frac{27 \lambda_2^3}{(1+\Lambda^{-2}\lambda_1)^3}~
\Lambda^{D-6}
-\frac{45 \lambda_2 \lambda_3/2}{(1+\Lambda^{-2}\lambda_1)^2} 
\Lambda^{-D}
+ O(\lambda_4)~\right)~,
\label{lambda3}\\
 & \vdots & ~
\ee
where we have defined  the dimensionful couplings
$\lambda_n$ as
\be
\lambda_n &=&\Lambda^{D(1-n)+2n}~\tilde{\lambda}_n~.
\ee
The set of the evolution equations above
systematically includes the effects of higher dimension operators,
and should be approximately valid for $\Lambda >> 1/R$
if some of the spatial dimensions are compactified.
In deriving the set of evolution equations
(\ref{lambda0})-(\ref{lambda3}),
 we have neglected the effect of
the compactification, but it is clear from the discussion above
that we could in principle introduce this
effect into the non-perturbative RG framework.
We leave this program to future work.

Recently,
 Dienes {\em et al.} \cite{dienes1} have suggested
a method to study 
low-energy physics of 
a Kaluza-Klein 
theory, and
their method is formulated
within a framework of  $D=4$ dimensional, renormalizable
theory.
Next we would like to see whether
 their result in the $R\to \infty$ limit goes over to 
what one finds in the non-perturbative RG approach.
To this end, we assume that we can neglect
$\lambda_1,\lambda_3,\dots$ in the evolution
of $\lambda_2$
--the scalar quartic coupling.  Then eq.  (\ref{lambda2})
 can be written as
\be
\Lambda\frac{d \lambda_2 }{d \Lambda}
&=&  A_D ~\frac{9}{2}~ \lambda_2^2 ~\Lambda^{D-4}~.
\label{lambda21}
\ee
Now assuming that  $\delta=D-4$ dimensions are compactified
on a circle of a fixed radius $R$, we find from eq. (\ref{lambda21})
that the evolution equation of the dimensionless
quartic coupling $\lambda=(2\pi R)^\delta\lambda_2$ of the 
four dimensional theory becomes
\be
\Lambda\frac{d \lambda }{d \Lambda}&=&
 \frac{1}{8\pi^2} \frac{9}{2(1+\delta/2)} 
X_\delta~\lambda^2~(R\Lambda)^{\delta}~,~
X_\delta=\frac{\pi^{\delta/2}}{\Gamma(1+\delta/2)}~,
\label{lambda22}
\ee
where $X_\delta$ (the volume of
a $\delta$ dimensional sphere of radius one) has been
introduced in ref. \cite{dienes1}.
We will compare this result with the one which we  obtain
by using the method of ref. \cite{dienes1}.
As we will see, they differ from each other.
To clarify the origin of the discrepancy, we first would like to
follow
 the method of ref. \cite{dienes1} for the present scalar
theory, and derive the evolution equation of $\lambda$.
We will then give an argument why two results are different and
motivate how to obtain the agreement.

As in ref. \cite{dienes1} we assume 
that  $\delta=D-4$ dimensions are compactified
on a circle of a fixed radius $R$, where $x$ and $y$ stand for
the $4$ and $\delta$ dimensional coordinates, respectively.
The scalar field satisfying the periodic boundary condition
\be
\phi(x,y) &=&\phi(x,y+2\pi R)~
\ee
can be expanded as
\be
\phi(x,y) &=&
\sum_{n_1=-\infty}^{\infty}
\sum_{n_2=-\infty}^{\infty}\cdots 
\sum_{n_\delta=-\infty}^{\infty}~
\phi_n(x)\exp (i n\cdot y/R)~,
\ee
where $n=(n_1,n_2,\dots,n_\delta)$ with
$n_i \in {\bf Z}$, and $n\cdot y=\sum_{i=1}^{\delta} n_i y_i$.
The starting Lagrangian is
\be
{\cal L}_D &=&
\frac{1}{2}(\partial_M \phi~\partial_M \phi
+m_0^2 \phi^2)+\frac{\lambda}{8}\phi^4~,
~M=1,\dots,4,5,\dots,4+\delta~.
\ee
To define the four dimensional theory, we  rescale
the field and the coupling $\lambda$ as
\be
\phi_n(x) &\to & (2\pi R)^{-\delta/2}\phi_n(x)~,~
\lambda \to (2\pi R)^{\delta} \lambda~.
\ee
The (four dimensional) mass squared of the Kaluza-Klein
modes $\phi_n(x)$ is given by
\be
m_n^2 &=& m_0^2+\frac{n\cdot n}{R^2}~,
\ee
where $m_0$ is the mass of the zero mode $\phi_0$,
which we would like to neglect as has been done in ref. 
\cite{dienes1}. 
For energies  above $\mu_0=R^{-1}$, the Kaluza-Klein excitations
are observable, and we may expect that in the $R \to \infty$ limit the
theory behaves as a full $4+\delta$ dimensional   theory.

Following ref. \cite{dienes1}, we  compute
the one-loop correction $\Pi^{(4)}_D$ to the four point vertex function
with zero external momenta to obtain the one-loop correction
to the coupling $\lambda$.
We find 
\be
\Pi^{(4)}_D &=&
\frac{9}{2}~\lambda^2~\int \frac{d^4 p}{(2\pi)^4}~
\sum_{n_1=-\infty}^{\infty}
\sum_{n_2=-\infty}^{\infty}\cdots 
\sum_{n_\delta=-\infty}^{\infty}~\frac{1}{(p^2+m_n^2)^2}\\
&=&\frac{\lambda^2}{8\pi^2}\frac{9}{2}
\int_0^\infty \frac{dt}{t}~(\frac{1}{2})~
[~\vartheta_3(i t/\pi R^2)~]^{\delta}~, 
\ee
where $\vartheta_3$ is one of the Jacobi theta functions
\be
\vartheta_3 (\tau) &=&
\sum_{n=-\infty}^{\infty}\exp (i \pi n^2)~,
\ee
and we have used
\be
\frac{1}{A^2} &=& \int_0^\infty dt t~\exp (-A t)~,~
\int~\frac{d^4 p}{(2\pi)^4}~\exp (-t p^2) =
\frac{1}{16 \pi^2 t^2}~.
\ee
Note that the $t$ integral is  ultraviolet as well as infrared
divergent (Dim$[t]=-2$).
Dienes {\em et al.} \cite{dienes1} introduced an ultraviolet and
infrared cut off to define the $t$ integral:
\be
\int_{0}^{\infty} dt &\to&
\int^{r \mu_{0}^{-2}}_{r\Lambda^{-2}} dt~,~
r=\pi(X_\delta)^{-2/\delta}~,
\label{t-integral}
\ee
where $X_\delta$ is defined in (\ref{lambda22}), and $\mu_0=1/R$.
We emphasize that the factor $r$ cannot be obtained
 within the framework
of the four dimensional theory, and so the explicit expression
given in (\ref{t-integral}) comes from an outside
information, to which we will come later.
Assuming that $t/R^2 << 1$ so that 
${\cal \vartheta}(it/\pi R^2)$ may be approximated
as $R\sqrt{\pi/t}$,
we perform the $t$ integration to obtain
\be
\Pi^{(4)}_D &=&
\frac{\lambda^2}{8\pi^2}\frac{9 X_\delta}{2\delta}
\left(~(\frac{\Lambda}{\mu_0})^\delta
-1\right)~.
\ee
Then we compute the $\beta$-function for $\lambda$:
\be
\Lambda \frac{d \lambda}{d \Lambda}~ &=&
\frac{\lambda^2}{8\pi^2}\frac{9 X_\delta}{2}
(\frac{\Lambda}{\mu_0})^\delta~,~
\lambda^{-1}(\Lambda) =\lambda^{-1}(\Lambda_0)
-(~\Pi^{(4)}_D(\Lambda)-\Pi^{(4)}_D(\Lambda_0)~)~.
\label{lambda23}
\ee
Comparing this result with the evolution equation
(\ref{lambda22}), we now
see that they differ by a factor $(1+\delta/2)$.

This difference may be understood in the following
way.
In the treatment  of ref. \cite{dienes1}, one has to 
define the infrared and ultraviolet cut off, i.e.,
the factor $r$ appearing in the $t$ integral
in (\ref{t-integral}).
They fix $r$ by interpreting that
the correction to the $\beta$-function
(coming from the massive excitations)
 is proportional to the number of
Kaluza-Klein excitations with masses smaller than $\Lambda$.
This number is approximately proportional
to the volume of $\delta$ dimensional sphere
of radius $\Lambda R$, 
that is, $X_\delta (\Lambda/\mu_0)^\delta$.
This interpretation does not lead to
a $\beta$-function that in the large 
$R$ limit approaches the corresponding 
$\beta$-function of the full $D$ dimensional theory
as we have seen above.
In the full theory, we have a $D$ dimensional integral
in the momentum space in the form
\be
\lim_{\Lambda \to \infty}
\Lambda \frac{\partial}{\partial \Lambda}\int 
\frac{d^D q}{(2\pi)^D}~ K(q^2)
=A_D ~\lim_{\Lambda \to \infty}~
\Lambda \frac{\partial}{\partial \Lambda}
\int^{\Lambda} d |q| |q|^{D-1}K(|q|^2)~,
\ee
where $A_D$ (the $D$ dimensional angular integral) is defined
in (\ref{ad}) \footnote{
This is the reason that the all $\beta$-functions 
in the full theory are proportional to $A_D$ 
(see (\ref{lambda21})).
Besides, there are only one-loop corrections in the
non-perturbative RG approach we are adopting 
\cite{wegner,hasen1,morris2}.}, whereas
the interpretation of Dienes {\em et al.} \cite{dienes1}
would correspond to the expression
\be
\lim_{\Lambda \to \infty} ~
\int \frac{d^{\delta} k}{(2\pi)^{\delta}}
\Lambda \frac{\partial}{\partial \Lambda}
\int \frac{d^4 p}{(2\pi)^4} K(p^2)
&=&A_\delta~A_4
~\lim_{\Lambda \to \infty}~\int^{\Lambda}
d |k| |k|^{\delta-1} \nn\\
& &\times
\Lambda \frac{\partial}{\partial \Lambda}~
\int^{\Lambda} d |p| |p|^{3} K(|p|^2)~.
\ee
Assuming that the function $K(x)$ has the form
$K(x)=x^{2 N}$ where $N$ is some arbitrary number,
we find that the difference of the two integrals 
above is exactly
the factor $(1+\delta/2)$, the same factor that appears 
between the $\beta$-functions 
(\ref{lambda22}) and (\ref{lambda23}).
This means that if one would multiply $r$ defined in 
(\ref{t-integral})
with the factor $(1+\delta/2)^{2/\delta}$,
 one would get $\beta$-functions that
in the large radius limit go over to those obtained in the
non-perturbative RG approach. So we would like to suggest to
rescale the cut off factor $r$ in the $t$ integration 
(\ref{t-integral}) as
\be
r &\to &(1+\delta/2)^{2/\delta}~r~,
\ee
or equivalently to replace $X_\delta$ according to
\be
X_\delta &\to& Y_\delta=\frac{\pi^{\delta/2}}{\Gamma(2+\delta/2)}~,
\label{Ydelta}
\ee
where $X_\delta$ is defined in eq. (\ref{lambda22}).

\subsection{Extension to the soft supersymmetry-breaking  sector}
We now discuss the running of the SSB parameters. To
this end, we
follow the notation of
 ref. \cite{jack3} and consider first a generic
$N=1$
supersymmetric gauge theory with the superpotential
\be
W(\Phi) &= &\frac{1}{6} Y^{ijk} \Phi_i \Phi_j \Phi_k + 
\frac{1}{2} \mu^{ij} \Phi_i\Phi_j ~,
\ee
and with the SSB part $L_{SSB}$ given by
 \cite{yamada1}
\be
L_{SSB}s(\Phi,W) &=& - \left( ~\int d^2\theta\eta (  \frac{1}{6} 
 h^{ijk} \Phi_i \Phi_j \Phi_k +  \frac{1}{2}  b^{ij} \Phi_i \Phi_j 
+  \frac{1}{2}  MW_A^\alpha W_{A\alpha} )+
\mbox{h.c.}~\right)\nn\\
& &-\int d^4\theta\tilde{\eta} \eta \overline{\Phi^j}                   
(m^2)^i_j(e^{2gV})_i^k \Phi_k~,
\ee
where $\eta = \theta^2$, 
$\tilde{\eta} = \tilde{\theta}^2$ are the external
spurion superfields and $\theta$, $\tilde{\theta}$ 
are the usual Grassmann
parameters, and $M$ is the gaugino mass.
The $\beta$-functions of the $M, h$ and $m^2$
parameters  can be computed from  \cite{jack3,avdeev1}:
\be
\beta_M &=& 2{\cal O}\left({\beta_g\over g}\right)~,
\label{betaM}\\
\beta_b^{ij}&=&\gamma^i{}_lb^{lj}+\gamma^j{}_lb^{il}
-2\gamma_1^i{}_l\mu^{lj}
-2\gamma_1^j{}_l\mu^{il}~,
\label{betab}\\
\beta_h^{ijk}&=&\gamma^i{}_lh^{ljk}+\gamma^j{}_lh^{ilk}
+\gamma^k{}_lh^{ijl}-2\gamma_1^i{}_lY^{ljk}
-2\gamma_1^j{}_lY^{ilk}-2\gamma_1^k{}_lY^{ijl}~,
\label{betah}\\
(\beta_{m^2})^i{}_j &=&\left[ \Delta 
+ X \frac{\partial}{\partial g}\right]\gamma^i{}_j~,
\label{betam2}\\
{\cal O} &=&\left(Mg^2{\partial\over{\partial g^2}}
-h^{lmn}{\partial
\over{\partial Y^{lmn}}}\right)~,
\label{diffo}\\
\Delta &=& 2{\cal O}{\cal O}^* +2|M|^2 g^2{\partial
\over{\partial g^2}} +\tilde{Y}_{lmn}
{\partial\over{\partial
Y_{lmn}}} +\tilde{Y}^{lmn}{\partial\over{\partial Y^{lmn}}}~,
\label{delta}
\ee
where $(\gamma_1)^i{}_j={\cal O}\gamma^i{}_j$, 
$Y_{lmn} = (Y^{lmn})^*$, and 
$\tilde{Y}^{ijk}=
(m^2)^i{}_lY^{ljk}+(m^2)^j{}_lY^{ilk}+(m^2)^k{}_lY^{ijl}$.

The formulae for the $\beta$-functions of the SSB parameters
(\ref{betaM})--(\ref{betam2}) have been derived from the 
observation that in a class of renormalization schemes
the divergent parts of the SSB parameters
are simply related to those in the symmetric theory,
$Y^{ijk}$ and $\mu^{ij}$.
It is, however, not exactly known in which class of renormalization
schemes these formulae have their validity.
In fact, the quantity $X\sim O(g^3)$ in eq. (\ref{betam2})
is  explicitly computed only in  two-loop order \cite{jack6}
and depends on the renormalization scheme employed \footnote{
There exists an indirect method (which
is based on a RG invariance argument) to
fix the exact form of $X$ \cite{kkz1,jack4}
in the Novikov-Shifman-Vainstein-Zakharov \cite{novikov1}
renormalization
scheme.}.

Since however we are interested only 
in the one-loop approximation to the $\beta$-functions,
the problem mentioned above is irrelevant because
the divergent parts in one-loop order are independent
of renormalization scheme.
Moreover, as we can see from the calculation of the contribution
coming from the massive Kaluza-Klein
excitations to the $\beta$-functions, 
these excitations do not disturb
the relations among
the divergent parts of the SSB parameters and 
$Y^{ijk}$ and $\mu^{ij}$ at least in one-loop order.
Therefore, we can easily compute the one-loop 
$\beta$-functions of the SSB
parameters above $\mu_0$ by applying the eqs.
(\ref{betaM})--(\ref{betam2}) 
on the $\beta$-functions and anomalous dimensions
which contain the one-loop
contributions coming from the massive
Kaluza-Klein  excitations.

To be more specific we consider the 
model of ref. \cite{dienes1}, the MSSM 
with a certain Kaluza-Klein towers. As ref. \cite{dienes1} we assume
that only the gauge boson and Higgs supermultiplets of the MSSM
have the towers of Kaluza-Klein states and that
the lepton and quark supermultiplets are stuck at a fixed point
of an orbifold on which the $\delta$ dimensional internal space is
compactified so that they have no  towers of Kaluza-Klein states.
Under these assumptions, 
the one-loop $\beta$-functions of the gauge couplings
and the one-loop anomalous dimensions  above and below $\mu_0$ 
become \cite{dienes1}:
\be
(16 \pi^2)\beta_1&=&\left\{
\begin{array}{c} g_1^3~(6+\frac{6}{5}(Y_\delta/2)
(\frac{\Lambda}{\mu_0})^{\delta}) \\
\frac{33}{5}~g_1^3\end{array} \right.~\\
(16 \pi^2)\beta_2&=&\left\{
\begin{array}{c} 
g_2^3~(4-6(Y_\delta/2) (\frac{\Lambda}{\mu_0})^{\delta}) \\
g_2^3\end{array} \right.~,\\
(16 \pi^2)\beta_3&=&\left\{
\begin{array}{c}
g_3^3~(3-12(Y_\delta/2) (\frac{\Lambda}{\mu_0})^{\delta})\\
-3~g_3^3\end{array} \right.~,\\
(16 \pi^2)\gamma_{t_L}&=& \left\{
\begin{array}{c}
Y_\delta ~(\frac{\Lambda}{\mu_0})^{\delta}
(g_t^2+g_b^2-( \frac{1}{30}g_1^2+ 
\frac{3}{2}g_2^2+\frac{8}{3}g_3^2 ))\\
g_t^2+g_b^2-( \frac{1}{30}g_1^2+ 
\frac{3}{2}g_2^2+\frac{8}{3}g_3^2 )
\end{array} \right.~,\\
(16 \pi^2)\gamma_{t_R}&=& \left\{
\begin{array}{c}
Y_\delta ~(\frac{\Lambda}{\mu_0})^{\delta}
(2 g_t^2-( \frac{8}{15}g_1^2+\frac{8}{3}g_3^2 ))\\
2 g_t^2-( \frac{8}{15}g_1^2+\frac{8}{3}g_3^2 )
\end{array} \right.~,\\
(16 \pi^2)\gamma_{b_R}&=& \left\{
\begin{array}{c}
Y_\delta~ (\frac{\Lambda}{\mu_0})^{\delta}
(2 g_b^2-( \frac{2}{15}g_1^2+\frac{8}{3}g_3^2 ))\\
2 g_b^2-( \frac{2}{15}g_1^2+\frac{8}{3}g_3^2 )
\end{array} \right.~,\\
(16 \pi^2)\gamma_{\tau_L}&=& \left\{
\begin{array}{c}
Y_\delta ~(\frac{\Lambda}{\mu_0})^{\delta}
(g_\tau^2-( \frac{3}{10}g_1^2+\frac{3}{2}g_2^2 ))\\
g_\tau^2-( \frac{3}{10}g_1^2+\frac{3}{2}g_2^2 )
\end{array} \right.~,\\
(16 \pi^2)\gamma_{\tau_R}&=&  \left\{
\begin{array}{c}
Y_\delta ~(\frac{\Lambda}{\mu_0})^{\delta}
(2 g_\tau^2-\frac{6}{5}g_1^2)\\
2 g_\tau^2-\frac{6}{5}g_1^2
\end{array} \right.~,\\
(16 \pi^2)\gamma_{H_u}&=& \left\{
\begin{array}{c}
3g_t^2-( \frac{3}{10}g_1^2+\frac{3}{2}g_2^2 )\\
3g_t^2-( \frac{3}{10}g_1^2+\frac{3}{2}g_2^2 )
\end{array} \right.~,\\
(16 \pi^2)\gamma_{H_d}&=& \left\{
\begin{array}{c}
3g_b^2+g_\tau^2-( \frac{3}{10}g_1^2+\frac{3}{2}g_2^2 )\\
3g_b^2+g_\tau^2-( \frac{3}{10}g_1^2+\frac{3}{2}g_2^2 )
\end{array} \right.~,
\ee
where $g_{t,b,\tau}$ are the Yukawa couplings for
the top, bottom and tau, respectively, 
we have neglected the Yukawa couplings of the
first and second generations.
($Y_\delta$ is defined in 
(\ref{Ydelta}).)
Here  we have used the fact \cite{dienes1} that 
in the model of \cite{dienes1}, the contributions of the excited
Kaluza-Klein states to the anomalous dimensions of the matter
supermultiplets have the same form as the massless mode contribution,
and that those of the Higgs supermultiplets due to $N=2$ 
supersymmetry in the excited sector vanish.

The one-loop $\beta$-functions for the Yukawa couplings
can be computed from
\be
\beta_{ijk} &=& g_{ijk}~(~\gamma_i+\gamma_j+\gamma_k~)~,
\ee
and we find that for energies above $\mu_0$
\be
(16 \pi^2)\beta_t&=&
g_t~[3 g_t^2-\frac{3}{10}g_1^2-\frac{3}{2} g_2^2\nn\\
& &+(Y_\delta/2)~ (\frac{\Lambda}{\mu_0})^{\delta}
(6g_t^2+2g_b^2
-\frac{17}{15}g_1^2-3g_2^2-\frac{32}{3}g_3^2)]~,\\
(16 \pi^2)\beta_b&=&
g_b~[3 g_b^2+g_\tau^2-\frac{3}{10}g_1^2-\frac{3}{2} g_2^2\nn\\
& &+ (Y_\delta/2)~ (\frac{\Lambda}{\mu_0})^{\delta}
(2g_t^2+6g_b^2
-\frac{1}{3}g_1^2-3g_2^2-\frac{32}{3}g_3^2)]~,\\
(16 \pi^2)\beta_\tau&=&
g_\tau~[3 g_b^2+g_\tau^2-\frac{3}{10}g_1^2-\frac{3}{2}g_2^2\nn\\
& &+ (Y_\delta/2) ~(\frac{\Lambda}{\mu_0})^{\delta}
(6g_\tau^2
-3g_1^2-3g_2^2)]~.
\ee
The gaugino mass $\beta$-functions are:
\be
(16 \pi^2)\beta_{M_1}&=&M_1~
g_1^2~(6+\frac{6}{5}(Y_\delta/2) 
(\frac{\Lambda}{\mu_0})^{\delta})~,\\\
(16 \pi^2)\beta_{M_2}&=&M_2~
g_2^2~(4-6(Y_\delta/2) (\frac{\Lambda}{\mu_0})^{\delta})~,\\
(16 \pi^2)\beta_{M_3}&=& M_3~
g_3^2~(3-12(Y_\delta/2) (\frac{\Lambda}{\mu_0})^{\delta})~,
\ee
where we have used eq. (\ref{betaM}).
One of the consequences of (\ref{betaM}) is that in one-loop
order  the relation
\be
\frac{M_1}{g_1^2} &=& \frac{M_2}{g_2^2}=\frac{M_3}{g_3^2}
\label{Moverg}
\ee
holds above as well as below $\mu_0$.
The $\beta$-functions of the trilinear couplings  above $\mu_0$ are:
\be
(16 \pi^2)\beta_{h_t}&=&
h_t~[9 g_t^2-\frac{3}{10}g_1^2-\frac{3}{2} g_2^2]
+g_t~[\frac{3}{5}g_1^2 M_1+3 g_2^2 M_2]
\nn\\
& &+(Y_\delta/2) ~(\frac{\Lambda}{\mu_0})^{\delta}
\left(
h_t~[18 g_t^2+2 g_b^2-\frac{17}{15}g_1^2-3 g_2^2
-\frac{32}{3}g_3^2]\right.\\
& &\left.
+g_t~[4 g_b h_b +\frac{34}{15}g_1^2 M_1+6 g_2^2 M_2
+\frac{64}{3}g_3^2 M_3]\right)~,\nn\\
(16 \pi^2)\beta_{h_b}&=&
h_b~[9 g_b^2+g_\tau^2-\frac{3}{10}g_1^2-\frac{3}{2} g_2^2]
+g_b~[2g_\tau h_\tau+\frac{3}{5}g_1^2 M_1+3 g_2^2 M_2]
\nn\\
& &+(Y_\delta/2)~ (\frac{\Lambda}{\mu_0})^{\delta}
\left(
h_b~[18 g_b^2+2 g_t^2-\frac{1}{3}g_1^2-3 g_2^2
-\frac{32}{3}g_3^2]\right.\\
& &\left.
+g_b~[4 g_t h_t +\frac{2}{3}g_1^2 M_1+6 g_2^2 M_2
+\frac{64}{3}g_3^2 M_3]\right)~,\nn\\
(16 \pi^2)\beta_{h_\tau}&=&
h_\tau~[3 g_b^2+3 g_\tau^2-\frac{3}{10}g_1^2-\frac{3}{2} g_2^2]
+g_\tau~[6g_b h_b+\frac{3}{5}g_1^2 M_1+3 g_2^2 M_2]
\nn\\
& &+(Y_\delta/2)~ (\frac{\Lambda}{\mu_0})^{\delta}
\left(
h_\tau~[18 g_\tau^2-3g_1^2-3 g_2^2]\right.\\
& &\left.+g_\tau~[6g_1^2 M_1+6 g_2^2 M_2]\right)~,\nn
\ee
where we have used eqs. (\ref{betaM})--(\ref{betam2}).

The $\beta$-functions for $\mu_H$ and $B$ are the same
 as those below
$\mu_0$,
because the anomalous dimensions for the Higgs superfields
$\gamma_{H_u},\gamma_{H_d}$ are the same.  From the same reason, 
the $\beta$-functions for
$m_{H_u}^{2},m_{H_d}^{2}$ are the same
 as those below
$\mu_0$.
To obtain the $\beta$-functions for the soft squared masses
of the leptons and quarks above $\mu_0$, 
we simply have to multiply their $\beta$-functions
below $\mu_0$ with the factor
\be
Y_\delta ~(\frac{\Lambda}{\mu_0})^{\delta}~,
\ee
where $Y_\delta$ is defined in (\ref{Ydelta}).

\section{Predictions from unification}
In the previous section we have
derived the matching condition 
and extended  the method  of  ref. \cite{dienes1}
to include the SSB sector.
In this section we would like to apply this result
to a model which has been considered in
 ref. \cite{dienes1}.
This model is nothing but 
the MSSM with the Kaluza-Klein
towers which are present only in the gauge supermultiplets and Higgs
supermultiplets \footnote{The proton is 
stable in this model \cite{dienes1}.}, and we have given
the one-loop  RG functions of this model
in the previous
section. We however will 
restrict ourselves to the case with $\delta =1$,
because the main feature of our results will not drastically  change
for $\delta  > 1$.
To simplify the situation, we assume  through out
our analyses a uniform SUSY threshold
$M_S$ and that  $M_S=1$ TeV.
Unless we notice explicitly, we study the evolution
of the dimensionless parameters such as gauge couplings
below $\mu_0 =R^{-1}$  at the two-loop level, along
with the experimental values,
\begin{equation}
M_\tau=1.777 ~{\rm GeV}, \quad M_Z=91.188 ~{\rm GeV},
\end{equation}
\begin{equation}
\alpha^{-1}_{\rm EM}(M_Z) = 127.9 +{8 \over 9 \pi} 
\log {M_t \over M_Z}~,
\end{equation}
\begin{equation}
\sin ^2 \theta_W(M_Z) =0.2319 - 3.03\times 10^{-5} T-8.4 \times 
10^{-8}T^2~,
\end{equation}
where $T=M_t/[{\rm GeV}]-165$.
Here $M_\tau$ and $M_t$  are the physical tau and top quark
masses, where  we take $M_t=174.1$ GeV in our
 analyses \footnote{The value quoted by Particle Data Group
\cite{pdg} is: $173.8\pm 5.2$ GeV.}.
(See ref. \cite{kmoz} for more details of the method
of our analyses.)
The evolution
of  all the parameters 
above $\mu_0 =R^{-1}$  as well as
the evolution of the SSB parameters
for the whole range of the energy scale will be
studied at the one-loop level.

In the following discussions we change our notation
for the GUT scale: $M_{\rm GUT} \to M_X$ while we use 
$M_Y$ for $M_{\rm GUT}$ when considering the level
of $U(1)_Y$ as free.

\subsection{Gauge coupling unification}

We begin by considering the unification of the gauge couplings
of the model.
Fig. 1 shows a representative
 example of the running  of the gauge couplings
$\alpha_a=g_a^2/4\pi$ ($a=1,2,3$) 
for $\mu_0=10^{11}$ GeV where we have assumed  $\alpha_3(M_Z)=0.117$.
Also shown is  the running  of the Yukawa couplings 
$g_i^2/4\pi$ ($i=t, b, \tau$) with $\tan \beta = 50$.
The initial value $g_b^2/4\pi (M_Z) $ has been calculated from
$m_b(M_Z)=3.4$ GeV, where $m_b(M_Z)$ is the $\overline{\mbox{MS}}$ 
bottom quark mass at $M_Z$ and we have not included
the MSSM superpartner correction (the so-called SUSY correction) to the bottom
mass.

\begin{center}
\input flow2.tex

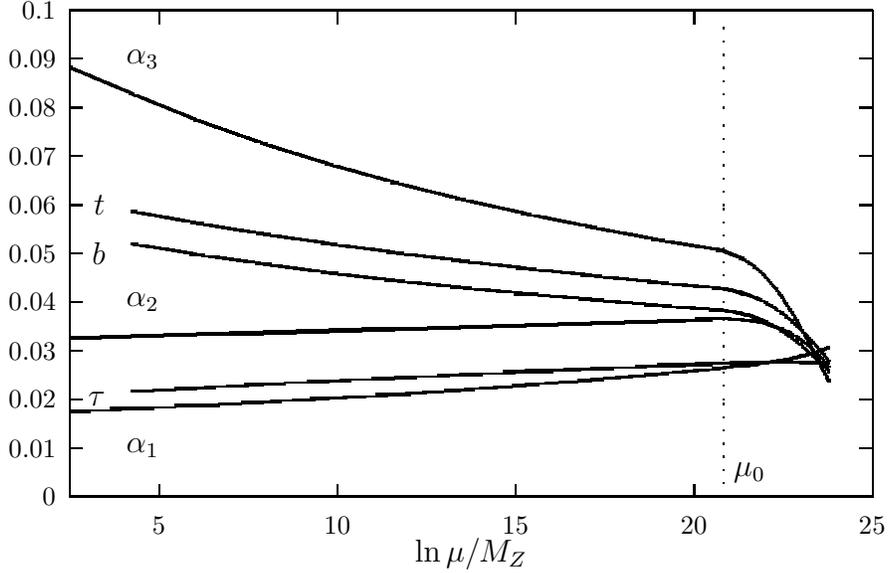
Fig. 1: Running of the gauge and Yukawa couplings for $\tan \beta =50$ 
and $\mu_0=10^{11}$ GeV.

\end{center}

Now we impose the  unification on the gauge couplings with
the conventional  level of
$U(1)_Y$, i.e.,  $k_Y=5/3$. Calculating the unification
scale $M_X$ and the unified coupling  $\alpha(M_X)$  from
the input data $\alpha_1(M_Z)$ and $\alpha_2(M_Z)$, we 
predict $\alpha_3(M_Z)$ as usual. Solid lines in Figs. 2, 3 and 4 show the
predicted values of $\alpha_3(M_Z)$,  
the unification scale $M_X$ and
the unified coupling $\alpha(M_X)$,  respectively.
The predicted value of $\alpha_3(M_Z)$ increases
as $\mu_0$ decreases, which has been observed
also in ref. \cite{ross1}.
Comparing this with the experimental value \cite{pdg}
\be
\alpha_3(M_Z) &=& 0.119\pm 0.002~,
\ee
we see that a lower $\mu_0$ might have a problem with the experimental
observation.
Of course, the threshold effects at $M_X$ will be very important 
to give  more precise values for $\alpha_3(M_Z)$.
But these effects cannot be estimated unless we fix a model above 
$M_X$, which is outside of the scope of the present paper.

\begin{center}
\input alpha3.tex

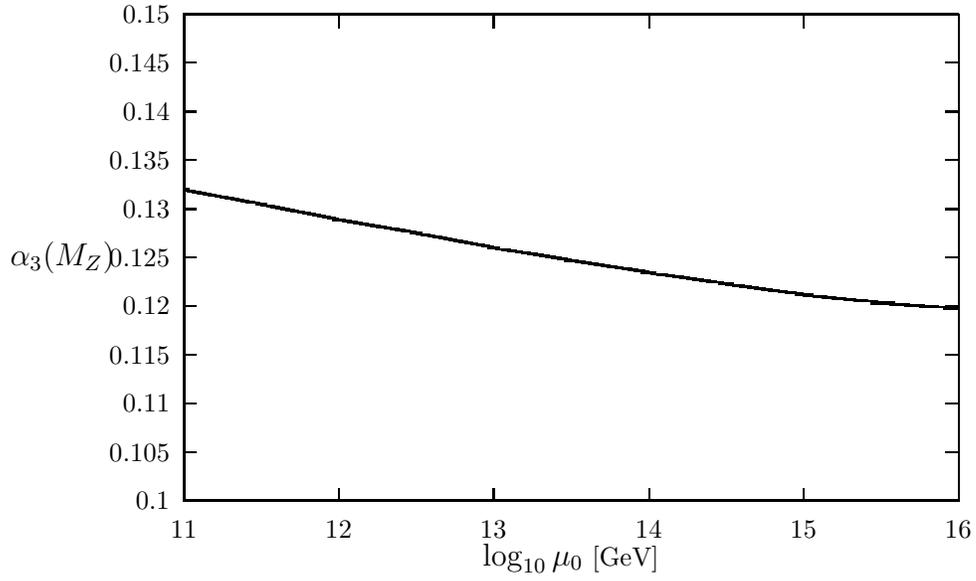
Fig. 2: Prediction of $\alpha_3(M_Z)$
\end{center}

\begin{center}
\input gut1.tex

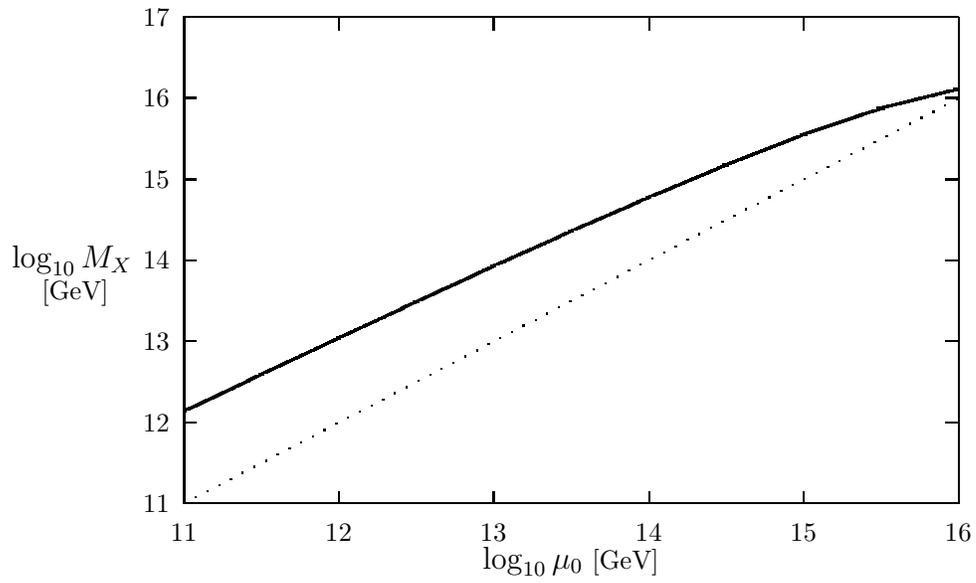
Fig. 3: The unification scale $M_X$
\end{center}

\begin{center}
\input alphax.tex

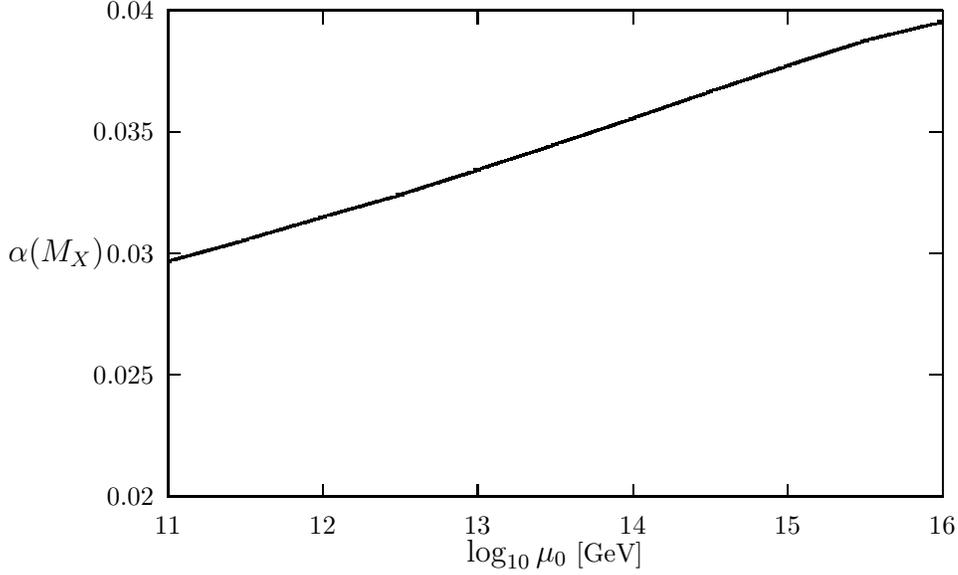
Fig. 4: The unified coupling $\alpha(M_X)$
\end{center}

The level of $U(1)_Y$ denoted by $k_Y$, which
can differ from the conventional value
 $5/3 \simeq 1.67$ in 
the framework of string unification \cite{level,benakli1},
could  also be responsible for the uncertainty in the
prediction of $\alpha_3(M_Z)$.
In this case, it is more appropriate 
to calculate
 the unification scale $M_Y$ and the
unified  coupling $\alpha(M_Y)$ from
 $\alpha_2(M_Z)$ and
$\alpha_3(M_Z)$ and then to predict $\alpha_1(M_Z)$.
Then from the ratio $\alpha_1(M_Z)/\alpha_Y(M_Z)$, 
we can obtain the level $k_Y$, which is shown in
Fig. 5, where the doted line correspond to the conventional level
$5/3$. 
The lower (upper) line in Fig. 5 
corresponds to $\alpha^{-1}_3(M_Z)=8.0$ (9.0).
Fig. 6 shows $M_Y$ for the given value of $\alpha_3(M_Z)=0.117$.
To obtain Fig. 5, we
 have used only the one-loop RG equations,
because we are interested in the qualitative change only.
As we can see from Fig. 5,
different values of $\alpha_3(M_Z)$ do not lead to 
a large difference
in  $k_Y$.
It is certainly an interesting approach to regard the level
 $k_Y$ as a free parameter in performing 
the analyses that will follow.  But we fix $k_Y$ at $5/3$ and use 
$M_X$ and $\alpha_3(M_Z)$ as well  as $\alpha(M_X)$ obtained 
in Figs. 2, 3 and 4 in the  following discussions.

\begin{center}
\input level.tex

Fig. 5: The level of $U(1)_Y$ $k_Y$

\end{center}

\begin{center}
\input gut2.tex

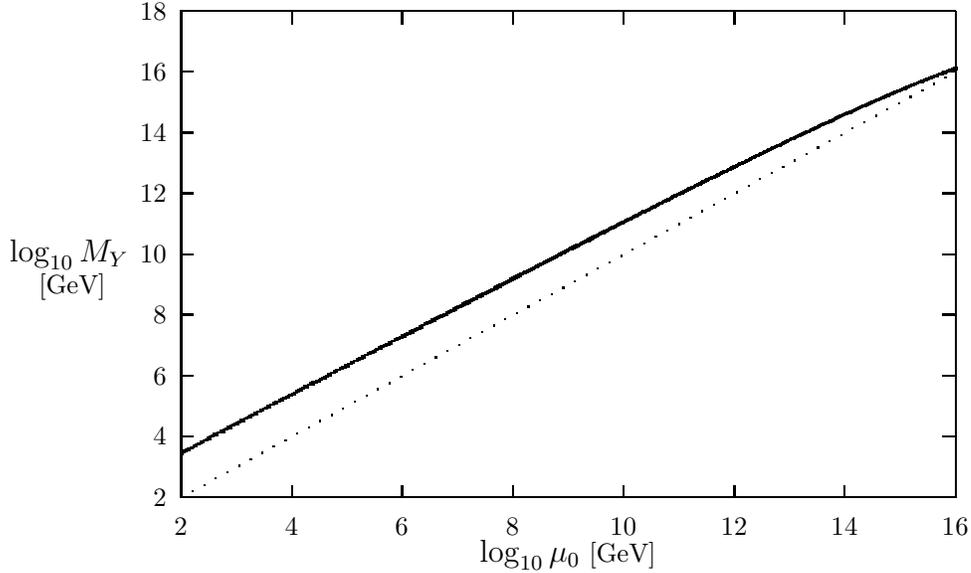
Fig. 6: The unification scale $M_Y$
\end{center}

\subsection{$b-\tau$ unification}

The $b-\tau$ Yukawa coupling unification is one of 
the important aspects 
in GUTs based on a gauge group like $SU(5)$ or $SO(10)$.
Under the assumption of
 the $b-\tau$ Yukawa unification at $M_X$,
the mass of the bottom quark becomes calculable, and
Fig. 7 shows the predicted value of the
$\overline{\mbox{MS}}$
mass $m_b(M_Z)$,
where (as before) we have not included the SUSY correction to
$m_b(M_Z)$. 
The upper and lower lines
correspond to $\tan \beta =2$ and 50,  respectively.
We have treated $\tan \beta$ as an independent parameter here,
although for certain GUTs like a $SO(10)$ GUT it
is no longer a free parameter.
As we can see from Fig. 7,
the predicted value for small $\tan \beta$
increases as $\mu_0$ decreases, while
it is relatively 
stable against the change of 
$\mu_0$ in most of the region for 
$\tan\beta =50$. For example, we have the predicted bottom mass
$m_b(M_Z)=3.4$ GeV  for $\mu_0=10^{11}$ GeV and $\tan \beta =50$.

\begin{center}
\input btau250.tex

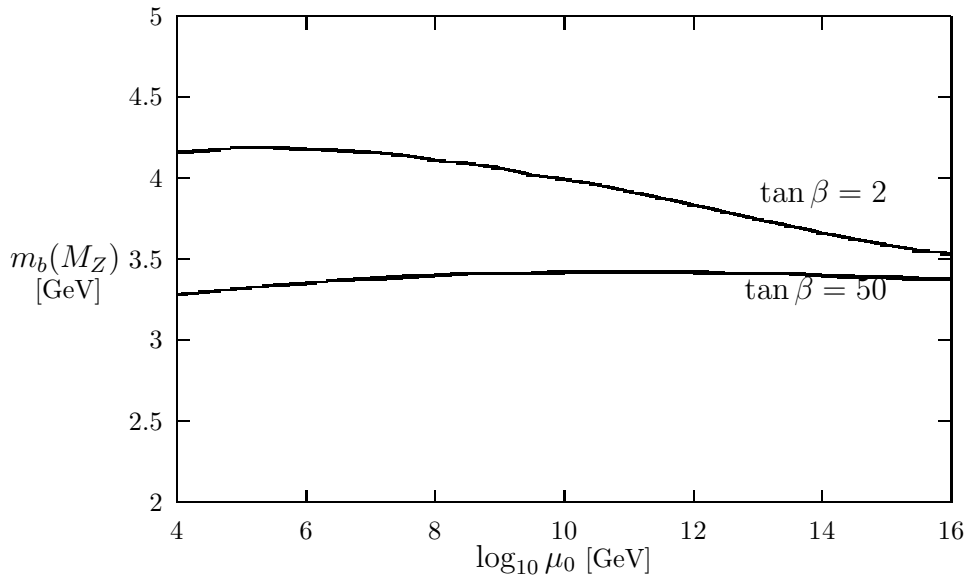
Fig. 7: The bottom mass $m_b(M_Z)$ under the $b-\tau$ Yukawa unification  
\end{center}

The present experimental value of the bottom mass 
contains large uncertainties:
Ref. \cite{bmass1},  for instance, gives
\begin{equation}
m_b(M_Z)= 2.67 \pm 0.50 \ {\rm GeV}~,
\label{mbex1}
\end{equation}
while the analysis of the $\Upsilon$ system \cite{bmass2} 
and the lattice result \cite{bmass3} give 
$m_b(m_b) = 4.13 \pm 0.06$ GeV 
and $4.15 \pm 0.20$ GeV, respectively, \footnote{See also 
ref. \cite{bmass4}.} which translate into
\begin{equation}
m_b(M_Z) = 2.8 \pm 0.2 \ {\rm GeV}.
\label{mbex2}
\end{equation}
It is known that the bottom mass can
receive a sizable SUSY correction 
in the large $\tan \beta$ scenario \cite{hall}.
For $\tan \beta =50$  it could amount to $O(20-30)\%$ \cite{COPW}
and its sign depends on the sign of $\mu_H$.
That is, large-radius compactifications prefer large 
$\tan \beta$.
Of course, this SUSY correction depends on the details of the
SUSY-mass spectrum, and  so   we leave the discussion on  it  to
future work.

\subsection{SSB sector and radiative breaking of $SU(2)\times U(1)$}

Here we use our result on the $\beta$-functions on the
SSB parameters in the previous and
 calculate the evolution of the SSB parameters.
For simplicity, we consider the universal SSB parameters at $M_X$, i.e. 
the universal gaugino masses $M_a(M_X)=M_0$,  
the universal soft scalar masses $m_i(M_X)=m_0$ and  
the universal $A$-parameters $A_i(M_X) \equiv h_i/g_i = A_0$.
On top of that, we assume that the relation
\be
A_0=-M_0
\ee
is satisfied
at the unification scale $M_X$, which 
is motivated  in the framework
of certain superstring theories as well as in  the RG invariance 
consideration \cite{BIM}-\cite{kkmz}.

Solid lines in Fig. 8 show gaugino masses at $M_S=1$ TeV, where
we have taken $M_0$=1 TeV at $M_X$.
All the gaugino masses increase as $\mu_0$ decreases, 
in accord with
the relation (\ref{Moverg}), i.e.,
$M_a/\alpha_a=M_0/\alpha(M_X)$, as well as 
with the $\mu_0$-dependence of $\alpha(M_X)$
which is shown in Fig. 4.
It should be noted that the gaugino masses in Fig. 8 are computed
by using the fixed level $k_Y=5/3$.
If we would regard the level as a free parameter and fix it
in such a way that all the gauge couplings
fit the experimental values at $M_Z$, we would find 
a change
\be
M_3(M_Z) \simeq 5.3 \mbox{TeV} & \to & 4.3 \mbox{TeV}~,~
M_1(M_Z) \simeq 0.83 \mbox{TeV}  \to  0.78 \mbox{TeV}
\ee
for $\mu_0=10$ TeV, for instance,
where $M_2 (M_Z)$ remains unchanged.
So, the net difference
compared with the case of the MSSM, is 
\be
\frac{M_{1}^{(k_Y)} }{M_{1}^{ (5/3)}} &\simeq & \frac{k_Y}{5/3}
\ee
for a fixed value of $M_2/\alpha_2  (M_Z)=M_3/\alpha_3  (M_Z)$,
where  $1.0  \gsim k_Y/(5/3) \gsim 0.9$ as we can see from Fig. 5.

\begin{center}
\input gaugino.tex

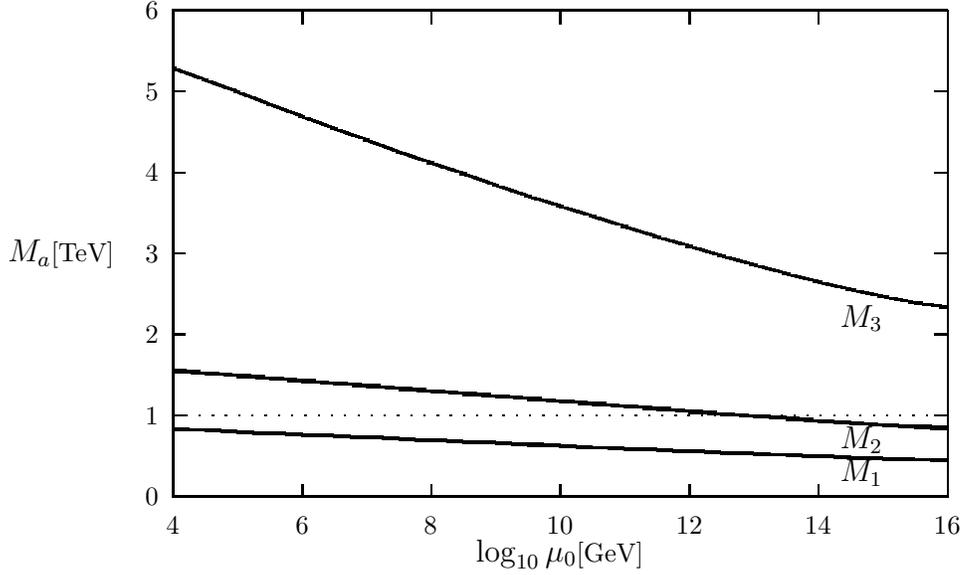
Fig. 8: Gaugino masses
\end{center}

Solid lines in Fig.9  show sfermion masses at the SUSY scale $M_S$.
In this figure, $\widetilde Q$, $\widetilde L$ and $\widetilde E$ stand
for
sfermions of the quark doublet, the lepton doublet and the lepton 
singlet for the first two families, for which we have neglected
the contribution of the Yukawa couplings to their
evolution.
We have taken $\tan \beta =2$, $M_0=1$ TeV and $m_0=0.8$ TeV.
We expect that the $\mu_0$-dependence of 
the squarks masses is large, because the
gaugino masses dominantly contribute to the evolution
of the squark masses of the first two generations.
In fact, the $\mu_0$-dependence of the gaugino masses shown
in Fig. 8 is reflected in  
 that as $\mu_0$ decreases,  the squarks 
become much heavier than
the  sleptons, as we can see in Fig.9 .
Similarly, we can calculate the sfermion masses of the third family.
These masses will be shown later after the discussion of radiative 
electroweak symmetry breaking.

\begin{center}
\input sfermi.tex

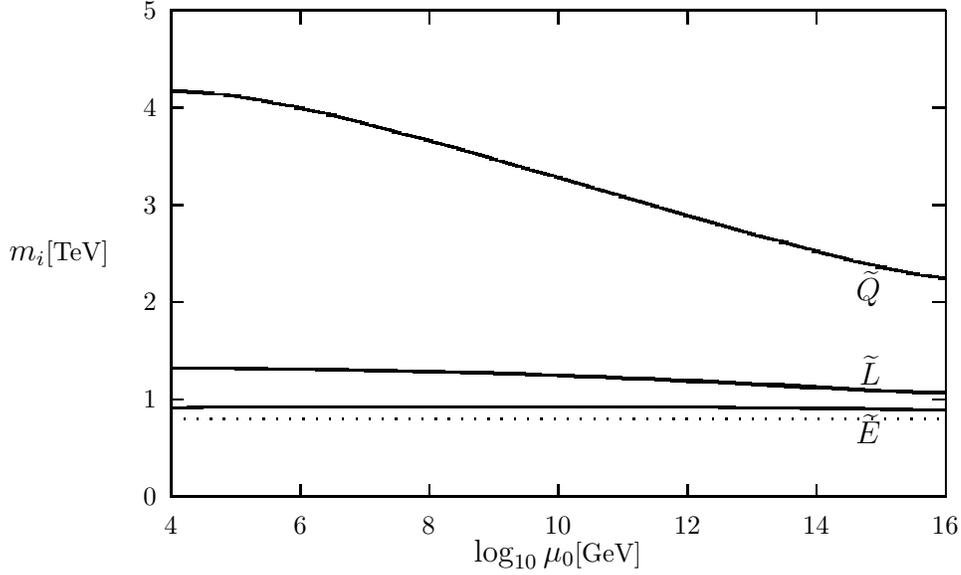
Fig.9: Sfermion masses
\end{center}

Now we come to discuss the radiative electroweak symmetry breaking.
We fix the values of $\mu_H$ and $B$ by using the
  two minimization conditions of the Higgs potential at the
weak scale, \begin{eqnarray}
&~& m_1^2+m_2^2 = -{2 \mu_H B \over \sin 2 \beta}, 
\label{mini1}\\
&~& m_1^2-m_2^2 = -\cos 2 \beta ~(M_Z^2+m_1^2+m_2^2), 
\label{mini2}
\end{eqnarray}
where $m_{1,2}^2=m_{H_d,H_u}^2+\mu_H^2$. 
For the desired electroweak 
symmetry breaking to occur, the condition 
\begin{eqnarray}
m_1^2 m_2^2 < |\mu_H B|^2
\label{SB}
\end{eqnarray}
should be satisfied, while
the bounded-from-below condition along 
the $D$-flat direction in the Higgs potential requires 
\begin{eqnarray}
m_1^2+m_2^2 >2 ~|\mu_H B|.
\label{BFB}
\end{eqnarray}

As we have noticed at the end of the previous section,
the $\beta$-functions for $\mu_H~,~B~,~
m_{H_u}^{2}$ and $m_{H_d}^{2} $ below 
$\mu_0$ do not change when passing the Kaluza-Klein threshold
$\mu_0$. We therefore expect that the existence of a 
large compactification  radius $R$  in the present model
has,
 through the other parameters
whose $\beta$-functions change at the threshold,
only an indirect influence
on the evolution
of these parameters.
We have in fact found that there exists 
 a wide region  in the parameter space $(M_0,m_0)$
 leading to the desired electroweak breaking .
For example,
the electroweak symmetry breaking always occurs 
for the region,
$m_0 \leq O(M_0)$, $\mu_0=10^5 \sim 10^{16}$ GeV and 
$\tan \beta =2 \sim 50$.

We have found that
the stau (mass)$^2$  becomes easily negative
in the large $\tan \beta$ scenario.
Similarly,  the stau mass becomes easily smaller than the lightest 
neutralino mass, in particular 
in the large $\tan \beta$ scenario.
In such a case the LSP is the stau which is electrically charged
and should have been observed 
if the $R$ parity is
 not violated\footnote{ In refs.\cite{kkmz,kks}
the experimental constraints  have been considered 
in details 
for (finite) $SU(5)$ GUTs and $SO(10)$ GUTs.}.
Therefore, it is  important to compare the masses of the lightest
neutralino and the stau.
Figs. 10 and 11 show the lightest stop $\tilde t$, sbottom $\tilde b$, 
stau $\tilde \tau$ and neutralino $\chi^0$ masses   
for $\tan \beta =2$ and 50, respectively.
We have taken $M_0=1$ TeV, $m_0=0.8$ TeV and $m_b(M_Z)=2.7$ GeV without 
assuming the $b-\tau$ Yukawa unification.
In the case considered above the stau becomes the LSP below
$\mu_0=10^{13.5}$ GeV if $\tan\beta=50$ (See Fig. 11).

\begin{center}
\input sfermi32.tex

Fig. 10: S-spectrum for $\tan \beta =2$
\end{center}

\begin{center}
\input sferm350.tex

Fig. 11: S-spectrum for $\tan \beta =50$
\end{center}

Figs. 12, 13 and 14 show the $m^2_{\tilde \tau}>0$ constraint  
for $\mu_0=10^{16}$, $10^{11}$ and $10^5$ GeV, where
we vary 
$\tan \beta$ from $\tan \beta =2$ to 50.
The parameter range in the $(M_0, m_0)$ space  shown in these figures 
always leads to a successful electroweak symmetry breaking.
The asterisks denote the region 
leading to a negative stau (mass)$^2$,
and the open squares stand for the region 
where the stau is lighter than
the  lightest neutralino.
These figures show a similar feature 
of the allowed range in the $(m_0,\tan\beta)$ space for 
a wide range of $\mu_0$.

\begin{center}
\input lsp13.tex

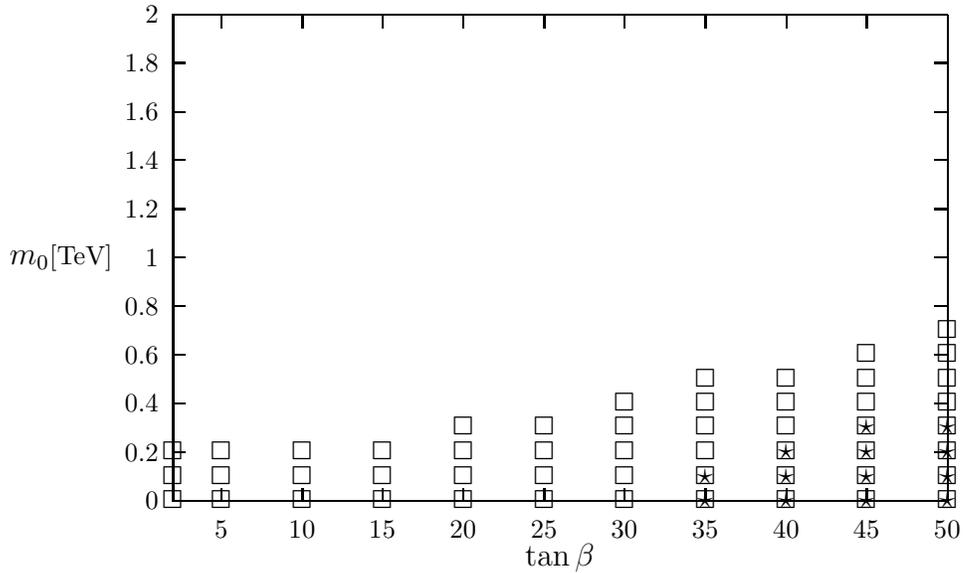
Fig. 12: The stau mass and the LSP for $\mu_0=10^{16}$ GeV
\end{center}

\begin{center}
\input lsp8.tex

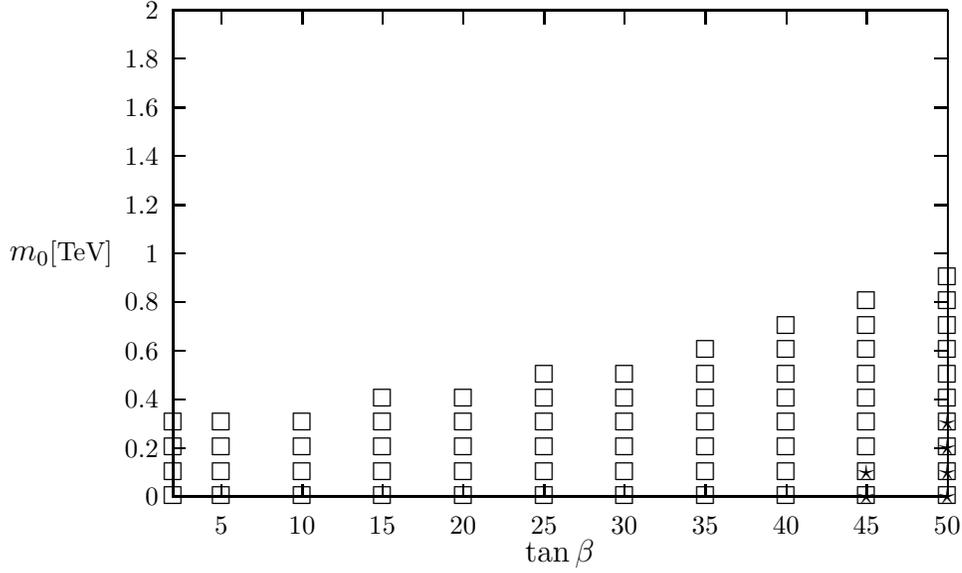
Fig. 13: The stau mass and the LSP for $\mu_0=10^{11}$ GeV
\end{center}
\begin{center}
\input lsp2.tex

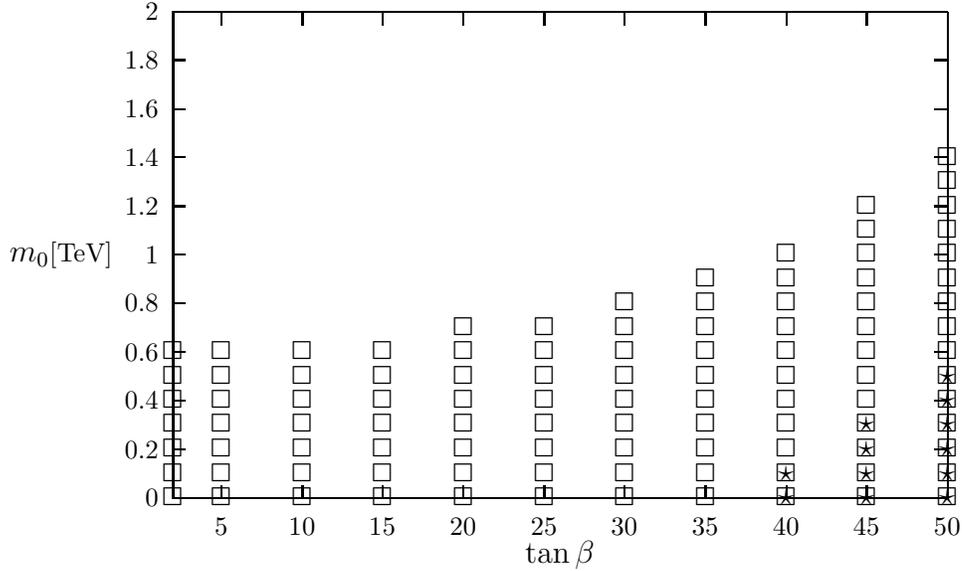
Fig. 14: The stau mass and the LSP for $\mu_0=10^{5}$ GeV
\end{center}

\section{Conclusion}

If one extrapolates the MSSM to higher dimensions so that 
the compactification scale is significantly smaller than
the ordinary SUSY-GUT scale $M_{\rm GUT} \simeq 2 \times 10^{16}$
GeV, the massive Kaluza-Klein states
modify the qualitative nature of the ordinary 
unification scenario 
\cite{antoniadis1}-\cite{hewett}. In this paper we have illustrated
how to apply the non-perturbative 
RG technique \cite{wilson1}-\cite{polchinski1} to
handle the problems associated with large-radius compactifications.
As an application we have
derived the matching condition 
between the effective, renormalizable and original, non-renormalizable
theories from the requirement that the $\beta$-functions calculated
in the effective theory go over 
in the large-compactification-radius limit to those calculated
using the non-perturbative RG technique with
 the assumption that the space-time dimensions 
are not compactified.
We have not  followed up this powerful RG technique further, 
but we would like to mention that 
the presence of higher-dimension operators, for instance, can be easily 
taken into account in this method.

Given the matching condition,  we have decided to stay 
in the renormalizable dimension, 
as Dienes {\em  et al}. \cite{dienes1}, because
if we ignore (provably higher order) effects
such as  those coming from higher-dimension operators,
their method is simple and convenient to calculate
the contributions of the massive Kaluza-Klein states
to the RG evolution of couplings.
Furthermore, as long as we want to stay in the one-loop approximation,
we can simply extend the treatment, due to the
recent development on renormalization
of the SSB parameters \cite{yamada1}-\cite{avdeev1},  to include
the SSB sector: We have computed the one-loop $\beta$-functions
of the SSB parameters in the MSSM which is extrapolated to
$D=4+\delta$ dimensions under the assumption that
the Kaluza-Klein towers exist only in the gauge supermultiplets
and Higgs supermultiplets \cite{dienes1}.

We have  addressed ourselves to various 
phenomenological issues in the model.
We first have confirmed the observation of Ghilencea and Ross that
the value of $\alpha_3(M_Z)$ predicted from the gauge coupling
unification increases as $\mu_0$ decreases, and could easily
exceed the experimental value $0.119 \pm 0.002$
if no other corrections such as the 
threshold corrections at $M_{\rm GUT}$ are taken into account.
In contrast to this, the $b-\tau$ unification
can be obtained in  a relatively wide range of  $\mu_0$
for a large $\tan\beta$, and we have found that
large-radius compactifications prefer large 
$\tan \beta$.
Another finding is that the relation
among the ratios
$M_i/g_i^2$ (\ref{Moverg}) holds in one-loop 
order in the model so that
if one takes account the level of $U(1)_Y$ appropriately,
 the model
with small $\mu_0$ differs
 from the conventional MSSM in
this sector of the SSB parameters.
We have also found that
the $\mu_0$-dependence of the sfermion masses, especially
those of the squarks, is quite large.

{}From our analyses in this paper, we would like to conclude
that there exist a certain chance to experimentally discriminate
the model from the MSSM even 
if the massive Kaluza-Klein states
are so heavy that they are 
not accessible in future collider experiments.

\vspace{1cm}
\noindent
{\bf Acknowledgment}

\vspace{0.3cm}
\noindent
Two of us (J.K and G.Z) would like to thank the Theory Group
of the Max-Planck-Institute for Physics, Munich and 
the Humboldt University, Berlin
for their hospitality,
where part of this work has been done.
J.K would like to thank H. Terao for useful discussions.

\end{document}

%% file: flow2.tex
\setlength{\unitlength}{0.240900pt}
\ifx\plotpoint\undefined\newsavebox{\plotpoint}\fi
\sbox{\plotpoint}{\rule[-0.200pt]{0.400pt}{0.400pt}}%
\begin{picture}(1500,900)(0,0)
\font\gnuplot=cmr10 at 10pt
\gnuplot
\sbox{\plotpoint}{\rule[-0.200pt]{0.400pt}{0.400pt}}%
\put(176.0,113.0){\rule[-0.200pt]{303.534pt}{0.400pt}}
\put(176.0,113.0){\rule[-0.200pt]{4.818pt}{0.400pt}}
\put(154,113){\makebox(0,0)[r]{0}}
\put(1416.0,113.0){\rule[-0.200pt]{4.818pt}{0.400pt}}
\put(176.0,189.0){\rule[-0.200pt]{4.818pt}{0.400pt}}
\put(154,189){\makebox(0,0)[r]{0.01}}
\put(1416.0,189.0){\rule[-0.200pt]{4.818pt}{0.400pt}}
\put(176.0,266.0){\rule[-0.200pt]{4.818pt}{0.400pt}}
\put(154,266){\makebox(0,0)[r]{0.02}}
\put(1416.0,266.0){\rule[-0.200pt]{4.818pt}{0.400pt}}
\put(176.0,342.0){\rule[-0.200pt]{4.818pt}{0.400pt}}
\put(154,342){\makebox(0,0)[r]{0.03}}
\put(1416.0,342.0){\rule[-0.200pt]{4.818pt}{0.400pt}}
\put(176.0,419.0){\rule[-0.200pt]{4.818pt}{0.400pt}}
\put(154,419){\makebox(0,0)[r]{0.04}}
\put(1416.0,419.0){\rule[-0.200pt]{4.818pt}{0.400pt}}
\put(176.0,495.0){\rule[-0.200pt]{4.818pt}{0.400pt}}
\put(154,495){\makebox(0,0)[r]{0.05}}
\put(1416.0,495.0){\rule[-0.200pt]{4.818pt}{0.400pt}}
\put(176.0,571.0){\rule[-0.200pt]{4.818pt}{0.400pt}}
\put(154,571){\makebox(0,0)[r]{0.06}}
\put(1416.0,571.0){\rule[-0.200pt]{4.818pt}{0.400pt}}
\put(176.0,648.0){\rule[-0.200pt]{4.818pt}{0.400pt}}
\put(154,648){\makebox(0,0)[r]{0.07}}
\put(1416.0,648.0){\rule[-0.200pt]{4.818pt}{0.400pt}}
\put(176.0,724.0){\rule[-0.200pt]{4.818pt}{0.400pt}}
\put(154,724){\makebox(0,0)[r]{0.08}}
\put(1416.0,724.0){\rule[-0.200pt]{4.818pt}{0.400pt}}
\put(176.0,801.0){\rule[-0.200pt]{4.818pt}{0.400pt}}
\put(154,801){\makebox(0,0)[r]{0.09}}
\put(1416.0,801.0){\rule[-0.200pt]{4.818pt}{0.400pt}}
\put(176.0,877.0){\rule[-0.200pt]{4.818pt}{0.400pt}}
\put(154,877){\makebox(0,0)[r]{0.1}}
\put(1416.0,877.0){\rule[-0.200pt]{4.818pt}{0.400pt}}
\put(316.0,113.0){\rule[-0.200pt]{0.400pt}{4.818pt}}
\put(316,68){\makebox(0,0){5}}
\put(316.0,857.0){\rule[-0.200pt]{0.400pt}{4.818pt}}
\put(596.0,113.0){\rule[-0.200pt]{0.400pt}{4.818pt}}
\put(596,68){\makebox(0,0){10}}
\put(596.0,857.0){\rule[-0.200pt]{0.400pt}{4.818pt}}
\put(876.0,113.0){\rule[-0.200pt]{0.400pt}{4.818pt}}
\put(876,68){\makebox(0,0){15}}
\put(876.0,857.0){\rule[-0.200pt]{0.400pt}{4.818pt}}
\put(1156.0,113.0){\rule[-0.200pt]{0.400pt}{4.818pt}}
\put(1156,68){\makebox(0,0){20}}
\put(1156.0,857.0){\rule[-0.200pt]{0.400pt}{4.818pt}}
\put(1436.0,113.0){\rule[-0.200pt]{0.400pt}{4.818pt}}
\put(1436,68){\makebox(0,0){25}}
\put(1436.0,857.0){\rule[-0.200pt]{0.400pt}{4.818pt}}
\put(176.0,113.0){\rule[-0.200pt]{303.534pt}{0.400pt}}
\put(1436.0,113.0){\rule[-0.200pt]{0.400pt}{184.048pt}}
\put(176.0,877.0){\rule[-0.200pt]{303.534pt}{0.400pt}}
\put(806,23){\makebox(0,0){$\ln \mu /M_Z$}}
\put(316,801){\makebox(0,0)[r]{$\alpha_3$}}
\put(316,419){\makebox(0,0)[r]{$\alpha_2$}}
\put(316,189){\makebox(0,0)[r]{$\alpha_1$}}
\put(232,571){\makebox(0,0)[r]{$t$}}
\put(232,495){\makebox(0,0)[r]{$b$}}
\put(232,266){\makebox(0,0)[r]{$\tau$}}
\put(1268,151){\makebox(0,0)[r]{$\mu_0$}}
\put(176.0,113.0){\rule[-0.200pt]{0.400pt}{184.048pt}}
\sbox{\plotpoint}{\rule[-0.400pt]{0.800pt}{0.800pt}}%
\multiput(176.00,247.38)(15.872,0.560){3}{\rule{15.720pt}{0.135pt}}
\multiput(176.00,244.34)(64.372,5.000){2}{\rule{7.860pt}{0.800pt}}
\multiput(273.00,252.38)(16.880,0.560){3}{\rule{16.680pt}{0.135pt}}
\multiput(273.00,249.34)(68.380,5.000){2}{\rule{8.340pt}{0.800pt}}
\multiput(376.00,257.38)(17.048,0.560){3}{\rule{16.840pt}{0.135pt}}
\multiput(376.00,254.34)(69.048,5.000){2}{\rule{8.420pt}{0.800pt}}
\multiput(480.00,262.39)(11.290,0.536){5}{\rule{13.933pt}{0.129pt}}
\multiput(480.00,259.34)(74.081,6.000){2}{\rule{6.967pt}{0.800pt}}
\multiput(583.00,268.39)(11.290,0.536){5}{\rule{13.933pt}{0.129pt}}
\multiput(583.00,265.34)(74.081,6.000){2}{\rule{6.967pt}{0.800pt}}
\multiput(686.00,274.40)(8.884,0.526){7}{\rule{11.971pt}{0.127pt}}
\multiput(686.00,271.34)(78.153,7.000){2}{\rule{5.986pt}{0.800pt}}
\multiput(789.00,281.40)(7.427,0.520){9}{\rule{10.500pt}{0.125pt}}
\multiput(789.00,278.34)(81.207,8.000){2}{\rule{5.250pt}{0.800pt}}
\multiput(892.00,289.40)(7.427,0.520){9}{\rule{10.500pt}{0.125pt}}
\multiput(892.00,286.34)(81.207,8.000){2}{\rule{5.250pt}{0.800pt}}
\multiput(995.00,297.40)(6.480,0.516){11}{\rule{9.444pt}{0.124pt}}
\multiput(995.00,294.34)(84.398,9.000){2}{\rule{4.722pt}{0.800pt}}
\multiput(1099.00,306.40)(5.664,0.514){13}{\rule{8.440pt}{0.124pt}}
\multiput(1099.00,303.34)(85.482,10.000){2}{\rule{4.220pt}{0.800pt}}
\put(1202,313.84){\rule{1.927pt}{0.800pt}}
\multiput(1202.00,313.34)(4.000,1.000){2}{\rule{0.964pt}{0.800pt}}
\put(1210,314.84){\rule{1.927pt}{0.800pt}}
\multiput(1210.00,314.34)(4.000,1.000){2}{\rule{0.964pt}{0.800pt}}
\put(1227,315.84){\rule{1.927pt}{0.800pt}}
\multiput(1227.00,315.34)(4.000,1.000){2}{\rule{0.964pt}{0.800pt}}
\put(1235,316.84){\rule{2.168pt}{0.800pt}}
\multiput(1235.00,316.34)(4.500,1.000){2}{\rule{1.084pt}{0.800pt}}
\put(1244,317.84){\rule{1.927pt}{0.800pt}}
\multiput(1244.00,317.34)(4.000,1.000){2}{\rule{0.964pt}{0.800pt}}
\put(1252,319.34){\rule{1.927pt}{0.800pt}}
\multiput(1252.00,318.34)(4.000,2.000){2}{\rule{0.964pt}{0.800pt}}
\put(1260,320.84){\rule{2.168pt}{0.800pt}}
\multiput(1260.00,320.34)(4.500,1.000){2}{\rule{1.084pt}{0.800pt}}
\put(1269,321.84){\rule{1.927pt}{0.800pt}}
\multiput(1269.00,321.34)(4.000,1.000){2}{\rule{0.964pt}{0.800pt}}
\put(1277,322.84){\rule{2.168pt}{0.800pt}}
\multiput(1277.00,322.34)(4.500,1.000){2}{\rule{1.084pt}{0.800pt}}
\put(1286,324.34){\rule{1.927pt}{0.800pt}}
\multiput(1286.00,323.34)(4.000,2.000){2}{\rule{0.964pt}{0.800pt}}
\put(1294,325.84){\rule{1.927pt}{0.800pt}}
\multiput(1294.00,325.34)(4.000,1.000){2}{\rule{0.964pt}{0.800pt}}
\put(1302,327.34){\rule{2.168pt}{0.800pt}}
\multiput(1302.00,326.34)(4.500,2.000){2}{\rule{1.084pt}{0.800pt}}
\put(1311,329.34){\rule{1.927pt}{0.800pt}}
\multiput(1311.00,328.34)(4.000,2.000){2}{\rule{0.964pt}{0.800pt}}
\put(1319,331.34){\rule{1.927pt}{0.800pt}}
\multiput(1319.00,330.34)(4.000,2.000){2}{\rule{0.964pt}{0.800pt}}
\put(1327,333.34){\rule{2.168pt}{0.800pt}}
\multiput(1327.00,332.34)(4.500,2.000){2}{\rule{1.084pt}{0.800pt}}
\put(1336,335.34){\rule{1.927pt}{0.800pt}}
\multiput(1336.00,334.34)(4.000,2.000){2}{\rule{0.964pt}{0.800pt}}
\put(1344,337.84){\rule{2.168pt}{0.800pt}}
\multiput(1344.00,336.34)(4.500,3.000){2}{\rule{1.084pt}{0.800pt}}
\put(1353,340.84){\rule{1.927pt}{0.800pt}}
\multiput(1353.00,339.34)(4.000,3.000){2}{\rule{0.964pt}{0.800pt}}
\put(1361,343.84){\rule{1.927pt}{0.800pt}}
\multiput(1361.00,342.34)(4.000,3.000){2}{\rule{0.964pt}{0.800pt}}
\put(1218.0,317.0){\rule[-0.400pt]{2.168pt}{0.800pt}}
\put(176,360.84){\rule{23.367pt}{0.800pt}}
\multiput(176.00,359.34)(48.500,3.000){2}{\rule{11.684pt}{0.800pt}}
\put(273,363.84){\rule{24.813pt}{0.800pt}}
\multiput(273.00,362.34)(51.500,3.000){2}{\rule{12.406pt}{0.800pt}}
\put(376,366.84){\rule{25.054pt}{0.800pt}}
\multiput(376.00,365.34)(52.000,3.000){2}{\rule{12.527pt}{0.800pt}}
\put(480,369.84){\rule{24.813pt}{0.800pt}}
\multiput(480.00,368.34)(51.500,3.000){2}{\rule{12.406pt}{0.800pt}}
\put(583,372.84){\rule{24.813pt}{0.800pt}}
\multiput(583.00,371.34)(51.500,3.000){2}{\rule{12.406pt}{0.800pt}}
\put(686,375.84){\rule{24.813pt}{0.800pt}}
\multiput(686.00,374.34)(51.500,3.000){2}{\rule{12.406pt}{0.800pt}}
\put(789,378.84){\rule{24.813pt}{0.800pt}}
\multiput(789.00,377.34)(51.500,3.000){2}{\rule{12.406pt}{0.800pt}}
\put(892,381.84){\rule{24.813pt}{0.800pt}}
\multiput(892.00,380.34)(51.500,3.000){2}{\rule{12.406pt}{0.800pt}}
\put(995,384.84){\rule{25.054pt}{0.800pt}}
\multiput(995.00,383.34)(52.000,3.000){2}{\rule{12.527pt}{0.800pt}}
\put(1099,388.34){\rule{20.800pt}{0.800pt}}
\multiput(1099.00,386.34)(59.829,4.000){2}{\rule{10.400pt}{0.800pt}}
\put(1210,389.84){\rule{1.927pt}{0.800pt}}
\multiput(1210.00,390.34)(4.000,-1.000){2}{\rule{0.964pt}{0.800pt}}
\put(1202.0,392.0){\rule[-0.400pt]{1.927pt}{0.800pt}}
\put(1227,388.84){\rule{1.927pt}{0.800pt}}
\multiput(1227.00,389.34)(4.000,-1.000){2}{\rule{0.964pt}{0.800pt}}
\put(1235,387.84){\rule{2.168pt}{0.800pt}}
\multiput(1235.00,388.34)(4.500,-1.000){2}{\rule{1.084pt}{0.800pt}}
\put(1244,386.84){\rule{1.927pt}{0.800pt}}
\multiput(1244.00,387.34)(4.000,-1.000){2}{\rule{0.964pt}{0.800pt}}
\put(1252,385.84){\rule{1.927pt}{0.800pt}}
\multiput(1252.00,386.34)(4.000,-1.000){2}{\rule{0.964pt}{0.800pt}}
\put(1260,384.34){\rule{2.168pt}{0.800pt}}
\multiput(1260.00,385.34)(4.500,-2.000){2}{\rule{1.084pt}{0.800pt}}
\put(1269,381.84){\rule{1.927pt}{0.800pt}}
\multiput(1269.00,383.34)(4.000,-3.000){2}{\rule{0.964pt}{0.800pt}}
\put(1277,378.84){\rule{2.168pt}{0.800pt}}
\multiput(1277.00,380.34)(4.500,-3.000){2}{\rule{1.084pt}{0.800pt}}
\put(1286,375.84){\rule{1.927pt}{0.800pt}}
\multiput(1286.00,377.34)(4.000,-3.000){2}{\rule{0.964pt}{0.800pt}}
\put(1294,372.34){\rule{1.800pt}{0.800pt}}
\multiput(1294.00,374.34)(4.264,-4.000){2}{\rule{0.900pt}{0.800pt}}
\put(1302,368.34){\rule{2.000pt}{0.800pt}}
\multiput(1302.00,370.34)(4.849,-4.000){2}{\rule{1.000pt}{0.800pt}}
\multiput(1311.00,366.07)(0.685,-0.536){5}{\rule{1.267pt}{0.129pt}}
\multiput(1311.00,366.34)(5.371,-6.000){2}{\rule{0.633pt}{0.800pt}}
\multiput(1319.00,360.06)(0.928,-0.560){3}{\rule{1.480pt}{0.135pt}}
\multiput(1319.00,360.34)(4.928,-5.000){2}{\rule{0.740pt}{0.800pt}}
\multiput(1327.00,355.08)(0.650,-0.526){7}{\rule{1.229pt}{0.127pt}}
\multiput(1327.00,355.34)(6.450,-7.000){2}{\rule{0.614pt}{0.800pt}}
\multiput(1336.00,348.08)(0.562,-0.526){7}{\rule{1.114pt}{0.127pt}}
\multiput(1336.00,348.34)(5.687,-7.000){2}{\rule{0.557pt}{0.800pt}}
\multiput(1344.00,341.08)(0.485,-0.516){11}{\rule{1.000pt}{0.124pt}}
\multiput(1344.00,341.34)(6.924,-9.000){2}{\rule{0.500pt}{0.800pt}}
\multiput(1353.00,332.08)(0.481,-0.520){9}{\rule{1.000pt}{0.125pt}}
\multiput(1353.00,332.34)(5.924,-8.000){2}{\rule{0.500pt}{0.800pt}}
\multiput(1362.40,321.02)(0.520,-0.627){9}{\rule{0.125pt}{1.200pt}}
\multiput(1359.34,323.51)(8.000,-7.509){2}{\rule{0.800pt}{0.600pt}}
\put(1218.0,391.0){\rule[-0.400pt]{2.168pt}{0.800pt}}
\multiput(176.00,785.09)(1.193,-0.502){75}{\rule{2.093pt}{0.121pt}}
\multiput(176.00,785.34)(92.657,-41.000){2}{\rule{1.046pt}{0.800pt}}
\multiput(273.00,744.09)(1.237,-0.502){77}{\rule{2.162pt}{0.121pt}}
\multiput(273.00,744.34)(98.513,-42.000){2}{\rule{1.081pt}{0.800pt}}
\multiput(376.00,702.09)(1.422,-0.503){67}{\rule{2.449pt}{0.121pt}}
\multiput(376.00,702.34)(98.918,-37.000){2}{\rule{1.224pt}{0.800pt}}
\multiput(480.00,665.09)(1.634,-0.503){57}{\rule{2.775pt}{0.121pt}}
\multiput(480.00,665.34)(97.240,-32.000){2}{\rule{1.388pt}{0.800pt}}
\multiput(583.00,633.09)(1.807,-0.504){51}{\rule{3.041pt}{0.121pt}}
\multiput(583.00,633.34)(96.687,-29.000){2}{\rule{1.521pt}{0.800pt}}
\multiput(686.00,604.09)(2.023,-0.504){45}{\rule{3.369pt}{0.121pt}}
\multiput(686.00,604.34)(96.007,-26.000){2}{\rule{1.685pt}{0.800pt}}
\multiput(789.00,578.09)(2.297,-0.505){39}{\rule{3.783pt}{0.122pt}}
\multiput(789.00,578.34)(95.149,-23.000){2}{\rule{1.891pt}{0.800pt}}
\multiput(892.00,555.09)(2.526,-0.505){35}{\rule{4.124pt}{0.122pt}}
\multiput(892.00,555.34)(94.441,-21.000){2}{\rule{2.062pt}{0.800pt}}
\multiput(995.00,534.09)(2.684,-0.505){33}{\rule{4.360pt}{0.122pt}}
\multiput(995.00,534.34)(94.951,-20.000){2}{\rule{2.180pt}{0.800pt}}
\multiput(1099.00,514.09)(3.156,-0.507){27}{\rule{5.047pt}{0.122pt}}
\multiput(1099.00,514.34)(92.525,-17.000){2}{\rule{2.524pt}{0.800pt}}
\put(1202,495.84){\rule{1.927pt}{0.800pt}}
\multiput(1202.00,497.34)(4.000,-3.000){2}{\rule{0.964pt}{0.800pt}}
\put(1210,492.84){\rule{1.927pt}{0.800pt}}
\multiput(1210.00,494.34)(4.000,-3.000){2}{\rule{0.964pt}{0.800pt}}
\put(1218,489.34){\rule{2.000pt}{0.800pt}}
\multiput(1218.00,491.34)(4.849,-4.000){2}{\rule{1.000pt}{0.800pt}}
\put(1227,485.34){\rule{1.800pt}{0.800pt}}
\multiput(1227.00,487.34)(4.264,-4.000){2}{\rule{0.900pt}{0.800pt}}
\multiput(1235.00,483.07)(0.797,-0.536){5}{\rule{1.400pt}{0.129pt}}
\multiput(1235.00,483.34)(6.094,-6.000){2}{\rule{0.700pt}{0.800pt}}
\multiput(1244.00,477.07)(0.685,-0.536){5}{\rule{1.267pt}{0.129pt}}
\multiput(1244.00,477.34)(5.371,-6.000){2}{\rule{0.633pt}{0.800pt}}
\multiput(1252.00,471.08)(0.562,-0.526){7}{\rule{1.114pt}{0.127pt}}
\multiput(1252.00,471.34)(5.687,-7.000){2}{\rule{0.557pt}{0.800pt}}
\multiput(1260.00,464.08)(0.554,-0.520){9}{\rule{1.100pt}{0.125pt}}
\multiput(1260.00,464.34)(6.717,-8.000){2}{\rule{0.550pt}{0.800pt}}
\multiput(1270.40,453.02)(0.520,-0.627){9}{\rule{0.125pt}{1.200pt}}
\multiput(1267.34,455.51)(8.000,-7.509){2}{\rule{0.800pt}{0.600pt}}
\multiput(1278.40,443.48)(0.516,-0.548){11}{\rule{0.124pt}{1.089pt}}
\multiput(1275.34,445.74)(9.000,-7.740){2}{\rule{0.800pt}{0.544pt}}
\multiput(1287.40,432.60)(0.520,-0.700){9}{\rule{0.125pt}{1.300pt}}
\multiput(1284.34,435.30)(8.000,-8.302){2}{\rule{0.800pt}{0.650pt}}
\multiput(1295.40,420.77)(0.520,-0.847){9}{\rule{0.125pt}{1.500pt}}
\multiput(1292.34,423.89)(8.000,-9.887){2}{\rule{0.800pt}{0.750pt}}
\multiput(1303.40,408.37)(0.516,-0.737){11}{\rule{0.124pt}{1.356pt}}
\multiput(1300.34,411.19)(9.000,-10.186){2}{\rule{0.800pt}{0.678pt}}
\multiput(1312.40,394.36)(0.520,-0.920){9}{\rule{0.125pt}{1.600pt}}
\multiput(1309.34,397.68)(8.000,-10.679){2}{\rule{0.800pt}{0.800pt}}
\multiput(1320.40,379.94)(0.520,-0.993){9}{\rule{0.125pt}{1.700pt}}
\multiput(1317.34,383.47)(8.000,-11.472){2}{\rule{0.800pt}{0.850pt}}
\multiput(1328.40,365.63)(0.516,-0.863){11}{\rule{0.124pt}{1.533pt}}
\multiput(1325.34,368.82)(9.000,-11.817){2}{\rule{0.800pt}{0.767pt}}
\multiput(1337.40,349.53)(0.520,-1.066){9}{\rule{0.125pt}{1.800pt}}
\multiput(1334.34,353.26)(8.000,-12.264){2}{\rule{0.800pt}{0.900pt}}
\multiput(1345.40,334.27)(0.516,-0.927){11}{\rule{0.124pt}{1.622pt}}
\multiput(1342.34,337.63)(9.000,-12.633){2}{\rule{0.800pt}{0.811pt}}
\multiput(1354.40,317.11)(0.520,-1.139){9}{\rule{0.125pt}{1.900pt}}
\multiput(1351.34,321.06)(8.000,-13.056){2}{\rule{0.800pt}{0.950pt}}
\multiput(1362.40,300.94)(0.520,-0.993){9}{\rule{0.125pt}{1.700pt}}
\multiput(1359.34,304.47)(8.000,-11.472){2}{\rule{0.800pt}{0.850pt}}
\put(273,561){\usebox{\plotpoint}}
\multiput(273.00,559.09)(2.806,-0.506){31}{\rule{4.537pt}{0.122pt}}
\multiput(273.00,559.34)(93.584,-19.000){2}{\rule{2.268pt}{0.800pt}}
\multiput(376.00,540.09)(3.187,-0.507){27}{\rule{5.094pt}{0.122pt}}
\multiput(376.00,540.34)(93.427,-17.000){2}{\rule{2.547pt}{0.800pt}}
\multiput(480.00,523.09)(3.608,-0.508){23}{\rule{5.693pt}{0.122pt}}
\multiput(480.00,523.34)(91.183,-15.000){2}{\rule{2.847pt}{0.800pt}}
\multiput(583.00,508.09)(3.887,-0.509){21}{\rule{6.086pt}{0.123pt}}
\multiput(583.00,508.34)(90.369,-14.000){2}{\rule{3.043pt}{0.800pt}}
\multiput(686.00,494.08)(4.214,-0.509){19}{\rule{6.538pt}{0.123pt}}
\multiput(686.00,494.34)(89.429,-13.000){2}{\rule{3.269pt}{0.800pt}}
\multiput(789.00,481.08)(4.604,-0.511){17}{\rule{7.067pt}{0.123pt}}
\multiput(789.00,481.34)(88.333,-12.000){2}{\rule{3.533pt}{0.800pt}}
\multiput(892.00,469.08)(5.077,-0.512){15}{\rule{7.691pt}{0.123pt}}
\multiput(892.00,469.34)(87.037,-11.000){2}{\rule{3.845pt}{0.800pt}}
\multiput(995.00,458.08)(5.127,-0.512){15}{\rule{7.764pt}{0.123pt}}
\multiput(995.00,458.34)(87.886,-11.000){2}{\rule{3.882pt}{0.800pt}}
\multiput(1099.00,447.08)(5.664,-0.514){13}{\rule{8.440pt}{0.124pt}}
\multiput(1099.00,447.34)(85.482,-10.000){2}{\rule{4.220pt}{0.800pt}}
\put(1202,436.84){\rule{1.927pt}{0.800pt}}
\multiput(1202.00,437.34)(4.000,-1.000){2}{\rule{0.964pt}{0.800pt}}
\put(1210,435.34){\rule{1.927pt}{0.800pt}}
\multiput(1210.00,436.34)(4.000,-2.000){2}{\rule{0.964pt}{0.800pt}}
\put(1218,433.34){\rule{2.168pt}{0.800pt}}
\multiput(1218.00,434.34)(4.500,-2.000){2}{\rule{1.084pt}{0.800pt}}
\put(1227,430.84){\rule{1.927pt}{0.800pt}}
\multiput(1227.00,432.34)(4.000,-3.000){2}{\rule{0.964pt}{0.800pt}}
\put(1235,428.34){\rule{2.168pt}{0.800pt}}
\multiput(1235.00,429.34)(4.500,-2.000){2}{\rule{1.084pt}{0.800pt}}
\put(1244,425.34){\rule{1.800pt}{0.800pt}}
\multiput(1244.00,427.34)(4.264,-4.000){2}{\rule{0.900pt}{0.800pt}}
\put(1252,421.84){\rule{1.927pt}{0.800pt}}
\multiput(1252.00,423.34)(4.000,-3.000){2}{\rule{0.964pt}{0.800pt}}
\multiput(1260.00,420.06)(1.096,-0.560){3}{\rule{1.640pt}{0.135pt}}
\multiput(1260.00,420.34)(5.596,-5.000){2}{\rule{0.820pt}{0.800pt}}
\put(1269,413.34){\rule{1.800pt}{0.800pt}}
\multiput(1269.00,415.34)(4.264,-4.000){2}{\rule{0.900pt}{0.800pt}}
\multiput(1277.00,411.07)(0.797,-0.536){5}{\rule{1.400pt}{0.129pt}}
\multiput(1277.00,411.34)(6.094,-6.000){2}{\rule{0.700pt}{0.800pt}}
\multiput(1286.00,405.07)(0.685,-0.536){5}{\rule{1.267pt}{0.129pt}}
\multiput(1286.00,405.34)(5.371,-6.000){2}{\rule{0.633pt}{0.800pt}}
\multiput(1294.00,399.07)(0.685,-0.536){5}{\rule{1.267pt}{0.129pt}}
\multiput(1294.00,399.34)(5.371,-6.000){2}{\rule{0.633pt}{0.800pt}}
\multiput(1302.00,393.08)(0.554,-0.520){9}{\rule{1.100pt}{0.125pt}}
\multiput(1302.00,393.34)(6.717,-8.000){2}{\rule{0.550pt}{0.800pt}}
\multiput(1311.00,385.08)(0.562,-0.526){7}{\rule{1.114pt}{0.127pt}}
\multiput(1311.00,385.34)(5.687,-7.000){2}{\rule{0.557pt}{0.800pt}}
\multiput(1319.00,378.08)(0.481,-0.520){9}{\rule{1.000pt}{0.125pt}}
\multiput(1319.00,378.34)(5.924,-8.000){2}{\rule{0.500pt}{0.800pt}}
\multiput(1327.00,370.08)(0.485,-0.516){11}{\rule{1.000pt}{0.124pt}}
\multiput(1327.00,370.34)(6.924,-9.000){2}{\rule{0.500pt}{0.800pt}}
\multiput(1337.40,358.43)(0.520,-0.554){9}{\rule{0.125pt}{1.100pt}}
\multiput(1334.34,360.72)(8.000,-6.717){2}{\rule{0.800pt}{0.550pt}}
\multiput(1345.40,349.48)(0.516,-0.548){11}{\rule{0.124pt}{1.089pt}}
\multiput(1342.34,351.74)(9.000,-7.740){2}{\rule{0.800pt}{0.544pt}}
\multiput(1354.40,339.43)(0.520,-0.554){9}{\rule{0.125pt}{1.100pt}}
\multiput(1351.34,341.72)(8.000,-6.717){2}{\rule{0.800pt}{0.550pt}}
\multiput(1362.40,330.02)(0.520,-0.627){9}{\rule{0.125pt}{1.200pt}}
\multiput(1359.34,332.51)(8.000,-7.509){2}{\rule{0.800pt}{0.600pt}}
\put(273,510){\usebox{\plotpoint}}
\multiput(273.00,508.09)(3.156,-0.507){27}{\rule{5.047pt}{0.122pt}}
\multiput(273.00,508.34)(92.525,-17.000){2}{\rule{2.524pt}{0.800pt}}
\multiput(376.00,491.09)(3.643,-0.508){23}{\rule{5.747pt}{0.122pt}}
\multiput(376.00,491.34)(92.073,-15.000){2}{\rule{2.873pt}{0.800pt}}
\multiput(480.00,476.09)(3.887,-0.509){21}{\rule{6.086pt}{0.123pt}}
\multiput(480.00,476.34)(90.369,-14.000){2}{\rule{3.043pt}{0.800pt}}
\multiput(583.00,462.08)(4.604,-0.511){17}{\rule{7.067pt}{0.123pt}}
\multiput(583.00,462.34)(88.333,-12.000){2}{\rule{3.533pt}{0.800pt}}
\multiput(686.00,450.08)(5.077,-0.512){15}{\rule{7.691pt}{0.123pt}}
\multiput(686.00,450.34)(87.037,-11.000){2}{\rule{3.845pt}{0.800pt}}
\multiput(789.00,439.08)(5.664,-0.514){13}{\rule{8.440pt}{0.124pt}}
\multiput(789.00,439.34)(85.482,-10.000){2}{\rule{4.220pt}{0.800pt}}
\multiput(892.00,429.08)(6.416,-0.516){11}{\rule{9.356pt}{0.124pt}}
\multiput(892.00,429.34)(83.582,-9.000){2}{\rule{4.678pt}{0.800pt}}
\multiput(995.00,420.08)(6.480,-0.516){11}{\rule{9.444pt}{0.124pt}}
\multiput(995.00,420.34)(84.398,-9.000){2}{\rule{4.722pt}{0.800pt}}
\multiput(1099.00,411.08)(7.427,-0.520){9}{\rule{10.500pt}{0.125pt}}
\multiput(1099.00,411.34)(81.207,-8.000){2}{\rule{5.250pt}{0.800pt}}
\put(1202,402.84){\rule{1.927pt}{0.800pt}}
\multiput(1202.00,403.34)(4.000,-1.000){2}{\rule{0.964pt}{0.800pt}}
\put(1210,401.84){\rule{1.927pt}{0.800pt}}
\multiput(1210.00,402.34)(4.000,-1.000){2}{\rule{0.964pt}{0.800pt}}
\put(1218,400.34){\rule{2.168pt}{0.800pt}}
\multiput(1218.00,401.34)(4.500,-2.000){2}{\rule{1.084pt}{0.800pt}}
\put(1227,398.34){\rule{1.927pt}{0.800pt}}
\multiput(1227.00,399.34)(4.000,-2.000){2}{\rule{0.964pt}{0.800pt}}
\put(1235,396.34){\rule{2.168pt}{0.800pt}}
\multiput(1235.00,397.34)(4.500,-2.000){2}{\rule{1.084pt}{0.800pt}}
\put(1244,393.84){\rule{1.927pt}{0.800pt}}
\multiput(1244.00,395.34)(4.000,-3.000){2}{\rule{0.964pt}{0.800pt}}
\put(1252,390.84){\rule{1.927pt}{0.800pt}}
\multiput(1252.00,392.34)(4.000,-3.000){2}{\rule{0.964pt}{0.800pt}}
\put(1260,387.34){\rule{2.000pt}{0.800pt}}
\multiput(1260.00,389.34)(4.849,-4.000){2}{\rule{1.000pt}{0.800pt}}
\put(1269,383.34){\rule{1.800pt}{0.800pt}}
\multiput(1269.00,385.34)(4.264,-4.000){2}{\rule{0.900pt}{0.800pt}}
\multiput(1277.00,381.06)(1.096,-0.560){3}{\rule{1.640pt}{0.135pt}}
\multiput(1277.00,381.34)(5.596,-5.000){2}{\rule{0.820pt}{0.800pt}}
\multiput(1286.00,376.06)(0.928,-0.560){3}{\rule{1.480pt}{0.135pt}}
\multiput(1286.00,376.34)(4.928,-5.000){2}{\rule{0.740pt}{0.800pt}}
\multiput(1294.00,371.07)(0.685,-0.536){5}{\rule{1.267pt}{0.129pt}}
\multiput(1294.00,371.34)(5.371,-6.000){2}{\rule{0.633pt}{0.800pt}}
\multiput(1302.00,365.07)(0.797,-0.536){5}{\rule{1.400pt}{0.129pt}}
\multiput(1302.00,365.34)(6.094,-6.000){2}{\rule{0.700pt}{0.800pt}}
\multiput(1311.00,359.07)(0.685,-0.536){5}{\rule{1.267pt}{0.129pt}}
\multiput(1311.00,359.34)(5.371,-6.000){2}{\rule{0.633pt}{0.800pt}}
\multiput(1319.00,353.08)(0.562,-0.526){7}{\rule{1.114pt}{0.127pt}}
\multiput(1319.00,353.34)(5.687,-7.000){2}{\rule{0.557pt}{0.800pt}}
\multiput(1327.00,346.08)(0.554,-0.520){9}{\rule{1.100pt}{0.125pt}}
\multiput(1327.00,346.34)(6.717,-8.000){2}{\rule{0.550pt}{0.800pt}}
\multiput(1336.00,338.08)(0.562,-0.526){7}{\rule{1.114pt}{0.127pt}}
\multiput(1336.00,338.34)(5.687,-7.000){2}{\rule{0.557pt}{0.800pt}}
\multiput(1344.00,331.08)(0.554,-0.520){9}{\rule{1.100pt}{0.125pt}}
\multiput(1344.00,331.34)(6.717,-8.000){2}{\rule{0.550pt}{0.800pt}}
\multiput(1353.00,323.08)(0.481,-0.520){9}{\rule{1.000pt}{0.125pt}}
\multiput(1353.00,323.34)(5.924,-8.000){2}{\rule{0.500pt}{0.800pt}}
\multiput(1361.00,315.08)(0.481,-0.520){9}{\rule{1.000pt}{0.125pt}}
\multiput(1361.00,315.34)(5.924,-8.000){2}{\rule{0.500pt}{0.800pt}}
\put(273,278){\usebox{\plotpoint}}
\multiput(273.00,279.38)(16.880,0.560){3}{\rule{16.680pt}{0.135pt}}
\multiput(273.00,276.34)(68.380,5.000){2}{\rule{8.340pt}{0.800pt}}
\multiput(376.00,284.39)(11.402,0.536){5}{\rule{14.067pt}{0.129pt}}
\multiput(376.00,281.34)(74.804,6.000){2}{\rule{7.033pt}{0.800pt}}
\multiput(480.00,290.38)(16.880,0.560){3}{\rule{16.680pt}{0.135pt}}
\multiput(480.00,287.34)(68.380,5.000){2}{\rule{8.340pt}{0.800pt}}
\multiput(583.00,295.38)(16.880,0.560){3}{\rule{16.680pt}{0.135pt}}
\multiput(583.00,292.34)(68.380,5.000){2}{\rule{8.340pt}{0.800pt}}
\multiput(686.00,300.38)(16.880,0.560){3}{\rule{16.680pt}{0.135pt}}
\multiput(686.00,297.34)(68.380,5.000){2}{\rule{8.340pt}{0.800pt}}
\multiput(789.00,305.38)(16.880,0.560){3}{\rule{16.680pt}{0.135pt}}
\multiput(789.00,302.34)(68.380,5.000){2}{\rule{8.340pt}{0.800pt}}
\put(892,309.34){\rule{20.800pt}{0.800pt}}
\multiput(892.00,307.34)(59.829,4.000){2}{\rule{10.400pt}{0.800pt}}
\multiput(995.00,314.38)(17.048,0.560){3}{\rule{16.840pt}{0.135pt}}
\multiput(995.00,311.34)(69.048,5.000){2}{\rule{8.420pt}{0.800pt}}
\put(1099,318.34){\rule{20.800pt}{0.800pt}}
\multiput(1099.00,316.34)(59.829,4.000){2}{\rule{10.400pt}{0.800pt}}
\put(1210,320.84){\rule{1.927pt}{0.800pt}}
\multiput(1210.00,320.34)(4.000,1.000){2}{\rule{0.964pt}{0.800pt}}
\put(1202.0,322.0){\rule[-0.400pt]{1.927pt}{0.800pt}}
\put(1244,321.84){\rule{1.927pt}{0.800pt}}
\multiput(1244.00,321.34)(4.000,1.000){2}{\rule{0.964pt}{0.800pt}}
\put(1218.0,323.0){\rule[-0.400pt]{6.263pt}{0.800pt}}
\put(1319,321.84){\rule{1.927pt}{0.800pt}}
\multiput(1319.00,322.34)(4.000,-1.000){2}{\rule{0.964pt}{0.800pt}}
\put(1252.0,324.0){\rule[-0.400pt]{16.140pt}{0.800pt}}
\put(1353,320.84){\rule{1.927pt}{0.800pt}}
\multiput(1353.00,321.34)(4.000,-1.000){2}{\rule{0.964pt}{0.800pt}}
\put(1327.0,323.0){\rule[-0.400pt]{6.263pt}{0.800pt}}
\put(1361.0,322.0){\rule[-0.400pt]{1.927pt}{0.800pt}}
\sbox{\plotpoint}{\rule[-0.200pt]{0.400pt}{0.400pt}}%
\put(1202,113){\usebox{\plotpoint}}
\multiput(1202,113)(0.000,20.756){8}{\usebox{\plotpoint}}
\multiput(1202,266)(0.000,20.756){7}{\usebox{\plotpoint}}
\multiput(1202,419)(0.000,20.756){8}{\usebox{\plotpoint}}
\multiput(1202,571)(0.000,20.756){7}{\usebox{\plotpoint}}
\multiput(1202,724)(0.000,20.756){7}{\usebox{\plotpoint}}
\put(1202,877){\usebox{\plotpoint}}
\end{picture}

%% file: alpha3.tex
\setlength{\unitlength}{0.240900pt}
\ifx\plotpoint\undefined\newsavebox{\plotpoint}\fi
\sbox{\plotpoint}{\rule[-0.200pt]{0.400pt}{0.400pt}}%
\begin{picture}(1500,900)(0,0)
\font\gnuplot=cmr10 at 10pt
\gnuplot
\sbox{\plotpoint}{\rule[-0.200pt]{0.400pt}{0.400pt}}%
\put(220.0,113.0){\rule[-0.200pt]{4.818pt}{0.400pt}}
\put(198,113){\makebox(0,0)[r]{0.1}}
\put(1416.0,113.0){\rule[-0.200pt]{4.818pt}{0.400pt}}
\put(220.0,189.0){\rule[-0.200pt]{4.818pt}{0.400pt}}
\put(198,189){\makebox(0,0)[r]{0.105}}
\put(1416.0,189.0){\rule[-0.200pt]{4.818pt}{0.400pt}}
\put(220.0,266.0){\rule[-0.200pt]{4.818pt}{0.400pt}}
\put(198,266){\makebox(0,0)[r]{0.11}}
\put(1416.0,266.0){\rule[-0.200pt]{4.818pt}{0.400pt}}
\put(220.0,342.0){\rule[-0.200pt]{4.818pt}{0.400pt}}
\put(198,342){\makebox(0,0)[r]{0.115}}
\put(1416.0,342.0){\rule[-0.200pt]{4.818pt}{0.400pt}}
\put(220.0,419.0){\rule[-0.200pt]{4.818pt}{0.400pt}}
\put(198,419){\makebox(0,0)[r]{0.12}}
\put(1416.0,419.0){\rule[-0.200pt]{4.818pt}{0.400pt}}
\put(220.0,495.0){\rule[-0.200pt]{4.818pt}{0.400pt}}
\put(198,495){\makebox(0,0)[r]{0.125}}
\put(1416.0,495.0){\rule[-0.200pt]{4.818pt}{0.400pt}}
\put(220.0,571.0){\rule[-0.200pt]{4.818pt}{0.400pt}}
\put(198,571){\makebox(0,0)[r]{0.13}}
\put(1416.0,571.0){\rule[-0.200pt]{4.818pt}{0.400pt}}
\put(220.0,648.0){\rule[-0.200pt]{4.818pt}{0.400pt}}
\put(198,648){\makebox(0,0)[r]{0.135}}
\put(1416.0,648.0){\rule[-0.200pt]{4.818pt}{0.400pt}}
\put(220.0,724.0){\rule[-0.200pt]{4.818pt}{0.400pt}}
\put(198,724){\makebox(0,0)[r]{0.14}}
\put(1416.0,724.0){\rule[-0.200pt]{4.818pt}{0.400pt}}
\put(220.0,801.0){\rule[-0.200pt]{4.818pt}{0.400pt}}
\put(198,801){\makebox(0,0)[r]{0.145}}
\put(1416.0,801.0){\rule[-0.200pt]{4.818pt}{0.400pt}}
\put(220.0,877.0){\rule[-0.200pt]{4.818pt}{0.400pt}}
\put(198,877){\makebox(0,0)[r]{0.15}}
\put(1416.0,877.0){\rule[-0.200pt]{4.818pt}{0.400pt}}
\put(220.0,113.0){\rule[-0.200pt]{0.400pt}{4.818pt}}
\put(220,68){\makebox(0,0){11}}
\put(220.0,857.0){\rule[-0.200pt]{0.400pt}{4.818pt}}
\put(463.0,113.0){\rule[-0.200pt]{0.400pt}{4.818pt}}
\put(463,68){\makebox(0,0){12}}
\put(463.0,857.0){\rule[-0.200pt]{0.400pt}{4.818pt}}
\put(706.0,113.0){\rule[-0.200pt]{0.400pt}{4.818pt}}
\put(706,68){\makebox(0,0){13}}
\put(706.0,857.0){\rule[-0.200pt]{0.400pt}{4.818pt}}
\put(950.0,113.0){\rule[-0.200pt]{0.400pt}{4.818pt}}
\put(950,68){\makebox(0,0){14}}
\put(950.0,857.0){\rule[-0.200pt]{0.400pt}{4.818pt}}
\put(1193.0,113.0){\rule[-0.200pt]{0.400pt}{4.818pt}}
\put(1193,68){\makebox(0,0){15}}
\put(1193.0,857.0){\rule[-0.200pt]{0.400pt}{4.818pt}}
\put(1436.0,113.0){\rule[-0.200pt]{0.400pt}{4.818pt}}
\put(1436,68){\makebox(0,0){16}}
\put(1436.0,857.0){\rule[-0.200pt]{0.400pt}{4.818pt}}
\put(220.0,113.0){\rule[-0.200pt]{292.934pt}{0.400pt}}
\put(1436.0,113.0){\rule[-0.200pt]{0.400pt}{184.048pt}}
\put(220.0,877.0){\rule[-0.200pt]{292.934pt}{0.400pt}}
\put(30,495){\makebox(0,0){$\alpha_3(M_Z)$}}
\put(828,23){\makebox(0,0){$\log_{10} \mu_0$ [GeV]}}
\put(220.0,113.0){\rule[-0.200pt]{0.400pt}{184.048pt}}
\sbox{\plotpoint}{\rule[-0.400pt]{0.800pt}{0.800pt}}%
\put(220,601){\usebox{\plotpoint}}
\multiput(220.00,599.09)(2.725,-0.505){39}{\rule{4.443pt}{0.122pt}}
\multiput(220.00,599.34)(112.777,-23.000){2}{\rule{2.222pt}{0.800pt}}
\multiput(342.00,576.09)(2.586,-0.504){41}{\rule{4.233pt}{0.122pt}}
\multiput(342.00,576.34)(112.214,-24.000){2}{\rule{2.117pt}{0.800pt}}
\multiput(463.00,552.09)(2.996,-0.505){35}{\rule{4.848pt}{0.122pt}}
\multiput(463.00,552.34)(111.939,-21.000){2}{\rule{2.424pt}{0.800pt}}
\multiput(585.00,531.09)(2.703,-0.505){39}{\rule{4.409pt}{0.122pt}}
\multiput(585.00,531.34)(111.850,-23.000){2}{\rule{2.204pt}{0.800pt}}
\multiput(706.00,508.09)(3.154,-0.505){33}{\rule{5.080pt}{0.122pt}}
\multiput(706.00,508.34)(111.456,-20.000){2}{\rule{2.540pt}{0.800pt}}
\multiput(828.00,488.09)(3.328,-0.506){31}{\rule{5.337pt}{0.122pt}}
\multiput(828.00,488.34)(110.923,-19.000){2}{\rule{2.668pt}{0.800pt}}
\multiput(950.00,469.09)(3.495,-0.506){29}{\rule{5.578pt}{0.122pt}}
\multiput(950.00,469.34)(109.423,-18.000){2}{\rule{2.789pt}{0.800pt}}
\multiput(1071.00,451.09)(3.744,-0.507){27}{\rule{5.941pt}{0.122pt}}
\multiput(1071.00,451.34)(109.669,-17.000){2}{\rule{2.971pt}{0.800pt}}
\multiput(1193.00,434.08)(4.959,-0.509){19}{\rule{7.646pt}{0.123pt}}
\multiput(1193.00,434.34)(105.130,-13.000){2}{\rule{3.823pt}{0.800pt}}
\multiput(1314.00,421.08)(7.615,-0.516){11}{\rule{11.044pt}{0.124pt}}
\multiput(1314.00,421.34)(99.077,-9.000){2}{\rule{5.522pt}{0.800pt}}
\end{picture}

%% file: gut1.tex
\setlength{\unitlength}{0.240900pt}
\ifx\plotpoint\undefined\newsavebox{\plotpoint}\fi
\sbox{\plotpoint}{\rule[-0.200pt]{0.400pt}{0.400pt}}%
\begin{picture}(1500,900)(0,0)
\font\gnuplot=cmr10 at 10pt
\gnuplot
\sbox{\plotpoint}{\rule[-0.200pt]{0.400pt}{0.400pt}}%
\put(220.0,113.0){\rule[-0.200pt]{4.818pt}{0.400pt}}
\put(198,113){\makebox(0,0)[r]{11}}
\put(1416.0,113.0){\rule[-0.200pt]{4.818pt}{0.400pt}}
\put(220.0,240.0){\rule[-0.200pt]{4.818pt}{0.400pt}}
\put(198,240){\makebox(0,0)[r]{12}}
\put(1416.0,240.0){\rule[-0.200pt]{4.818pt}{0.400pt}}
\put(220.0,368.0){\rule[-0.200pt]{4.818pt}{0.400pt}}
\put(198,368){\makebox(0,0)[r]{13}}
\put(1416.0,368.0){\rule[-0.200pt]{4.818pt}{0.400pt}}
\put(220.0,495.0){\rule[-0.200pt]{4.818pt}{0.400pt}}
\put(198,495){\makebox(0,0)[r]{14}}
\put(1416.0,495.0){\rule[-0.200pt]{4.818pt}{0.400pt}}
\put(220.0,622.0){\rule[-0.200pt]{4.818pt}{0.400pt}}
\put(198,622){\makebox(0,0)[r]{15}}
\put(1416.0,622.0){\rule[-0.200pt]{4.818pt}{0.400pt}}
\put(220.0,750.0){\rule[-0.200pt]{4.818pt}{0.400pt}}
\put(198,750){\makebox(0,0)[r]{16}}
\put(1416.0,750.0){\rule[-0.200pt]{4.818pt}{0.400pt}}
\put(220.0,877.0){\rule[-0.200pt]{4.818pt}{0.400pt}}
\put(198,877){\makebox(0,0)[r]{17}}
\put(1416.0,877.0){\rule[-0.200pt]{4.818pt}{0.400pt}}
\put(220.0,113.0){\rule[-0.200pt]{0.400pt}{4.818pt}}
\put(220,68){\makebox(0,0){11}}
\put(220.0,857.0){\rule[-0.200pt]{0.400pt}{4.818pt}}
\put(463.0,113.0){\rule[-0.200pt]{0.400pt}{4.818pt}}
\put(463,68){\makebox(0,0){12}}
\put(463.0,857.0){\rule[-0.200pt]{0.400pt}{4.818pt}}
\put(706.0,113.0){\rule[-0.200pt]{0.400pt}{4.818pt}}
\put(706,68){\makebox(0,0){13}}
\put(706.0,857.0){\rule[-0.200pt]{0.400pt}{4.818pt}}
\put(950.0,113.0){\rule[-0.200pt]{0.400pt}{4.818pt}}
\put(950,68){\makebox(0,0){14}}
\put(950.0,857.0){\rule[-0.200pt]{0.400pt}{4.818pt}}
\put(1193.0,113.0){\rule[-0.200pt]{0.400pt}{4.818pt}}
\put(1193,68){\makebox(0,0){15}}
\put(1193.0,857.0){\rule[-0.200pt]{0.400pt}{4.818pt}}
\put(1436.0,113.0){\rule[-0.200pt]{0.400pt}{4.818pt}}
\put(1436,68){\makebox(0,0){16}}
\put(1436.0,857.0){\rule[-0.200pt]{0.400pt}{4.818pt}}
\put(220.0,113.0){\rule[-0.200pt]{292.934pt}{0.400pt}}
\put(1436.0,113.0){\rule[-0.200pt]{0.400pt}{184.048pt}}
\put(220.0,877.0){\rule[-0.200pt]{292.934pt}{0.400pt}}
\put(45,495){\makebox(0,0){$\log_{10} M_X$}}
\put(50,445){\makebox(0,0){[GeV]}}
\put(828,23){\makebox(0,0){$\log_{10} \mu_0$ [GeV]}}
\put(220.0,113.0){\rule[-0.200pt]{0.400pt}{184.048pt}}
\sbox{\plotpoint}{\rule[-0.400pt]{0.800pt}{0.800pt}}%
\put(220,258){\usebox{\plotpoint}}
\multiput(220.00,259.41)(1.057,0.502){109}{\rule{1.883pt}{0.121pt}}
\multiput(220.00,256.34)(118.092,58.000){2}{\rule{0.941pt}{0.800pt}}
\multiput(342.00,317.41)(1.067,0.502){107}{\rule{1.898pt}{0.121pt}}
\multiput(342.00,314.34)(117.060,57.000){2}{\rule{0.949pt}{0.800pt}}
\multiput(463.00,374.41)(1.076,0.502){107}{\rule{1.912pt}{0.121pt}}
\multiput(463.00,371.34)(118.031,57.000){2}{\rule{0.956pt}{0.800pt}}
\multiput(585.00,431.41)(1.086,0.502){105}{\rule{1.929pt}{0.121pt}}
\multiput(585.00,428.34)(116.997,56.000){2}{\rule{0.964pt}{0.800pt}}
\multiput(706.00,487.41)(1.116,0.502){103}{\rule{1.975pt}{0.121pt}}
\multiput(706.00,484.34)(117.902,55.000){2}{\rule{0.987pt}{0.800pt}}
\multiput(828.00,542.41)(1.158,0.502){99}{\rule{2.042pt}{0.121pt}}
\multiput(828.00,539.34)(117.763,53.000){2}{\rule{1.021pt}{0.800pt}}
\multiput(950.00,595.41)(1.195,0.502){95}{\rule{2.098pt}{0.121pt}}
\multiput(950.00,592.34)(116.645,51.000){2}{\rule{1.049pt}{0.800pt}}
\multiput(1071.00,646.41)(1.281,0.502){89}{\rule{2.233pt}{0.121pt}}
\multiput(1071.00,643.34)(117.365,48.000){2}{\rule{1.117pt}{0.800pt}}
\multiput(1193.00,694.41)(1.491,0.502){75}{\rule{2.561pt}{0.121pt}}
\multiput(1193.00,691.34)(115.685,41.000){2}{\rule{1.280pt}{0.800pt}}
\multiput(1314.00,735.41)(2.002,0.503){55}{\rule{3.348pt}{0.121pt}}
\multiput(1314.00,732.34)(115.050,31.000){2}{\rule{1.674pt}{0.800pt}}
\sbox{\plotpoint}{\rule[-0.200pt]{0.400pt}{0.400pt}}%
\put(220,113){\usebox{\plotpoint}}
\put(220.00,113.00){\usebox{\plotpoint}}
\put(238.46,122.48){\usebox{\plotpoint}}
\put(256.92,131.96){\usebox{\plotpoint}}
\multiput(257,132)(17.928,10.458){0}{\usebox{\plotpoint}}
\put(275.06,142.03){\usebox{\plotpoint}}
\put(293.43,151.69){\usebox{\plotpoint}}
\multiput(294,152)(18.564,9.282){0}{\usebox{\plotpoint}}
\put(311.98,160.99){\usebox{\plotpoint}}
\put(330.35,170.65){\usebox{\plotpoint}}
\multiput(331,171)(18.564,9.282){0}{\usebox{\plotpoint}}
\put(348.70,180.33){\usebox{\plotpoint}}
\multiput(355,184)(18.564,9.282){0}{\usebox{\plotpoint}}
\put(367.04,190.02){\usebox{\plotpoint}}
\put(385.40,199.70){\usebox{\plotpoint}}
\put(403.97,208.98){\usebox{\plotpoint}}
\multiput(404,209)(18.275,9.840){0}{\usebox{\plotpoint}}
\put(422.33,218.66){\usebox{\plotpoint}}
\put(440.48,228.70){\usebox{\plotpoint}}
\multiput(441,229)(18.564,9.282){0}{\usebox{\plotpoint}}
\put(458.93,238.20){\usebox{\plotpoint}}
\put(477.39,247.69){\usebox{\plotpoint}}
\multiput(478,248)(18.564,9.282){0}{\usebox{\plotpoint}}
\put(495.86,257.15){\usebox{\plotpoint}}
\put(514.31,266.65){\usebox{\plotpoint}}
\multiput(515,267)(17.928,10.458){0}{\usebox{\plotpoint}}
\put(532.45,276.72){\usebox{\plotpoint}}
\put(550.82,286.37){\usebox{\plotpoint}}
\multiput(552,287)(18.564,9.282){0}{\usebox{\plotpoint}}
\put(569.37,295.68){\usebox{\plotpoint}}
\put(587.53,305.72){\usebox{\plotpoint}}
\multiput(588,306)(18.845,8.698){0}{\usebox{\plotpoint}}
\put(606.09,314.97){\usebox{\plotpoint}}
\put(624.40,324.70){\usebox{\plotpoint}}
\multiput(625,325)(18.275,9.840){0}{\usebox{\plotpoint}}
\put(642.76,334.38){\usebox{\plotpoint}}
\put(660.94,344.38){\usebox{\plotpoint}}
\multiput(662,345)(18.564,9.282){0}{\usebox{\plotpoint}}
\put(679.55,353.56){\usebox{\plotpoint}}
\put(697.84,363.32){\usebox{\plotpoint}}
\multiput(699,364)(18.564,9.282){0}{\usebox{\plotpoint}}
\put(716.28,372.84){\usebox{\plotpoint}}
\put(734.72,382.36){\usebox{\plotpoint}}
\multiput(736,383)(17.928,10.458){0}{\usebox{\plotpoint}}
\put(752.86,392.43){\usebox{\plotpoint}}
\put(771.60,401.35){\usebox{\plotpoint}}
\multiput(773,402)(17.928,10.458){0}{\usebox{\plotpoint}}
\put(789.76,411.38){\usebox{\plotpoint}}
\put(808.14,421.00){\usebox{\plotpoint}}
\multiput(810,422)(18.564,9.282){0}{\usebox{\plotpoint}}
\put(826.52,430.64){\usebox{\plotpoint}}
\put(844.82,440.41){\usebox{\plotpoint}}
\multiput(846,441)(18.845,8.698){0}{\usebox{\plotpoint}}
\put(863.42,449.58){\usebox{\plotpoint}}
\put(881.71,459.36){\usebox{\plotpoint}}
\multiput(883,460)(18.275,9.840){0}{\usebox{\plotpoint}}
\put(900.07,469.04){\usebox{\plotpoint}}
\put(918.27,478.99){\usebox{\plotpoint}}
\multiput(920,480)(18.564,9.282){0}{\usebox{\plotpoint}}
\put(936.85,488.24){\usebox{\plotpoint}}
\put(955.17,497.93){\usebox{\plotpoint}}
\multiput(957,499)(18.564,9.282){0}{\usebox{\plotpoint}}
\put(973.60,507.48){\usebox{\plotpoint}}
\put(992.03,517.01){\usebox{\plotpoint}}
\multiput(994,518)(17.928,10.458){0}{\usebox{\plotpoint}}
\put(1010.17,527.08){\usebox{\plotpoint}}
\put(1028.89,536.03){\usebox{\plotpoint}}
\multiput(1031,537)(17.928,10.458){0}{\usebox{\plotpoint}}
\put(1047.06,546.03){\usebox{\plotpoint}}
\put(1065.46,555.63){\usebox{\plotpoint}}
\multiput(1068,557)(18.564,9.282){0}{\usebox{\plotpoint}}
\put(1083.85,565.25){\usebox{\plotpoint}}
\put(1102.12,575.06){\usebox{\plotpoint}}
\multiput(1104,576)(18.845,8.698){0}{\usebox{\plotpoint}}
\put(1120.75,584.19){\usebox{\plotpoint}}
\put(1139.02,594.01){\usebox{\plotpoint}}
\multiput(1141,595)(17.928,10.458){0}{\usebox{\plotpoint}}
\put(1157.22,603.95){\usebox{\plotpoint}}
\put(1175.58,613.59){\usebox{\plotpoint}}
\multiput(1178,615)(18.564,9.282){0}{\usebox{\plotpoint}}
\put(1194.12,622.90){\usebox{\plotpoint}}
\put(1212.48,632.53){\usebox{\plotpoint}}
\multiput(1215,634)(18.564,9.282){0}{\usebox{\plotpoint}}
\put(1230.82,642.23){\usebox{\plotpoint}}
\put(1249.24,651.73){\usebox{\plotpoint}}
\multiput(1252,653)(17.928,10.458){0}{\usebox{\plotpoint}}
\put(1267.42,661.71){\usebox{\plotpoint}}
\put(1286.14,670.68){\usebox{\plotpoint}}
\multiput(1289,672)(17.928,10.458){0}{\usebox{\plotpoint}}
\put(1304.32,680.66){\usebox{\plotpoint}}
\put(1322.55,690.57){\usebox{\plotpoint}}
\multiput(1325,692)(18.845,8.698){0}{\usebox{\plotpoint}}
\put(1341.11,699.81){\usebox{\plotpoint}}
\put(1359.36,709.68){\usebox{\plotpoint}}
\multiput(1362,711)(18.275,9.840){0}{\usebox{\plotpoint}}
\put(1377.71,719.36){\usebox{\plotpoint}}
\put(1396.28,728.64){\usebox{\plotpoint}}
\multiput(1399,730)(17.928,10.458){0}{\usebox{\plotpoint}}
\put(1414.47,738.60){\usebox{\plotpoint}}
\put(1432.86,748.17){\usebox{\plotpoint}}
\put(1436,750){\usebox{\plotpoint}}
\end{picture}

%% file: alphax.tex
\setlength{\unitlength}{0.240900pt}
\ifx\plotpoint\undefined\newsavebox{\plotpoint}\fi
\sbox{\plotpoint}{\rule[-0.200pt]{0.400pt}{0.400pt}}%
\begin{picture}(1500,900)(0,0)
\font\gnuplot=cmr10 at 10pt
\gnuplot
\sbox{\plotpoint}{\rule[-0.200pt]{0.400pt}{0.400pt}}%
\put(220.0,113.0){\rule[-0.200pt]{4.818pt}{0.400pt}}
\put(198,113){\makebox(0,0)[r]{0.02}}
\put(1416.0,113.0){\rule[-0.200pt]{4.818pt}{0.400pt}}
\put(220.0,304.0){\rule[-0.200pt]{4.818pt}{0.400pt}}
\put(198,304){\makebox(0,0)[r]{0.025}}
\put(1416.0,304.0){\rule[-0.200pt]{4.818pt}{0.400pt}}
\put(220.0,495.0){\rule[-0.200pt]{4.818pt}{0.400pt}}
\put(198,495){\makebox(0,0)[r]{0.03}}
\put(1416.0,495.0){\rule[-0.200pt]{4.818pt}{0.400pt}}
\put(220.0,686.0){\rule[-0.200pt]{4.818pt}{0.400pt}}
\put(198,686){\makebox(0,0)[r]{0.035}}
\put(1416.0,686.0){\rule[-0.200pt]{4.818pt}{0.400pt}}
\put(220.0,877.0){\rule[-0.200pt]{4.818pt}{0.400pt}}
\put(198,877){\makebox(0,0)[r]{0.04}}
\put(1416.0,877.0){\rule[-0.200pt]{4.818pt}{0.400pt}}
\put(220.0,113.0){\rule[-0.200pt]{0.400pt}{4.818pt}}
\put(220,68){\makebox(0,0){11}}
\put(220.0,857.0){\rule[-0.200pt]{0.400pt}{4.818pt}}
\put(463.0,113.0){\rule[-0.200pt]{0.400pt}{4.818pt}}
\put(463,68){\makebox(0,0){12}}
\put(463.0,857.0){\rule[-0.200pt]{0.400pt}{4.818pt}}
\put(706.0,113.0){\rule[-0.200pt]{0.400pt}{4.818pt}}
\put(706,68){\makebox(0,0){13}}
\put(706.0,857.0){\rule[-0.200pt]{0.400pt}{4.818pt}}
\put(950.0,113.0){\rule[-0.200pt]{0.400pt}{4.818pt}}
\put(950,68){\makebox(0,0){14}}
\put(950.0,857.0){\rule[-0.200pt]{0.400pt}{4.818pt}}
\put(1193.0,113.0){\rule[-0.200pt]{0.400pt}{4.818pt}}
\put(1193,68){\makebox(0,0){15}}
\put(1193.0,857.0){\rule[-0.200pt]{0.400pt}{4.818pt}}
\put(1436.0,113.0){\rule[-0.200pt]{0.400pt}{4.818pt}}
\put(1436,68){\makebox(0,0){16}}
\put(1436.0,857.0){\rule[-0.200pt]{0.400pt}{4.818pt}}
\put(220.0,113.0){\rule[-0.200pt]{292.934pt}{0.400pt}}
\put(1436.0,113.0){\rule[-0.200pt]{0.400pt}{184.048pt}}
\put(220.0,877.0){\rule[-0.200pt]{292.934pt}{0.400pt}}
\put(45,495){\makebox(0,0){$\alpha(M_X)$}}
\put(828,23){\makebox(0,0){$\log_{10} \mu_0$ [GeV]}}
\put(220.0,113.0){\rule[-0.200pt]{0.400pt}{184.048pt}}
\sbox{\plotpoint}{\rule[-0.400pt]{0.800pt}{0.800pt}}%
\put(220,482){\usebox{\plotpoint}}
\multiput(220.00,483.41)(1.821,0.503){61}{\rule{3.071pt}{0.121pt}}
\multiput(220.00,480.34)(115.627,34.000){2}{\rule{1.535pt}{0.800pt}}
\multiput(342.00,517.41)(1.703,0.503){65}{\rule{2.889pt}{0.121pt}}
\multiput(342.00,514.34)(115.004,36.000){2}{\rule{1.444pt}{0.800pt}}
\multiput(463.00,553.41)(1.768,0.503){63}{\rule{2.989pt}{0.121pt}}
\multiput(463.00,550.34)(115.797,35.000){2}{\rule{1.494pt}{0.800pt}}
\multiput(585.00,588.41)(1.570,0.503){71}{\rule{2.682pt}{0.121pt}}
\multiput(585.00,585.34)(115.433,39.000){2}{\rule{1.341pt}{0.800pt}}
\multiput(706.00,627.41)(1.542,0.502){73}{\rule{2.640pt}{0.121pt}}
\multiput(706.00,624.34)(116.521,40.000){2}{\rule{1.320pt}{0.800pt}}
\multiput(828.00,667.41)(1.504,0.502){75}{\rule{2.580pt}{0.121pt}}
\multiput(828.00,664.34)(116.644,41.000){2}{\rule{1.290pt}{0.800pt}}
\multiput(950.00,708.41)(1.455,0.502){77}{\rule{2.505pt}{0.121pt}}
\multiput(950.00,705.34)(115.801,42.000){2}{\rule{1.252pt}{0.800pt}}
\multiput(1071.00,750.41)(1.504,0.502){75}{\rule{2.580pt}{0.121pt}}
\multiput(1071.00,747.34)(116.644,41.000){2}{\rule{1.290pt}{0.800pt}}
\multiput(1193.00,791.41)(1.570,0.503){71}{\rule{2.682pt}{0.121pt}}
\multiput(1193.00,788.34)(115.433,39.000){2}{\rule{1.341pt}{0.800pt}}
\multiput(1314.00,830.41)(2.070,0.503){53}{\rule{3.453pt}{0.121pt}}
\multiput(1314.00,827.34)(114.832,30.000){2}{\rule{1.727pt}{0.800pt}}
\end{picture}

%% file: gut2.tex
\setlength{\unitlength}{0.240900pt}
\ifx\plotpoint\undefined\newsavebox{\plotpoint}\fi
\sbox{\plotpoint}{\rule[-0.200pt]{0.400pt}{0.400pt}}%
\begin{picture}(1500,900)(0,0)
\font\gnuplot=cmr10 at 10pt
\gnuplot
\sbox{\plotpoint}{\rule[-0.200pt]{0.400pt}{0.400pt}}%
\put(220.0,113.0){\rule[-0.200pt]{4.818pt}{0.400pt}}
\put(198,113){\makebox(0,0)[r]{2}}
\put(1416.0,113.0){\rule[-0.200pt]{4.818pt}{0.400pt}}
\put(220.0,209.0){\rule[-0.200pt]{4.818pt}{0.400pt}}
\put(198,209){\makebox(0,0)[r]{4}}
\put(1416.0,209.0){\rule[-0.200pt]{4.818pt}{0.400pt}}
\put(220.0,304.0){\rule[-0.200pt]{4.818pt}{0.400pt}}
\put(198,304){\makebox(0,0)[r]{6}}
\put(1416.0,304.0){\rule[-0.200pt]{4.818pt}{0.400pt}}
\put(220.0,400.0){\rule[-0.200pt]{4.818pt}{0.400pt}}
\put(198,400){\makebox(0,0)[r]{8}}
\put(1416.0,400.0){\rule[-0.200pt]{4.818pt}{0.400pt}}
\put(220.0,495.0){\rule[-0.200pt]{4.818pt}{0.400pt}}
\put(198,495){\makebox(0,0)[r]{10}}
\put(1416.0,495.0){\rule[-0.200pt]{4.818pt}{0.400pt}}
\put(220.0,591.0){\rule[-0.200pt]{4.818pt}{0.400pt}}
\put(198,591){\makebox(0,0)[r]{12}}
\put(1416.0,591.0){\rule[-0.200pt]{4.818pt}{0.400pt}}
\put(220.0,686.0){\rule[-0.200pt]{4.818pt}{0.400pt}}
\put(198,686){\makebox(0,0)[r]{14}}
\put(1416.0,686.0){\rule[-0.200pt]{4.818pt}{0.400pt}}
\put(220.0,782.0){\rule[-0.200pt]{4.818pt}{0.400pt}}
\put(198,782){\makebox(0,0)[r]{16}}
\put(1416.0,782.0){\rule[-0.200pt]{4.818pt}{0.400pt}}
\put(220.0,877.0){\rule[-0.200pt]{4.818pt}{0.400pt}}
\put(198,877){\makebox(0,0)[r]{18}}
\put(1416.0,877.0){\rule[-0.200pt]{4.818pt}{0.400pt}}
\put(220.0,113.0){\rule[-0.200pt]{0.400pt}{4.818pt}}
\put(220,68){\makebox(0,0){2}}
\put(220.0,857.0){\rule[-0.200pt]{0.400pt}{4.818pt}}
\put(394.0,113.0){\rule[-0.200pt]{0.400pt}{4.818pt}}
\put(394,68){\makebox(0,0){4}}
\put(394.0,857.0){\rule[-0.200pt]{0.400pt}{4.818pt}}
\put(567.0,113.0){\rule[-0.200pt]{0.400pt}{4.818pt}}
\put(567,68){\makebox(0,0){6}}
\put(567.0,857.0){\rule[-0.200pt]{0.400pt}{4.818pt}}
\put(741.0,113.0){\rule[-0.200pt]{0.400pt}{4.818pt}}
\put(741,68){\makebox(0,0){8}}
\put(741.0,857.0){\rule[-0.200pt]{0.400pt}{4.818pt}}
\put(915.0,113.0){\rule[-0.200pt]{0.400pt}{4.818pt}}
\put(915,68){\makebox(0,0){10}}
\put(915.0,857.0){\rule[-0.200pt]{0.400pt}{4.818pt}}
\put(1089.0,113.0){\rule[-0.200pt]{0.400pt}{4.818pt}}
\put(1089,68){\makebox(0,0){12}}
\put(1089.0,857.0){\rule[-0.200pt]{0.400pt}{4.818pt}}
\put(1262.0,113.0){\rule[-0.200pt]{0.400pt}{4.818pt}}
\put(1262,68){\makebox(0,0){14}}
\put(1262.0,857.0){\rule[-0.200pt]{0.400pt}{4.818pt}}
\put(1436.0,113.0){\rule[-0.200pt]{0.400pt}{4.818pt}}
\put(1436,68){\makebox(0,0){16}}
\put(1436.0,857.0){\rule[-0.200pt]{0.400pt}{4.818pt}}
\put(220.0,113.0){\rule[-0.200pt]{292.934pt}{0.400pt}}
\put(1436.0,113.0){\rule[-0.200pt]{0.400pt}{184.048pt}}
\put(220.0,877.0){\rule[-0.200pt]{292.934pt}{0.400pt}}
\put(45,495){\makebox(0,0){$\log_{10} M_Y$}}
\put(50,445){\makebox(0,0){[GeV]}}
\put(828,23){\makebox(0,0){$\log_{10} \mu_0$ [GeV]}}
\put(220.0,113.0){\rule[-0.200pt]{0.400pt}{184.048pt}}
\sbox{\plotpoint}{\rule[-0.400pt]{0.800pt}{0.800pt}}%
\put(1436,787){\usebox{\plotpoint}}
\put(1433,784.84){\rule{0.241pt}{0.800pt}}
\multiput(1433.50,785.34)(-0.500,-1.000){2}{\rule{0.120pt}{0.800pt}}
\put(1434.0,787.0){\usebox{\plotpoint}}
\put(1431,783.84){\rule{0.241pt}{0.800pt}}
\multiput(1431.50,784.34)(-0.500,-1.000){2}{\rule{0.120pt}{0.800pt}}
\put(1432.0,786.0){\usebox{\plotpoint}}
\put(1427,782.84){\rule{0.241pt}{0.800pt}}
\multiput(1427.50,783.34)(-0.500,-1.000){2}{\rule{0.120pt}{0.800pt}}
\put(1428.0,785.0){\usebox{\plotpoint}}
\put(1425,781.84){\rule{0.241pt}{0.800pt}}
\multiput(1425.50,782.34)(-0.500,-1.000){2}{\rule{0.120pt}{0.800pt}}
\put(1426.0,784.0){\usebox{\plotpoint}}
\put(1422,780.84){\rule{0.482pt}{0.800pt}}
\multiput(1423.00,781.34)(-1.000,-1.000){2}{\rule{0.241pt}{0.800pt}}
\put(1424.0,783.0){\usebox{\plotpoint}}
\put(1420,779.84){\rule{0.241pt}{0.800pt}}
\multiput(1420.50,780.34)(-0.500,-1.000){2}{\rule{0.120pt}{0.800pt}}
\put(1421.0,782.0){\usebox{\plotpoint}}
\put(1417,778.84){\rule{0.482pt}{0.800pt}}
\multiput(1418.00,779.34)(-1.000,-1.000){2}{\rule{0.241pt}{0.800pt}}
\put(1419.0,781.0){\usebox{\plotpoint}}
\put(1415,777.84){\rule{0.241pt}{0.800pt}}
\multiput(1415.50,778.34)(-0.500,-1.000){2}{\rule{0.120pt}{0.800pt}}
\put(1416.0,780.0){\usebox{\plotpoint}}
\put(1412,776.84){\rule{0.482pt}{0.800pt}}
\multiput(1413.00,777.34)(-1.000,-1.000){2}{\rule{0.241pt}{0.800pt}}
\put(1414.0,779.0){\usebox{\plotpoint}}
\put(1409,775.84){\rule{0.241pt}{0.800pt}}
\multiput(1409.50,776.34)(-0.500,-1.000){2}{\rule{0.120pt}{0.800pt}}
\put(1407,774.84){\rule{0.482pt}{0.800pt}}
\multiput(1408.00,775.34)(-1.000,-1.000){2}{\rule{0.241pt}{0.800pt}}
\put(1410.0,778.0){\usebox{\plotpoint}}
\put(1405,773.84){\rule{0.241pt}{0.800pt}}
\multiput(1405.50,774.34)(-0.500,-1.000){2}{\rule{0.120pt}{0.800pt}}
\put(1406.0,776.0){\usebox{\plotpoint}}
\put(1402,772.84){\rule{0.241pt}{0.800pt}}
\multiput(1402.50,773.34)(-0.500,-1.000){2}{\rule{0.120pt}{0.800pt}}
\put(1403.0,775.0){\usebox{\plotpoint}}
\put(1399,771.84){\rule{0.482pt}{0.800pt}}
\multiput(1400.00,772.34)(-1.000,-1.000){2}{\rule{0.241pt}{0.800pt}}
\put(1401.0,774.0){\usebox{\plotpoint}}
\put(1397,770.84){\rule{0.241pt}{0.800pt}}
\multiput(1397.50,771.34)(-0.500,-1.000){2}{\rule{0.120pt}{0.800pt}}
\put(1398.0,773.0){\usebox{\plotpoint}}
\put(1394,769.84){\rule{0.241pt}{0.800pt}}
\multiput(1394.50,770.34)(-0.500,-1.000){2}{\rule{0.120pt}{0.800pt}}
\put(1395.0,772.0){\usebox{\plotpoint}}
\put(1391,768.84){\rule{0.482pt}{0.800pt}}
\multiput(1392.00,769.34)(-1.000,-1.000){2}{\rule{0.241pt}{0.800pt}}
\put(1393.0,771.0){\usebox{\plotpoint}}
\put(1389,767.84){\rule{0.241pt}{0.800pt}}
\multiput(1389.50,768.34)(-0.500,-1.000){2}{\rule{0.120pt}{0.800pt}}
\put(1390.0,770.0){\usebox{\plotpoint}}
\put(1386,766.84){\rule{0.241pt}{0.800pt}}
\multiput(1386.50,767.34)(-0.500,-1.000){2}{\rule{0.120pt}{0.800pt}}
\put(1384,765.84){\rule{0.482pt}{0.800pt}}
\multiput(1385.00,766.34)(-1.000,-1.000){2}{\rule{0.241pt}{0.800pt}}
\put(1387.0,769.0){\usebox{\plotpoint}}
\put(1381,764.84){\rule{0.482pt}{0.800pt}}
\multiput(1382.00,765.34)(-1.000,-1.000){2}{\rule{0.241pt}{0.800pt}}
\put(1383.0,767.0){\usebox{\plotpoint}}
\put(1379,763.84){\rule{0.241pt}{0.800pt}}
\multiput(1379.50,764.34)(-0.500,-1.000){2}{\rule{0.120pt}{0.800pt}}
\put(1377,762.84){\rule{0.482pt}{0.800pt}}
\multiput(1378.00,763.34)(-1.000,-1.000){2}{\rule{0.241pt}{0.800pt}}
\put(1380.0,766.0){\usebox{\plotpoint}}
\put(1374,761.84){\rule{0.482pt}{0.800pt}}
\multiput(1375.00,762.34)(-1.000,-1.000){2}{\rule{0.241pt}{0.800pt}}
\put(1376.0,764.0){\usebox{\plotpoint}}
\put(1371,760.84){\rule{0.482pt}{0.800pt}}
\multiput(1372.00,761.34)(-1.000,-1.000){2}{\rule{0.241pt}{0.800pt}}
\put(1370,759.84){\rule{0.241pt}{0.800pt}}
\multiput(1370.50,760.34)(-0.500,-1.000){2}{\rule{0.120pt}{0.800pt}}
\put(1373.0,763.0){\usebox{\plotpoint}}
\put(1367,758.84){\rule{0.241pt}{0.800pt}}
\multiput(1367.50,759.34)(-0.500,-1.000){2}{\rule{0.120pt}{0.800pt}}
\put(1368.0,761.0){\usebox{\plotpoint}}
\put(1364,757.84){\rule{0.241pt}{0.800pt}}
\multiput(1364.50,758.34)(-0.500,-1.000){2}{\rule{0.120pt}{0.800pt}}
\put(1362,756.84){\rule{0.482pt}{0.800pt}}
\multiput(1363.00,757.34)(-1.000,-1.000){2}{\rule{0.241pt}{0.800pt}}
\put(1365.0,760.0){\usebox{\plotpoint}}
\put(1359,755.84){\rule{0.482pt}{0.800pt}}
\multiput(1360.00,756.34)(-1.000,-1.000){2}{\rule{0.241pt}{0.800pt}}
\put(1358,754.84){\rule{0.241pt}{0.800pt}}
\multiput(1358.50,755.34)(-0.500,-1.000){2}{\rule{0.120pt}{0.800pt}}
\put(1361.0,758.0){\usebox{\plotpoint}}
\put(1354,753.84){\rule{0.482pt}{0.800pt}}
\multiput(1355.00,754.34)(-1.000,-1.000){2}{\rule{0.241pt}{0.800pt}}
\put(1353,752.84){\rule{0.241pt}{0.800pt}}
\multiput(1353.50,753.34)(-0.500,-1.000){2}{\rule{0.120pt}{0.800pt}}
\put(1356.0,756.0){\usebox{\plotpoint}}
\put(1350,751.84){\rule{0.241pt}{0.800pt}}
\multiput(1350.50,752.34)(-0.500,-1.000){2}{\rule{0.120pt}{0.800pt}}
\put(1348,750.84){\rule{0.482pt}{0.800pt}}
\multiput(1349.00,751.34)(-1.000,-1.000){2}{\rule{0.241pt}{0.800pt}}
\put(1351.0,754.0){\usebox{\plotpoint}}
\put(1345,749.84){\rule{0.241pt}{0.800pt}}
\multiput(1345.50,750.34)(-0.500,-1.000){2}{\rule{0.120pt}{0.800pt}}
\put(1343,748.84){\rule{0.482pt}{0.800pt}}
\multiput(1344.00,749.34)(-1.000,-1.000){2}{\rule{0.241pt}{0.800pt}}
\put(1346.0,752.0){\usebox{\plotpoint}}
\put(1340,747.84){\rule{0.482pt}{0.800pt}}
\multiput(1341.00,748.34)(-1.000,-1.000){2}{\rule{0.241pt}{0.800pt}}
\put(1338,746.84){\rule{0.482pt}{0.800pt}}
\multiput(1339.00,747.34)(-1.000,-1.000){2}{\rule{0.241pt}{0.800pt}}
\put(1342.0,750.0){\usebox{\plotpoint}}
\put(1335,745.84){\rule{0.482pt}{0.800pt}}
\multiput(1336.00,746.34)(-1.000,-1.000){2}{\rule{0.241pt}{0.800pt}}
\put(1333,744.84){\rule{0.482pt}{0.800pt}}
\multiput(1334.00,745.34)(-1.000,-1.000){2}{\rule{0.241pt}{0.800pt}}
\put(1331,743.84){\rule{0.482pt}{0.800pt}}
\multiput(1332.00,744.34)(-1.000,-1.000){2}{\rule{0.241pt}{0.800pt}}
\put(1337.0,748.0){\usebox{\plotpoint}}
\put(1328,742.84){\rule{0.482pt}{0.800pt}}
\multiput(1329.00,743.34)(-1.000,-1.000){2}{\rule{0.241pt}{0.800pt}}
\put(1326,741.84){\rule{0.482pt}{0.800pt}}
\multiput(1327.00,742.34)(-1.000,-1.000){2}{\rule{0.241pt}{0.800pt}}
\put(1324,740.84){\rule{0.482pt}{0.800pt}}
\multiput(1325.00,741.34)(-1.000,-1.000){2}{\rule{0.241pt}{0.800pt}}
\put(1330.0,745.0){\usebox{\plotpoint}}
\put(1321,739.84){\rule{0.482pt}{0.800pt}}
\multiput(1322.00,740.34)(-1.000,-1.000){2}{\rule{0.241pt}{0.800pt}}
\put(1319,738.84){\rule{0.482pt}{0.800pt}}
\multiput(1320.00,739.34)(-1.000,-1.000){2}{\rule{0.241pt}{0.800pt}}
\put(1317,737.84){\rule{0.482pt}{0.800pt}}
\multiput(1318.00,738.34)(-1.000,-1.000){2}{\rule{0.241pt}{0.800pt}}
\put(1316,736.84){\rule{0.241pt}{0.800pt}}
\multiput(1316.50,737.34)(-0.500,-1.000){2}{\rule{0.120pt}{0.800pt}}
\put(1323.0,742.0){\usebox{\plotpoint}}
\put(1312,735.84){\rule{0.482pt}{0.800pt}}
\multiput(1313.00,736.34)(-1.000,-1.000){2}{\rule{0.241pt}{0.800pt}}
\put(1310,734.84){\rule{0.482pt}{0.800pt}}
\multiput(1311.00,735.34)(-1.000,-1.000){2}{\rule{0.241pt}{0.800pt}}
\put(1308,733.84){\rule{0.482pt}{0.800pt}}
\multiput(1309.00,734.34)(-1.000,-1.000){2}{\rule{0.241pt}{0.800pt}}
\put(1306,732.84){\rule{0.482pt}{0.800pt}}
\multiput(1307.00,733.34)(-1.000,-1.000){2}{\rule{0.241pt}{0.800pt}}
\put(1314.0,738.0){\usebox{\plotpoint}}
\put(1303,731.84){\rule{0.241pt}{0.800pt}}
\multiput(1303.50,732.34)(-0.500,-1.000){2}{\rule{0.120pt}{0.800pt}}
\put(1301,730.84){\rule{0.482pt}{0.800pt}}
\multiput(1302.00,731.34)(-1.000,-1.000){2}{\rule{0.241pt}{0.800pt}}
\put(1299,729.84){\rule{0.482pt}{0.800pt}}
\multiput(1300.00,730.34)(-1.000,-1.000){2}{\rule{0.241pt}{0.800pt}}
\put(1297,728.84){\rule{0.482pt}{0.800pt}}
\multiput(1298.00,729.34)(-1.000,-1.000){2}{\rule{0.241pt}{0.800pt}}
\put(1295,727.84){\rule{0.482pt}{0.800pt}}
\multiput(1296.00,728.34)(-1.000,-1.000){2}{\rule{0.241pt}{0.800pt}}
\put(1304.0,734.0){\usebox{\plotpoint}}
\put(1291,726.84){\rule{0.482pt}{0.800pt}}
\multiput(1292.00,727.34)(-1.000,-1.000){2}{\rule{0.241pt}{0.800pt}}
\put(1289,725.84){\rule{0.482pt}{0.800pt}}
\multiput(1290.00,726.34)(-1.000,-1.000){2}{\rule{0.241pt}{0.800pt}}
\put(1287,724.84){\rule{0.482pt}{0.800pt}}
\multiput(1288.00,725.34)(-1.000,-1.000){2}{\rule{0.241pt}{0.800pt}}
\put(1285,723.84){\rule{0.482pt}{0.800pt}}
\multiput(1286.00,724.34)(-1.000,-1.000){2}{\rule{0.241pt}{0.800pt}}
\put(1283,722.84){\rule{0.482pt}{0.800pt}}
\multiput(1284.00,723.34)(-1.000,-1.000){2}{\rule{0.241pt}{0.800pt}}
\put(1281,721.84){\rule{0.482pt}{0.800pt}}
\multiput(1282.00,722.34)(-1.000,-1.000){2}{\rule{0.241pt}{0.800pt}}
\put(1279,720.84){\rule{0.482pt}{0.800pt}}
\multiput(1280.00,721.34)(-1.000,-1.000){2}{\rule{0.241pt}{0.800pt}}
\put(1277,719.84){\rule{0.482pt}{0.800pt}}
\multiput(1278.00,720.34)(-1.000,-1.000){2}{\rule{0.241pt}{0.800pt}}
\put(1275,718.84){\rule{0.482pt}{0.800pt}}
\multiput(1276.00,719.34)(-1.000,-1.000){2}{\rule{0.241pt}{0.800pt}}
\put(1273,717.84){\rule{0.482pt}{0.800pt}}
\multiput(1274.00,718.34)(-1.000,-1.000){2}{\rule{0.241pt}{0.800pt}}
\put(1270,716.84){\rule{0.723pt}{0.800pt}}
\multiput(1271.50,717.34)(-1.500,-1.000){2}{\rule{0.361pt}{0.800pt}}
\put(1293.0,729.0){\usebox{\plotpoint}}
\put(1266,715.84){\rule{0.482pt}{0.800pt}}
\multiput(1267.00,716.34)(-1.000,-1.000){2}{\rule{0.241pt}{0.800pt}}
\put(1264,714.84){\rule{0.482pt}{0.800pt}}
\multiput(1265.00,715.34)(-1.000,-1.000){2}{\rule{0.241pt}{0.800pt}}
\put(1262,713.84){\rule{0.482pt}{0.800pt}}
\multiput(1263.00,714.34)(-1.000,-1.000){2}{\rule{0.241pt}{0.800pt}}
\put(1260,712.84){\rule{0.482pt}{0.800pt}}
\multiput(1261.00,713.34)(-1.000,-1.000){2}{\rule{0.241pt}{0.800pt}}
\put(1257,711.84){\rule{0.723pt}{0.800pt}}
\multiput(1258.50,712.34)(-1.500,-1.000){2}{\rule{0.361pt}{0.800pt}}
\put(1255,710.84){\rule{0.482pt}{0.800pt}}
\multiput(1256.00,711.34)(-1.000,-1.000){2}{\rule{0.241pt}{0.800pt}}
\put(1253,709.84){\rule{0.482pt}{0.800pt}}
\multiput(1254.00,710.34)(-1.000,-1.000){2}{\rule{0.241pt}{0.800pt}}
\put(1251,708.84){\rule{0.482pt}{0.800pt}}
\multiput(1252.00,709.34)(-1.000,-1.000){2}{\rule{0.241pt}{0.800pt}}
\put(1249,707.84){\rule{0.482pt}{0.800pt}}
\multiput(1250.00,708.34)(-1.000,-1.000){2}{\rule{0.241pt}{0.800pt}}
\put(1246,706.34){\rule{0.723pt}{0.800pt}}
\multiput(1247.50,707.34)(-1.500,-2.000){2}{\rule{0.361pt}{0.800pt}}
\put(1244,704.84){\rule{0.482pt}{0.800pt}}
\multiput(1245.00,705.34)(-1.000,-1.000){2}{\rule{0.241pt}{0.800pt}}
\put(1242,703.84){\rule{0.482pt}{0.800pt}}
\multiput(1243.00,704.34)(-1.000,-1.000){2}{\rule{0.241pt}{0.800pt}}
\put(1239,702.84){\rule{0.723pt}{0.800pt}}
\multiput(1240.50,703.34)(-1.500,-1.000){2}{\rule{0.361pt}{0.800pt}}
\put(1237,701.84){\rule{0.482pt}{0.800pt}}
\multiput(1238.00,702.34)(-1.000,-1.000){2}{\rule{0.241pt}{0.800pt}}
\put(1235,700.84){\rule{0.482pt}{0.800pt}}
\multiput(1236.00,701.34)(-1.000,-1.000){2}{\rule{0.241pt}{0.800pt}}
\put(1232,699.84){\rule{0.723pt}{0.800pt}}
\multiput(1233.50,700.34)(-1.500,-1.000){2}{\rule{0.361pt}{0.800pt}}
\put(1230,698.84){\rule{0.482pt}{0.800pt}}
\multiput(1231.00,699.34)(-1.000,-1.000){2}{\rule{0.241pt}{0.800pt}}
\put(1227,697.84){\rule{0.723pt}{0.800pt}}
\multiput(1228.50,698.34)(-1.500,-1.000){2}{\rule{0.361pt}{0.800pt}}
\put(1225,696.84){\rule{0.482pt}{0.800pt}}
\multiput(1226.00,697.34)(-1.000,-1.000){2}{\rule{0.241pt}{0.800pt}}
\put(1223,695.84){\rule{0.482pt}{0.800pt}}
\multiput(1224.00,696.34)(-1.000,-1.000){2}{\rule{0.241pt}{0.800pt}}
\put(1220,694.34){\rule{0.723pt}{0.800pt}}
\multiput(1221.50,695.34)(-1.500,-2.000){2}{\rule{0.361pt}{0.800pt}}
\put(1218,692.84){\rule{0.482pt}{0.800pt}}
\multiput(1219.00,693.34)(-1.000,-1.000){2}{\rule{0.241pt}{0.800pt}}
\put(1215,691.84){\rule{0.723pt}{0.800pt}}
\multiput(1216.50,692.34)(-1.500,-1.000){2}{\rule{0.361pt}{0.800pt}}
\put(1213,690.84){\rule{0.482pt}{0.800pt}}
\multiput(1214.00,691.34)(-1.000,-1.000){2}{\rule{0.241pt}{0.800pt}}
\put(1210,689.84){\rule{0.723pt}{0.800pt}}
\multiput(1211.50,690.34)(-1.500,-1.000){2}{\rule{0.361pt}{0.800pt}}
\put(1208,688.34){\rule{0.482pt}{0.800pt}}
\multiput(1209.00,689.34)(-1.000,-2.000){2}{\rule{0.241pt}{0.800pt}}
\put(1205,686.84){\rule{0.723pt}{0.800pt}}
\multiput(1206.50,687.34)(-1.500,-1.000){2}{\rule{0.361pt}{0.800pt}}
\put(1202,685.84){\rule{0.723pt}{0.800pt}}
\multiput(1203.50,686.34)(-1.500,-1.000){2}{\rule{0.361pt}{0.800pt}}
\put(1200,684.84){\rule{0.482pt}{0.800pt}}
\multiput(1201.00,685.34)(-1.000,-1.000){2}{\rule{0.241pt}{0.800pt}}
\put(1197,683.84){\rule{0.723pt}{0.800pt}}
\multiput(1198.50,684.34)(-1.500,-1.000){2}{\rule{0.361pt}{0.800pt}}
\put(1194,682.34){\rule{0.723pt}{0.800pt}}
\multiput(1195.50,683.34)(-1.500,-2.000){2}{\rule{0.361pt}{0.800pt}}
\put(1192,680.84){\rule{0.482pt}{0.800pt}}
\multiput(1193.00,681.34)(-1.000,-1.000){2}{\rule{0.241pt}{0.800pt}}
\put(1189,679.84){\rule{0.723pt}{0.800pt}}
\multiput(1190.50,680.34)(-1.500,-1.000){2}{\rule{0.361pt}{0.800pt}}
\put(1186,678.84){\rule{0.723pt}{0.800pt}}
\multiput(1187.50,679.34)(-1.500,-1.000){2}{\rule{0.361pt}{0.800pt}}
\put(1184,677.34){\rule{0.482pt}{0.800pt}}
\multiput(1185.00,678.34)(-1.000,-2.000){2}{\rule{0.241pt}{0.800pt}}
\put(1181,675.84){\rule{0.723pt}{0.800pt}}
\multiput(1182.50,676.34)(-1.500,-1.000){2}{\rule{0.361pt}{0.800pt}}
\put(1178,674.84){\rule{0.723pt}{0.800pt}}
\multiput(1179.50,675.34)(-1.500,-1.000){2}{\rule{0.361pt}{0.800pt}}
\put(1175,673.34){\rule{0.723pt}{0.800pt}}
\multiput(1176.50,674.34)(-1.500,-2.000){2}{\rule{0.361pt}{0.800pt}}
\put(1172,671.84){\rule{0.723pt}{0.800pt}}
\multiput(1173.50,672.34)(-1.500,-1.000){2}{\rule{0.361pt}{0.800pt}}
\put(1170,670.84){\rule{0.482pt}{0.800pt}}
\multiput(1171.00,671.34)(-1.000,-1.000){2}{\rule{0.241pt}{0.800pt}}
\put(1167,669.34){\rule{0.723pt}{0.800pt}}
\multiput(1168.50,670.34)(-1.500,-2.000){2}{\rule{0.361pt}{0.800pt}}
\put(1164,667.84){\rule{0.723pt}{0.800pt}}
\multiput(1165.50,668.34)(-1.500,-1.000){2}{\rule{0.361pt}{0.800pt}}
\put(1161,666.34){\rule{0.723pt}{0.800pt}}
\multiput(1162.50,667.34)(-1.500,-2.000){2}{\rule{0.361pt}{0.800pt}}
\put(1158,664.84){\rule{0.723pt}{0.800pt}}
\multiput(1159.50,665.34)(-1.500,-1.000){2}{\rule{0.361pt}{0.800pt}}
\put(1155,663.84){\rule{0.723pt}{0.800pt}}
\multiput(1156.50,664.34)(-1.500,-1.000){2}{\rule{0.361pt}{0.800pt}}
\put(1152,662.34){\rule{0.723pt}{0.800pt}}
\multiput(1153.50,663.34)(-1.500,-2.000){2}{\rule{0.361pt}{0.800pt}}
\put(1149,660.84){\rule{0.723pt}{0.800pt}}
\multiput(1150.50,661.34)(-1.500,-1.000){2}{\rule{0.361pt}{0.800pt}}
\put(1146,659.34){\rule{0.723pt}{0.800pt}}
\multiput(1147.50,660.34)(-1.500,-2.000){2}{\rule{0.361pt}{0.800pt}}
\put(1143,657.84){\rule{0.723pt}{0.800pt}}
\multiput(1144.50,658.34)(-1.500,-1.000){2}{\rule{0.361pt}{0.800pt}}
\put(1140,656.34){\rule{0.723pt}{0.800pt}}
\multiput(1141.50,657.34)(-1.500,-2.000){2}{\rule{0.361pt}{0.800pt}}
\put(1137,654.84){\rule{0.723pt}{0.800pt}}
\multiput(1138.50,655.34)(-1.500,-1.000){2}{\rule{0.361pt}{0.800pt}}
\put(1134,653.34){\rule{0.723pt}{0.800pt}}
\multiput(1135.50,654.34)(-1.500,-2.000){2}{\rule{0.361pt}{0.800pt}}
\put(1131,651.84){\rule{0.723pt}{0.800pt}}
\multiput(1132.50,652.34)(-1.500,-1.000){2}{\rule{0.361pt}{0.800pt}}
\put(1127,650.34){\rule{0.964pt}{0.800pt}}
\multiput(1129.00,651.34)(-2.000,-2.000){2}{\rule{0.482pt}{0.800pt}}
\put(1124,648.84){\rule{0.723pt}{0.800pt}}
\multiput(1125.50,649.34)(-1.500,-1.000){2}{\rule{0.361pt}{0.800pt}}
\put(1121,647.34){\rule{0.723pt}{0.800pt}}
\multiput(1122.50,648.34)(-1.500,-2.000){2}{\rule{0.361pt}{0.800pt}}
\put(1118,645.84){\rule{0.723pt}{0.800pt}}
\multiput(1119.50,646.34)(-1.500,-1.000){2}{\rule{0.361pt}{0.800pt}}
\put(1115,644.34){\rule{0.723pt}{0.800pt}}
\multiput(1116.50,645.34)(-1.500,-2.000){2}{\rule{0.361pt}{0.800pt}}
\put(1111,642.34){\rule{0.964pt}{0.800pt}}
\multiput(1113.00,643.34)(-2.000,-2.000){2}{\rule{0.482pt}{0.800pt}}
\put(1108,640.84){\rule{0.723pt}{0.800pt}}
\multiput(1109.50,641.34)(-1.500,-1.000){2}{\rule{0.361pt}{0.800pt}}
\put(1105,639.34){\rule{0.723pt}{0.800pt}}
\multiput(1106.50,640.34)(-1.500,-2.000){2}{\rule{0.361pt}{0.800pt}}
\put(1101,637.34){\rule{0.964pt}{0.800pt}}
\multiput(1103.00,638.34)(-2.000,-2.000){2}{\rule{0.482pt}{0.800pt}}
\put(1098,635.84){\rule{0.723pt}{0.800pt}}
\multiput(1099.50,636.34)(-1.500,-1.000){2}{\rule{0.361pt}{0.800pt}}
\put(1094,634.34){\rule{0.964pt}{0.800pt}}
\multiput(1096.00,635.34)(-2.000,-2.000){2}{\rule{0.482pt}{0.800pt}}
\put(1091,632.34){\rule{0.723pt}{0.800pt}}
\multiput(1092.50,633.34)(-1.500,-2.000){2}{\rule{0.361pt}{0.800pt}}
\put(1087,630.84){\rule{0.964pt}{0.800pt}}
\multiput(1089.00,631.34)(-2.000,-1.000){2}{\rule{0.482pt}{0.800pt}}
\put(1084,629.34){\rule{0.723pt}{0.800pt}}
\multiput(1085.50,630.34)(-1.500,-2.000){2}{\rule{0.361pt}{0.800pt}}
\put(1080,627.34){\rule{0.964pt}{0.800pt}}
\multiput(1082.00,628.34)(-2.000,-2.000){2}{\rule{0.482pt}{0.800pt}}
\put(1077,625.34){\rule{0.723pt}{0.800pt}}
\multiput(1078.50,626.34)(-1.500,-2.000){2}{\rule{0.361pt}{0.800pt}}
\put(1073,623.84){\rule{0.964pt}{0.800pt}}
\multiput(1075.00,624.34)(-2.000,-1.000){2}{\rule{0.482pt}{0.800pt}}
\put(1069,622.34){\rule{0.964pt}{0.800pt}}
\multiput(1071.00,623.34)(-2.000,-2.000){2}{\rule{0.482pt}{0.800pt}}
\put(1066,620.34){\rule{0.723pt}{0.800pt}}
\multiput(1067.50,621.34)(-1.500,-2.000){2}{\rule{0.361pt}{0.800pt}}
\put(1062,618.34){\rule{0.964pt}{0.800pt}}
\multiput(1064.00,619.34)(-2.000,-2.000){2}{\rule{0.482pt}{0.800pt}}
\put(1058,616.34){\rule{0.964pt}{0.800pt}}
\multiput(1060.00,617.34)(-2.000,-2.000){2}{\rule{0.482pt}{0.800pt}}
\put(1055,614.84){\rule{0.723pt}{0.800pt}}
\multiput(1056.50,615.34)(-1.500,-1.000){2}{\rule{0.361pt}{0.800pt}}
\put(1051,613.34){\rule{0.964pt}{0.800pt}}
\multiput(1053.00,614.34)(-2.000,-2.000){2}{\rule{0.482pt}{0.800pt}}
\put(1047,611.34){\rule{0.964pt}{0.800pt}}
\multiput(1049.00,612.34)(-2.000,-2.000){2}{\rule{0.482pt}{0.800pt}}
\put(1043,609.34){\rule{0.964pt}{0.800pt}}
\multiput(1045.00,610.34)(-2.000,-2.000){2}{\rule{0.482pt}{0.800pt}}
\put(1039,607.34){\rule{0.964pt}{0.800pt}}
\multiput(1041.00,608.34)(-2.000,-2.000){2}{\rule{0.482pt}{0.800pt}}
\put(1035,605.34){\rule{0.964pt}{0.800pt}}
\multiput(1037.00,606.34)(-2.000,-2.000){2}{\rule{0.482pt}{0.800pt}}
\put(1031,603.34){\rule{0.964pt}{0.800pt}}
\multiput(1033.00,604.34)(-2.000,-2.000){2}{\rule{0.482pt}{0.800pt}}
\put(1027,601.34){\rule{0.964pt}{0.800pt}}
\multiput(1029.00,602.34)(-2.000,-2.000){2}{\rule{0.482pt}{0.800pt}}
\put(1023,599.34){\rule{0.964pt}{0.800pt}}
\multiput(1025.00,600.34)(-2.000,-2.000){2}{\rule{0.482pt}{0.800pt}}
\put(1019,597.34){\rule{0.964pt}{0.800pt}}
\multiput(1021.00,598.34)(-2.000,-2.000){2}{\rule{0.482pt}{0.800pt}}
\put(1015,595.34){\rule{0.964pt}{0.800pt}}
\multiput(1017.00,596.34)(-2.000,-2.000){2}{\rule{0.482pt}{0.800pt}}
\put(1011,593.34){\rule{0.964pt}{0.800pt}}
\multiput(1013.00,594.34)(-2.000,-2.000){2}{\rule{0.482pt}{0.800pt}}
\put(1007,591.34){\rule{0.964pt}{0.800pt}}
\multiput(1009.00,592.34)(-2.000,-2.000){2}{\rule{0.482pt}{0.800pt}}
\put(1003,589.34){\rule{0.964pt}{0.800pt}}
\multiput(1005.00,590.34)(-2.000,-2.000){2}{\rule{0.482pt}{0.800pt}}
\put(999,587.34){\rule{0.964pt}{0.800pt}}
\multiput(1001.00,588.34)(-2.000,-2.000){2}{\rule{0.482pt}{0.800pt}}
\put(994,585.34){\rule{1.204pt}{0.800pt}}
\multiput(996.50,586.34)(-2.500,-2.000){2}{\rule{0.602pt}{0.800pt}}
\put(990,582.84){\rule{0.964pt}{0.800pt}}
\multiput(992.00,584.34)(-2.000,-3.000){2}{\rule{0.482pt}{0.800pt}}
\put(986,580.34){\rule{0.964pt}{0.800pt}}
\multiput(988.00,581.34)(-2.000,-2.000){2}{\rule{0.482pt}{0.800pt}}
\put(981,578.34){\rule{1.204pt}{0.800pt}}
\multiput(983.50,579.34)(-2.500,-2.000){2}{\rule{0.602pt}{0.800pt}}
\put(977,576.34){\rule{0.964pt}{0.800pt}}
\multiput(979.00,577.34)(-2.000,-2.000){2}{\rule{0.482pt}{0.800pt}}
\put(973,574.34){\rule{0.964pt}{0.800pt}}
\multiput(975.00,575.34)(-2.000,-2.000){2}{\rule{0.482pt}{0.800pt}}
\put(968,571.84){\rule{1.204pt}{0.800pt}}
\multiput(970.50,573.34)(-2.500,-3.000){2}{\rule{0.602pt}{0.800pt}}
\put(964,569.34){\rule{0.964pt}{0.800pt}}
\multiput(966.00,570.34)(-2.000,-2.000){2}{\rule{0.482pt}{0.800pt}}
\put(959,567.34){\rule{1.204pt}{0.800pt}}
\multiput(961.50,568.34)(-2.500,-2.000){2}{\rule{0.602pt}{0.800pt}}
\put(954,564.84){\rule{1.204pt}{0.800pt}}
\multiput(956.50,566.34)(-2.500,-3.000){2}{\rule{0.602pt}{0.800pt}}
\put(950,562.34){\rule{0.964pt}{0.800pt}}
\multiput(952.00,563.34)(-2.000,-2.000){2}{\rule{0.482pt}{0.800pt}}
\put(945,560.34){\rule{1.204pt}{0.800pt}}
\multiput(947.50,561.34)(-2.500,-2.000){2}{\rule{0.602pt}{0.800pt}}
\put(940,557.84){\rule{1.204pt}{0.800pt}}
\multiput(942.50,559.34)(-2.500,-3.000){2}{\rule{0.602pt}{0.800pt}}
\put(936,555.34){\rule{0.964pt}{0.800pt}}
\multiput(938.00,556.34)(-2.000,-2.000){2}{\rule{0.482pt}{0.800pt}}
\put(931,553.34){\rule{1.204pt}{0.800pt}}
\multiput(933.50,554.34)(-2.500,-2.000){2}{\rule{0.602pt}{0.800pt}}
\put(926,550.84){\rule{1.204pt}{0.800pt}}
\multiput(928.50,552.34)(-2.500,-3.000){2}{\rule{0.602pt}{0.800pt}}
\put(921,548.34){\rule{1.204pt}{0.800pt}}
\multiput(923.50,549.34)(-2.500,-2.000){2}{\rule{0.602pt}{0.800pt}}
\put(916,545.84){\rule{1.204pt}{0.800pt}}
\multiput(918.50,547.34)(-2.500,-3.000){2}{\rule{0.602pt}{0.800pt}}
\put(911,543.34){\rule{1.204pt}{0.800pt}}
\multiput(913.50,544.34)(-2.500,-2.000){2}{\rule{0.602pt}{0.800pt}}
\put(906,540.84){\rule{1.204pt}{0.800pt}}
\multiput(908.50,542.34)(-2.500,-3.000){2}{\rule{0.602pt}{0.800pt}}
\put(901,537.84){\rule{1.204pt}{0.800pt}}
\multiput(903.50,539.34)(-2.500,-3.000){2}{\rule{0.602pt}{0.800pt}}
\put(896,535.34){\rule{1.204pt}{0.800pt}}
\multiput(898.50,536.34)(-2.500,-2.000){2}{\rule{0.602pt}{0.800pt}}
\put(891,532.84){\rule{1.204pt}{0.800pt}}
\multiput(893.50,534.34)(-2.500,-3.000){2}{\rule{0.602pt}{0.800pt}}
\put(886,530.34){\rule{1.204pt}{0.800pt}}
\multiput(888.50,531.34)(-2.500,-2.000){2}{\rule{0.602pt}{0.800pt}}
\put(880,527.84){\rule{1.445pt}{0.800pt}}
\multiput(883.00,529.34)(-3.000,-3.000){2}{\rule{0.723pt}{0.800pt}}
\put(875,524.84){\rule{1.204pt}{0.800pt}}
\multiput(877.50,526.34)(-2.500,-3.000){2}{\rule{0.602pt}{0.800pt}}
\put(870,522.34){\rule{1.204pt}{0.800pt}}
\multiput(872.50,523.34)(-2.500,-2.000){2}{\rule{0.602pt}{0.800pt}}
\put(865,519.84){\rule{1.204pt}{0.800pt}}
\multiput(867.50,521.34)(-2.500,-3.000){2}{\rule{0.602pt}{0.800pt}}
\put(859,516.84){\rule{1.445pt}{0.800pt}}
\multiput(862.00,518.34)(-3.000,-3.000){2}{\rule{0.723pt}{0.800pt}}
\put(854,513.84){\rule{1.204pt}{0.800pt}}
\multiput(856.50,515.34)(-2.500,-3.000){2}{\rule{0.602pt}{0.800pt}}
\put(848,510.84){\rule{1.445pt}{0.800pt}}
\multiput(851.00,512.34)(-3.000,-3.000){2}{\rule{0.723pt}{0.800pt}}
\put(843,508.34){\rule{1.204pt}{0.800pt}}
\multiput(845.50,509.34)(-2.500,-2.000){2}{\rule{0.602pt}{0.800pt}}
\put(837,505.84){\rule{1.445pt}{0.800pt}}
\multiput(840.00,507.34)(-3.000,-3.000){2}{\rule{0.723pt}{0.800pt}}
\put(831,502.84){\rule{1.445pt}{0.800pt}}
\multiput(834.00,504.34)(-3.000,-3.000){2}{\rule{0.723pt}{0.800pt}}
\put(826,499.84){\rule{1.204pt}{0.800pt}}
\multiput(828.50,501.34)(-2.500,-3.000){2}{\rule{0.602pt}{0.800pt}}
\put(820,496.84){\rule{1.445pt}{0.800pt}}
\multiput(823.00,498.34)(-3.000,-3.000){2}{\rule{0.723pt}{0.800pt}}
\put(814,493.84){\rule{1.445pt}{0.800pt}}
\multiput(817.00,495.34)(-3.000,-3.000){2}{\rule{0.723pt}{0.800pt}}
\put(808,490.84){\rule{1.445pt}{0.800pt}}
\multiput(811.00,492.34)(-3.000,-3.000){2}{\rule{0.723pt}{0.800pt}}
\put(802,487.84){\rule{1.445pt}{0.800pt}}
\multiput(805.00,489.34)(-3.000,-3.000){2}{\rule{0.723pt}{0.800pt}}
\put(796,484.84){\rule{1.445pt}{0.800pt}}
\multiput(799.00,486.34)(-3.000,-3.000){2}{\rule{0.723pt}{0.800pt}}
\put(790,481.84){\rule{1.445pt}{0.800pt}}
\multiput(793.00,483.34)(-3.000,-3.000){2}{\rule{0.723pt}{0.800pt}}
\put(784,478.34){\rule{1.400pt}{0.800pt}}
\multiput(787.09,480.34)(-3.094,-4.000){2}{\rule{0.700pt}{0.800pt}}
\put(778,474.84){\rule{1.445pt}{0.800pt}}
\multiput(781.00,476.34)(-3.000,-3.000){2}{\rule{0.723pt}{0.800pt}}
\put(772,471.84){\rule{1.445pt}{0.800pt}}
\multiput(775.00,473.34)(-3.000,-3.000){2}{\rule{0.723pt}{0.800pt}}
\put(766,468.84){\rule{1.445pt}{0.800pt}}
\multiput(769.00,470.34)(-3.000,-3.000){2}{\rule{0.723pt}{0.800pt}}
\put(759,465.84){\rule{1.686pt}{0.800pt}}
\multiput(762.50,467.34)(-3.500,-3.000){2}{\rule{0.843pt}{0.800pt}}
\put(753,462.34){\rule{1.400pt}{0.800pt}}
\multiput(756.09,464.34)(-3.094,-4.000){2}{\rule{0.700pt}{0.800pt}}
\put(746,458.84){\rule{1.686pt}{0.800pt}}
\multiput(749.50,460.34)(-3.500,-3.000){2}{\rule{0.843pt}{0.800pt}}
\put(740,455.84){\rule{1.445pt}{0.800pt}}
\multiput(743.00,457.34)(-3.000,-3.000){2}{\rule{0.723pt}{0.800pt}}
\put(733,452.34){\rule{1.600pt}{0.800pt}}
\multiput(736.68,454.34)(-3.679,-4.000){2}{\rule{0.800pt}{0.800pt}}
\put(727,448.84){\rule{1.445pt}{0.800pt}}
\multiput(730.00,450.34)(-3.000,-3.000){2}{\rule{0.723pt}{0.800pt}}
\put(720,445.34){\rule{1.600pt}{0.800pt}}
\multiput(723.68,447.34)(-3.679,-4.000){2}{\rule{0.800pt}{0.800pt}}
\put(714,441.84){\rule{1.445pt}{0.800pt}}
\multiput(717.00,443.34)(-3.000,-3.000){2}{\rule{0.723pt}{0.800pt}}
\put(707,438.34){\rule{1.600pt}{0.800pt}}
\multiput(710.68,440.34)(-3.679,-4.000){2}{\rule{0.800pt}{0.800pt}}
\put(700,434.84){\rule{1.686pt}{0.800pt}}
\multiput(703.50,436.34)(-3.500,-3.000){2}{\rule{0.843pt}{0.800pt}}
\put(693,431.34){\rule{1.600pt}{0.800pt}}
\multiput(696.68,433.34)(-3.679,-4.000){2}{\rule{0.800pt}{0.800pt}}
\put(686,427.84){\rule{1.686pt}{0.800pt}}
\multiput(689.50,429.34)(-3.500,-3.000){2}{\rule{0.843pt}{0.800pt}}
\put(679,424.34){\rule{1.600pt}{0.800pt}}
\multiput(682.68,426.34)(-3.679,-4.000){2}{\rule{0.800pt}{0.800pt}}
\put(672,420.34){\rule{1.600pt}{0.800pt}}
\multiput(675.68,422.34)(-3.679,-4.000){2}{\rule{0.800pt}{0.800pt}}
\put(665,416.84){\rule{1.686pt}{0.800pt}}
\multiput(668.50,418.34)(-3.500,-3.000){2}{\rule{0.843pt}{0.800pt}}
\put(658,413.34){\rule{1.600pt}{0.800pt}}
\multiput(661.68,415.34)(-3.679,-4.000){2}{\rule{0.800pt}{0.800pt}}
\put(650,409.34){\rule{1.800pt}{0.800pt}}
\multiput(654.26,411.34)(-4.264,-4.000){2}{\rule{0.900pt}{0.800pt}}
\put(643,405.34){\rule{1.600pt}{0.800pt}}
\multiput(646.68,407.34)(-3.679,-4.000){2}{\rule{0.800pt}{0.800pt}}
\put(636,401.34){\rule{1.600pt}{0.800pt}}
\multiput(639.68,403.34)(-3.679,-4.000){2}{\rule{0.800pt}{0.800pt}}
\put(628,397.34){\rule{1.800pt}{0.800pt}}
\multiput(632.26,399.34)(-4.264,-4.000){2}{\rule{0.900pt}{0.800pt}}
\put(621,393.84){\rule{1.686pt}{0.800pt}}
\multiput(624.50,395.34)(-3.500,-3.000){2}{\rule{0.843pt}{0.800pt}}
\put(613,390.34){\rule{1.800pt}{0.800pt}}
\multiput(617.26,392.34)(-4.264,-4.000){2}{\rule{0.900pt}{0.800pt}}
\put(605,386.34){\rule{1.800pt}{0.800pt}}
\multiput(609.26,388.34)(-4.264,-4.000){2}{\rule{0.900pt}{0.800pt}}
\multiput(598.86,384.06)(-0.928,-0.560){3}{\rule{1.480pt}{0.135pt}}
\multiput(601.93,384.34)(-4.928,-5.000){2}{\rule{0.740pt}{0.800pt}}
\put(590,377.34){\rule{1.600pt}{0.800pt}}
\multiput(593.68,379.34)(-3.679,-4.000){2}{\rule{0.800pt}{0.800pt}}
\put(582,373.34){\rule{1.800pt}{0.800pt}}
\multiput(586.26,375.34)(-4.264,-4.000){2}{\rule{0.900pt}{0.800pt}}
\put(574,369.34){\rule{1.800pt}{0.800pt}}
\multiput(578.26,371.34)(-4.264,-4.000){2}{\rule{0.900pt}{0.800pt}}
\put(566,365.34){\rule{1.800pt}{0.800pt}}
\multiput(570.26,367.34)(-4.264,-4.000){2}{\rule{0.900pt}{0.800pt}}
\put(558,361.34){\rule{1.800pt}{0.800pt}}
\multiput(562.26,363.34)(-4.264,-4.000){2}{\rule{0.900pt}{0.800pt}}
\multiput(551.19,359.06)(-1.096,-0.560){3}{\rule{1.640pt}{0.135pt}}
\multiput(554.60,359.34)(-5.596,-5.000){2}{\rule{0.820pt}{0.800pt}}
\put(541,352.34){\rule{1.800pt}{0.800pt}}
\multiput(545.26,354.34)(-4.264,-4.000){2}{\rule{0.900pt}{0.800pt}}
\put(533,348.34){\rule{1.800pt}{0.800pt}}
\multiput(537.26,350.34)(-4.264,-4.000){2}{\rule{0.900pt}{0.800pt}}
\multiput(526.19,346.06)(-1.096,-0.560){3}{\rule{1.640pt}{0.135pt}}
\multiput(529.60,346.34)(-5.596,-5.000){2}{\rule{0.820pt}{0.800pt}}
\put(516,339.34){\rule{1.800pt}{0.800pt}}
\multiput(520.26,341.34)(-4.264,-4.000){2}{\rule{0.900pt}{0.800pt}}
\multiput(509.19,337.06)(-1.096,-0.560){3}{\rule{1.640pt}{0.135pt}}
\multiput(512.60,337.34)(-5.596,-5.000){2}{\rule{0.820pt}{0.800pt}}
\put(499,330.34){\rule{1.800pt}{0.800pt}}
\multiput(503.26,332.34)(-4.264,-4.000){2}{\rule{0.900pt}{0.800pt}}
\multiput(492.19,328.06)(-1.096,-0.560){3}{\rule{1.640pt}{0.135pt}}
\multiput(495.60,328.34)(-5.596,-5.000){2}{\rule{0.820pt}{0.800pt}}
\multiput(483.19,323.06)(-1.096,-0.560){3}{\rule{1.640pt}{0.135pt}}
\multiput(486.60,323.34)(-5.596,-5.000){2}{\rule{0.820pt}{0.800pt}}
\put(472,316.34){\rule{2.000pt}{0.800pt}}
\multiput(476.85,318.34)(-4.849,-4.000){2}{\rule{1.000pt}{0.800pt}}
\multiput(465.19,314.06)(-1.096,-0.560){3}{\rule{1.640pt}{0.135pt}}
\multiput(468.60,314.34)(-5.596,-5.000){2}{\rule{0.820pt}{0.800pt}}
\multiput(456.19,309.06)(-1.096,-0.560){3}{\rule{1.640pt}{0.135pt}}
\multiput(459.60,309.34)(-5.596,-5.000){2}{\rule{0.820pt}{0.800pt}}
\multiput(447.19,304.06)(-1.096,-0.560){3}{\rule{1.640pt}{0.135pt}}
\multiput(450.60,304.34)(-5.596,-5.000){2}{\rule{0.820pt}{0.800pt}}
\multiput(438.19,299.06)(-1.096,-0.560){3}{\rule{1.640pt}{0.135pt}}
\multiput(441.60,299.34)(-5.596,-5.000){2}{\rule{0.820pt}{0.800pt}}
\put(426,292.34){\rule{2.200pt}{0.800pt}}
\multiput(431.43,294.34)(-5.434,-4.000){2}{\rule{1.100pt}{0.800pt}}
\multiput(419.19,290.06)(-1.096,-0.560){3}{\rule{1.640pt}{0.135pt}}
\multiput(422.60,290.34)(-5.596,-5.000){2}{\rule{0.820pt}{0.800pt}}
\multiput(409.53,285.06)(-1.264,-0.560){3}{\rule{1.800pt}{0.135pt}}
\multiput(413.26,285.34)(-6.264,-5.000){2}{\rule{0.900pt}{0.800pt}}
\multiput(401.19,280.07)(-0.797,-0.536){5}{\rule{1.400pt}{0.129pt}}
\multiput(404.09,280.34)(-6.094,-6.000){2}{\rule{0.700pt}{0.800pt}}
\multiput(390.53,274.06)(-1.264,-0.560){3}{\rule{1.800pt}{0.135pt}}
\multiput(394.26,274.34)(-6.264,-5.000){2}{\rule{0.900pt}{0.800pt}}
\multiput(380.53,269.06)(-1.264,-0.560){3}{\rule{1.800pt}{0.135pt}}
\multiput(384.26,269.34)(-6.264,-5.000){2}{\rule{0.900pt}{0.800pt}}
\multiput(370.53,264.06)(-1.264,-0.560){3}{\rule{1.800pt}{0.135pt}}
\multiput(374.26,264.34)(-6.264,-5.000){2}{\rule{0.900pt}{0.800pt}}
\multiput(360.53,259.06)(-1.264,-0.560){3}{\rule{1.800pt}{0.135pt}}
\multiput(364.26,259.34)(-6.264,-5.000){2}{\rule{0.900pt}{0.800pt}}
\multiput(351.63,254.07)(-0.909,-0.536){5}{\rule{1.533pt}{0.129pt}}
\multiput(354.82,254.34)(-6.817,-6.000){2}{\rule{0.767pt}{0.800pt}}
\multiput(340.53,248.06)(-1.264,-0.560){3}{\rule{1.800pt}{0.135pt}}
\multiput(344.26,248.34)(-6.264,-5.000){2}{\rule{0.900pt}{0.800pt}}
\multiput(330.53,243.06)(-1.264,-0.560){3}{\rule{1.800pt}{0.135pt}}
\multiput(334.26,243.34)(-6.264,-5.000){2}{\rule{0.900pt}{0.800pt}}
\multiput(321.63,238.07)(-0.909,-0.536){5}{\rule{1.533pt}{0.129pt}}
\multiput(324.82,238.34)(-6.817,-6.000){2}{\rule{0.767pt}{0.800pt}}
\multiput(309.86,232.06)(-1.432,-0.560){3}{\rule{1.960pt}{0.135pt}}
\multiput(313.93,232.34)(-6.932,-5.000){2}{\rule{0.980pt}{0.800pt}}
\multiput(300.63,227.07)(-0.909,-0.536){5}{\rule{1.533pt}{0.129pt}}
\multiput(303.82,227.34)(-6.817,-6.000){2}{\rule{0.767pt}{0.800pt}}
\multiput(290.08,221.07)(-1.020,-0.536){5}{\rule{1.667pt}{0.129pt}}
\multiput(293.54,221.34)(-7.541,-6.000){2}{\rule{0.833pt}{0.800pt}}
\multiput(277.86,215.06)(-1.432,-0.560){3}{\rule{1.960pt}{0.135pt}}
\multiput(281.93,215.34)(-6.932,-5.000){2}{\rule{0.980pt}{0.800pt}}
\multiput(268.08,210.07)(-1.020,-0.536){5}{\rule{1.667pt}{0.129pt}}
\multiput(271.54,210.34)(-7.541,-6.000){2}{\rule{0.833pt}{0.800pt}}
\multiput(257.63,204.07)(-0.909,-0.536){5}{\rule{1.533pt}{0.129pt}}
\multiput(260.82,204.34)(-6.817,-6.000){2}{\rule{0.767pt}{0.800pt}}
\multiput(247.08,198.07)(-1.020,-0.536){5}{\rule{1.667pt}{0.129pt}}
\multiput(250.54,198.34)(-7.541,-6.000){2}{\rule{0.833pt}{0.800pt}}
\multiput(235.53,192.07)(-1.132,-0.536){5}{\rule{1.800pt}{0.129pt}}
\multiput(239.26,192.34)(-8.264,-6.000){2}{\rule{0.900pt}{0.800pt}}
\multiput(224.08,186.07)(-1.020,-0.536){5}{\rule{1.667pt}{0.129pt}}
\multiput(227.54,186.34)(-7.541,-6.000){2}{\rule{0.833pt}{0.800pt}}
\put(1268.0,718.0){\usebox{\plotpoint}}
\put(220,182){\usebox{\plotpoint}}
\sbox{\plotpoint}{\rule[-0.200pt]{0.400pt}{0.400pt}}%
\put(220,113){\usebox{\plotpoint}}
\put(220.00,113.00){\usebox{\plotpoint}}
\put(238.04,123.25){\usebox{\plotpoint}}
\put(256.50,132.75){\usebox{\plotpoint}}
\multiput(257,133)(17.928,10.458){0}{\usebox{\plotpoint}}
\put(274.44,143.17){\usebox{\plotpoint}}
\put(292.59,153.24){\usebox{\plotpoint}}
\multiput(294,154)(18.564,9.282){0}{\usebox{\plotpoint}}
\put(310.96,162.89){\usebox{\plotpoint}}
\put(329.09,172.97){\usebox{\plotpoint}}
\multiput(331,174)(17.928,10.458){0}{\usebox{\plotpoint}}
\put(347.20,183.10){\usebox{\plotpoint}}
\put(365.40,193.07){\usebox{\plotpoint}}
\multiput(367,194)(18.275,9.840){0}{\usebox{\plotpoint}}
\put(383.57,203.08){\usebox{\plotpoint}}
\put(401.84,212.92){\usebox{\plotpoint}}
\multiput(404,214)(18.275,9.840){0}{\usebox{\plotpoint}}
\put(420.09,222.80){\usebox{\plotpoint}}
\put(438.02,233.26){\usebox{\plotpoint}}
\multiput(441,235)(18.564,9.282){0}{\usebox{\plotpoint}}
\put(456.42,242.84){\usebox{\plotpoint}}
\put(474.53,252.98){\usebox{\plotpoint}}
\multiput(478,255)(17.928,10.458){0}{\usebox{\plotpoint}}
\put(492.58,263.19){\usebox{\plotpoint}}
\put(511.02,272.68){\usebox{\plotpoint}}
\multiput(515,275)(17.928,10.458){0}{\usebox{\plotpoint}}
\put(528.95,283.14){\usebox{\plotpoint}}
\put(547.28,292.82){\usebox{\plotpoint}}
\multiput(552,295)(17.928,10.458){0}{\usebox{\plotpoint}}
\put(565.43,302.84){\usebox{\plotpoint}}
\put(583.36,313.30){\usebox{\plotpoint}}
\multiput(588,316)(18.845,8.698){0}{\usebox{\plotpoint}}
\put(601.92,322.54){\usebox{\plotpoint}}
\put(619.85,333.00){\usebox{\plotpoint}}
\multiput(625,336)(18.275,9.840){0}{\usebox{\plotpoint}}
\put(638.03,343.01){\usebox{\plotpoint}}
\put(656.37,352.71){\usebox{\plotpoint}}
\multiput(662,356)(17.928,10.458){0}{\usebox{\plotpoint}}
\put(674.30,363.16){\usebox{\plotpoint}}
\put(692.66,372.83){\usebox{\plotpoint}}
\put(710.81,382.89){\usebox{\plotpoint}}
\multiput(711,383)(18.275,9.840){0}{\usebox{\plotpoint}}
\put(728.98,392.91){\usebox{\plotpoint}}
\put(747.30,402.65){\usebox{\plotpoint}}
\multiput(748,403)(17.928,10.458){0}{\usebox{\plotpoint}}
\put(765.35,412.88){\usebox{\plotpoint}}
\put(783.42,423.08){\usebox{\plotpoint}}
\multiput(785,424)(18.564,9.282){0}{\usebox{\plotpoint}}
\put(801.86,432.61){\usebox{\plotpoint}}
\put(819.94,442.80){\usebox{\plotpoint}}
\multiput(822,444)(17.928,10.458){0}{\usebox{\plotpoint}}
\put(838.00,453.00){\usebox{\plotpoint}}
\put(856.40,462.60){\usebox{\plotpoint}}
\multiput(859,464)(17.928,10.458){0}{\usebox{\plotpoint}}
\put(874.38,472.97){\usebox{\plotpoint}}
\put(892.79,482.52){\usebox{\plotpoint}}
\multiput(896,484)(17.928,10.458){0}{\usebox{\plotpoint}}
\put(910.87,492.67){\usebox{\plotpoint}}
\put(928.80,503.13){\usebox{\plotpoint}}
\multiput(932,505)(18.845,8.698){0}{\usebox{\plotpoint}}
\put(947.36,512.38){\usebox{\plotpoint}}
\put(965.29,522.83){\usebox{\plotpoint}}
\multiput(969,525)(18.275,9.840){0}{\usebox{\plotpoint}}
\put(983.51,532.76){\usebox{\plotpoint}}
\put(1001.80,542.55){\usebox{\plotpoint}}
\multiput(1006,545)(17.928,10.458){0}{\usebox{\plotpoint}}
\put(1019.76,552.95){\usebox{\plotpoint}}
\put(1038.15,562.57){\usebox{\plotpoint}}
\multiput(1043,565)(17.928,10.458){0}{\usebox{\plotpoint}}
\put(1056.27,572.68){\usebox{\plotpoint}}
\put(1074.42,582.74){\usebox{\plotpoint}}
\multiput(1080,586)(18.564,9.282){0}{\usebox{\plotpoint}}
\put(1092.76,592.44){\usebox{\plotpoint}}
\put(1110.81,602.67){\usebox{\plotpoint}}
\put(1128.86,612.92){\usebox{\plotpoint}}
\multiput(1129,613)(18.564,9.282){0}{\usebox{\plotpoint}}
\put(1147.20,622.62){\usebox{\plotpoint}}
\put(1165.36,632.66){\usebox{\plotpoint}}
\multiput(1166,633)(17.928,10.458){0}{\usebox{\plotpoint}}
\put(1183.49,642.74){\usebox{\plotpoint}}
\put(1201.87,652.39){\usebox{\plotpoint}}
\multiput(1203,653)(17.928,10.458){0}{\usebox{\plotpoint}}
\put(1219.82,662.81){\usebox{\plotpoint}}
\put(1238.13,672.56){\usebox{\plotpoint}}
\multiput(1239,673)(18.275,9.840){0}{\usebox{\plotpoint}}
\put(1256.33,682.53){\usebox{\plotpoint}}
\put(1274.26,692.98){\usebox{\plotpoint}}
\multiput(1276,694)(18.845,8.698){0}{\usebox{\plotpoint}}
\put(1292.82,702.23){\usebox{\plotpoint}}
\put(1310.75,712.69){\usebox{\plotpoint}}
\multiput(1313,714)(17.928,10.458){0}{\usebox{\plotpoint}}
\put(1328.86,722.78){\usebox{\plotpoint}}
\put(1347.24,732.39){\usebox{\plotpoint}}
\multiput(1350,734)(17.928,10.458){0}{\usebox{\plotpoint}}
\put(1365.23,742.74){\usebox{\plotpoint}}
\put(1383.63,752.32){\usebox{\plotpoint}}
\multiput(1387,754)(17.928,10.458){0}{\usebox{\plotpoint}}
\put(1401.68,762.56){\usebox{\plotpoint}}
\put(1419.77,772.72){\usebox{\plotpoint}}
\multiput(1424,775)(17.928,10.458){0}{\usebox{\plotpoint}}
\put(1436,782){\usebox{\plotpoint}}
\end{picture}

%% file: btau250.tex
\setlength{\unitlength}{0.240900pt}
\ifx\plotpoint\undefined\newsavebox{\plotpoint}\fi
\sbox{\plotpoint}{\rule[-0.200pt]{0.400pt}{0.400pt}}%
\begin{picture}(1500,900)(0,0)
\font\gnuplot=cmr10 at 10pt
\gnuplot
\sbox{\plotpoint}{\rule[-0.200pt]{0.400pt}{0.400pt}}%
\put(220.0,113.0){\rule[-0.200pt]{4.818pt}{0.400pt}}
\put(198,113){\makebox(0,0)[r]{2}}
\put(1416.0,113.0){\rule[-0.200pt]{4.818pt}{0.400pt}}
\put(220.0,240.0){\rule[-0.200pt]{4.818pt}{0.400pt}}
\put(198,240){\makebox(0,0)[r]{2.5}}
\put(1416.0,240.0){\rule[-0.200pt]{4.818pt}{0.400pt}}
\put(220.0,368.0){\rule[-0.200pt]{4.818pt}{0.400pt}}
\put(198,368){\makebox(0,0)[r]{3}}
\put(1416.0,368.0){\rule[-0.200pt]{4.818pt}{0.400pt}}
\put(220.0,495.0){\rule[-0.200pt]{4.818pt}{0.400pt}}
\put(198,495){\makebox(0,0)[r]{3.5}}
\put(1416.0,495.0){\rule[-0.200pt]{4.818pt}{0.400pt}}
\put(220.0,622.0){\rule[-0.200pt]{4.818pt}{0.400pt}}
\put(198,622){\makebox(0,0)[r]{4}}
\put(1416.0,622.0){\rule[-0.200pt]{4.818pt}{0.400pt}}
\put(220.0,750.0){\rule[-0.200pt]{4.818pt}{0.400pt}}
\put(198,750){\makebox(0,0)[r]{4.5}}
\put(1416.0,750.0){\rule[-0.200pt]{4.818pt}{0.400pt}}
\put(220.0,877.0){\rule[-0.200pt]{4.818pt}{0.400pt}}
\put(198,877){\makebox(0,0)[r]{5}}
\put(1416.0,877.0){\rule[-0.200pt]{4.818pt}{0.400pt}}
\put(220.0,113.0){\rule[-0.200pt]{0.400pt}{4.818pt}}
\put(220,68){\makebox(0,0){4}}
\put(220.0,857.0){\rule[-0.200pt]{0.400pt}{4.818pt}}
\put(423.0,113.0){\rule[-0.200pt]{0.400pt}{4.818pt}}
\put(423,68){\makebox(0,0){6}}
\put(423.0,857.0){\rule[-0.200pt]{0.400pt}{4.818pt}}
\put(625.0,113.0){\rule[-0.200pt]{0.400pt}{4.818pt}}
\put(625,68){\makebox(0,0){8}}
\put(625.0,857.0){\rule[-0.200pt]{0.400pt}{4.818pt}}
\put(828.0,113.0){\rule[-0.200pt]{0.400pt}{4.818pt}}
\put(828,68){\makebox(0,0){10}}
\put(828.0,857.0){\rule[-0.200pt]{0.400pt}{4.818pt}}
\put(1031.0,113.0){\rule[-0.200pt]{0.400pt}{4.818pt}}
\put(1031,68){\makebox(0,0){12}}
\put(1031.0,857.0){\rule[-0.200pt]{0.400pt}{4.818pt}}
\put(1233.0,113.0){\rule[-0.200pt]{0.400pt}{4.818pt}}
\put(1233,68){\makebox(0,0){14}}
\put(1233.0,857.0){\rule[-0.200pt]{0.400pt}{4.818pt}}
\put(1436.0,113.0){\rule[-0.200pt]{0.400pt}{4.818pt}}
\put(1436,68){\makebox(0,0){16}}
\put(1436.0,857.0){\rule[-0.200pt]{0.400pt}{4.818pt}}
\put(220.0,113.0){\rule[-0.200pt]{292.934pt}{0.400pt}}
\put(1436.0,113.0){\rule[-0.200pt]{0.400pt}{184.048pt}}
\put(220.0,877.0){\rule[-0.200pt]{292.934pt}{0.400pt}}
\put(45,495){\makebox(0,0){$m_b(M_Z)$}}
\put(50,445){\makebox(0,0){[GeV]}}
\put(828,23){\makebox(0,0){$\log_{10} \mu_0$ [GeV]}}
\put(1335,597){\makebox(0,0)[r]{$\tan \beta =2$}}
\put(1335,444){\makebox(0,0)[r]{$\tan \beta =50$}}
\put(220.0,113.0){\rule[-0.200pt]{0.400pt}{184.048pt}}
\sbox{\plotpoint}{\rule[-0.400pt]{0.800pt}{0.800pt}}%
\put(220,663){\usebox{\plotpoint}}
\put(220,662.84){\rule{12.286pt}{0.800pt}}
\multiput(220.00,661.34)(25.500,3.000){2}{\rule{6.143pt}{0.800pt}}
\multiput(271.00,667.38)(7.980,0.560){3}{\rule{8.200pt}{0.135pt}}
\multiput(271.00,664.34)(32.980,5.000){2}{\rule{4.100pt}{0.800pt}}
\put(372,667.84){\rule{12.286pt}{0.800pt}}
\multiput(372.00,669.34)(25.500,-3.000){2}{\rule{6.143pt}{0.800pt}}
\put(423,665.34){\rule{12.045pt}{0.800pt}}
\multiput(423.00,666.34)(25.000,-2.000){2}{\rule{6.022pt}{0.800pt}}
\put(473,662.84){\rule{12.286pt}{0.800pt}}
\multiput(473.00,664.34)(25.500,-3.000){2}{\rule{6.143pt}{0.800pt}}
\multiput(524.00,661.06)(8.148,-0.560){3}{\rule{8.360pt}{0.135pt}}
\multiput(524.00,661.34)(33.648,-5.000){2}{\rule{4.180pt}{0.800pt}}
\multiput(575.00,656.08)(3.552,-0.520){9}{\rule{5.200pt}{0.125pt}}
\multiput(575.00,656.34)(39.207,-8.000){2}{\rule{2.600pt}{0.800pt}}
\multiput(625.00,648.06)(8.148,-0.560){3}{\rule{8.360pt}{0.135pt}}
\multiput(625.00,648.34)(33.648,-5.000){2}{\rule{4.180pt}{0.800pt}}
\multiput(676.00,643.08)(4.329,-0.526){7}{\rule{6.029pt}{0.127pt}}
\multiput(676.00,643.34)(38.487,-7.000){2}{\rule{3.014pt}{0.800pt}}
\multiput(727.00,636.08)(2.434,-0.512){15}{\rule{3.836pt}{0.123pt}}
\multiput(727.00,636.34)(42.037,-11.000){2}{\rule{1.918pt}{0.800pt}}
\multiput(777.00,625.08)(4.329,-0.526){7}{\rule{6.029pt}{0.127pt}}
\multiput(777.00,625.34)(38.487,-7.000){2}{\rule{3.014pt}{0.800pt}}
\multiput(828.00,618.08)(3.625,-0.520){9}{\rule{5.300pt}{0.125pt}}
\multiput(828.00,618.34)(40.000,-8.000){2}{\rule{2.650pt}{0.800pt}}
\multiput(879.00,610.08)(2.434,-0.512){15}{\rule{3.836pt}{0.123pt}}
\multiput(879.00,610.34)(42.037,-11.000){2}{\rule{1.918pt}{0.800pt}}
\multiput(929.00,599.08)(2.769,-0.514){13}{\rule{4.280pt}{0.124pt}}
\multiput(929.00,599.34)(42.117,-10.000){2}{\rule{2.140pt}{0.800pt}}
\multiput(980.00,589.08)(2.484,-0.512){15}{\rule{3.909pt}{0.123pt}}
\multiput(980.00,589.34)(42.886,-11.000){2}{\rule{1.955pt}{0.800pt}}
\multiput(1031.00,578.08)(2.434,-0.512){15}{\rule{3.836pt}{0.123pt}}
\multiput(1031.00,578.34)(42.037,-11.000){2}{\rule{1.918pt}{0.800pt}}
\multiput(1081.00,567.08)(2.254,-0.511){17}{\rule{3.600pt}{0.123pt}}
\multiput(1081.00,567.34)(43.528,-12.000){2}{\rule{1.800pt}{0.800pt}}
\multiput(1132.00,555.08)(2.769,-0.514){13}{\rule{4.280pt}{0.124pt}}
\multiput(1132.00,555.34)(42.117,-10.000){2}{\rule{2.140pt}{0.800pt}}
\multiput(1183.00,545.08)(2.434,-0.512){15}{\rule{3.836pt}{0.123pt}}
\multiput(1183.00,545.34)(42.037,-11.000){2}{\rule{1.918pt}{0.800pt}}
\multiput(1233.00,534.08)(2.769,-0.514){13}{\rule{4.280pt}{0.124pt}}
\multiput(1233.00,534.34)(42.117,-10.000){2}{\rule{2.140pt}{0.800pt}}
\multiput(1284.00,524.08)(3.135,-0.516){11}{\rule{4.733pt}{0.124pt}}
\multiput(1284.00,524.34)(41.176,-9.000){2}{\rule{2.367pt}{0.800pt}}
\multiput(1335.00,515.08)(3.072,-0.516){11}{\rule{4.644pt}{0.124pt}}
\multiput(1335.00,515.34)(40.360,-9.000){2}{\rule{2.322pt}{0.800pt}}
\multiput(1385.00,506.06)(8.148,-0.560){3}{\rule{8.360pt}{0.135pt}}
\multiput(1385.00,506.34)(33.648,-5.000){2}{\rule{4.180pt}{0.800pt}}
\put(321.0,671.0){\rule[-0.400pt]{12.286pt}{0.800pt}}
\put(220,439){\usebox{\plotpoint}}
\multiput(220.00,440.38)(8.148,0.560){3}{\rule{8.360pt}{0.135pt}}
\multiput(220.00,437.34)(33.648,5.000){2}{\rule{4.180pt}{0.800pt}}
\multiput(271.00,445.38)(7.980,0.560){3}{\rule{8.200pt}{0.135pt}}
\multiput(271.00,442.34)(32.980,5.000){2}{\rule{4.100pt}{0.800pt}}
\multiput(321.00,450.38)(8.148,0.560){3}{\rule{8.360pt}{0.135pt}}
\multiput(321.00,447.34)(33.648,5.000){2}{\rule{4.180pt}{0.800pt}}
\put(372,453.84){\rule{12.286pt}{0.800pt}}
\multiput(372.00,452.34)(25.500,3.000){2}{\rule{6.143pt}{0.800pt}}
\multiput(423.00,458.38)(7.980,0.560){3}{\rule{8.200pt}{0.135pt}}
\multiput(423.00,455.34)(32.980,5.000){2}{\rule{4.100pt}{0.800pt}}
\put(473,461.34){\rule{12.286pt}{0.800pt}}
\multiput(473.00,460.34)(25.500,2.000){2}{\rule{6.143pt}{0.800pt}}
\put(524,463.84){\rule{12.286pt}{0.800pt}}
\multiput(524.00,462.34)(25.500,3.000){2}{\rule{6.143pt}{0.800pt}}
\put(575,466.84){\rule{12.045pt}{0.800pt}}
\multiput(575.00,465.34)(25.000,3.000){2}{\rule{6.022pt}{0.800pt}}
\put(625,469.34){\rule{12.286pt}{0.800pt}}
\multiput(625.00,468.34)(25.500,2.000){2}{\rule{6.143pt}{0.800pt}}
\put(777,471.84){\rule{12.286pt}{0.800pt}}
\multiput(777.00,470.34)(25.500,3.000){2}{\rule{6.143pt}{0.800pt}}
\put(676.0,472.0){\rule[-0.400pt]{24.331pt}{0.800pt}}
\put(1031,471.84){\rule{12.045pt}{0.800pt}}
\multiput(1031.00,473.34)(25.000,-3.000){2}{\rule{6.022pt}{0.800pt}}
\put(828.0,475.0){\rule[-0.400pt]{48.903pt}{0.800pt}}
\put(1183,469.34){\rule{12.045pt}{0.800pt}}
\multiput(1183.00,470.34)(25.000,-2.000){2}{\rule{6.022pt}{0.800pt}}
\put(1233,466.84){\rule{12.286pt}{0.800pt}}
\multiput(1233.00,468.34)(25.500,-3.000){2}{\rule{6.143pt}{0.800pt}}
\put(1081.0,472.0){\rule[-0.400pt]{24.572pt}{0.800pt}}
\put(1335,463.84){\rule{12.045pt}{0.800pt}}
\multiput(1335.00,465.34)(25.000,-3.000){2}{\rule{6.022pt}{0.800pt}}
\put(1284.0,467.0){\rule[-0.400pt]{12.286pt}{0.800pt}}
\put(1385.0,464.0){\rule[-0.400pt]{12.286pt}{0.800pt}}
\end{picture}

%% file: gaugino.tex
\setlength{\unitlength}{0.240900pt}
\ifx\plotpoint\undefined\newsavebox{\plotpoint}\fi
\sbox{\plotpoint}{\rule[-0.200pt]{0.400pt}{0.400pt}}%
\begin{picture}(1500,900)(0,0)
\font\gnuplot=cmr10 at 10pt
\gnuplot
\sbox{\plotpoint}{\rule[-0.200pt]{0.400pt}{0.400pt}}%
\put(220.0,113.0){\rule[-0.200pt]{292.934pt}{0.400pt}}
\put(220.0,113.0){\rule[-0.200pt]{4.818pt}{0.400pt}}
\put(198,113){\makebox(0,0)[r]{0}}
\put(1416.0,113.0){\rule[-0.200pt]{4.818pt}{0.400pt}}
\put(220.0,240.0){\rule[-0.200pt]{4.818pt}{0.400pt}}
\put(198,240){\makebox(0,0)[r]{1}}
\put(1416.0,240.0){\rule[-0.200pt]{4.818pt}{0.400pt}}
\put(220.0,368.0){\rule[-0.200pt]{4.818pt}{0.400pt}}
\put(198,368){\makebox(0,0)[r]{2}}
\put(1416.0,368.0){\rule[-0.200pt]{4.818pt}{0.400pt}}
\put(220.0,495.0){\rule[-0.200pt]{4.818pt}{0.400pt}}
\put(198,495){\makebox(0,0)[r]{3}}
\put(1416.0,495.0){\rule[-0.200pt]{4.818pt}{0.400pt}}
\put(220.0,622.0){\rule[-0.200pt]{4.818pt}{0.400pt}}
\put(198,622){\makebox(0,0)[r]{4}}
\put(1416.0,622.0){\rule[-0.200pt]{4.818pt}{0.400pt}}
\put(220.0,750.0){\rule[-0.200pt]{4.818pt}{0.400pt}}
\put(198,750){\makebox(0,0)[r]{5}}
\put(1416.0,750.0){\rule[-0.200pt]{4.818pt}{0.400pt}}
\put(220.0,877.0){\rule[-0.200pt]{4.818pt}{0.400pt}}
\put(198,877){\makebox(0,0)[r]{6}}
\put(1416.0,877.0){\rule[-0.200pt]{4.818pt}{0.400pt}}
\put(220.0,113.0){\rule[-0.200pt]{0.400pt}{4.818pt}}
\put(220,68){\makebox(0,0){4}}
\put(220.0,857.0){\rule[-0.200pt]{0.400pt}{4.818pt}}
\put(423.0,113.0){\rule[-0.200pt]{0.400pt}{4.818pt}}
\put(423,68){\makebox(0,0){6}}
\put(423.0,857.0){\rule[-0.200pt]{0.400pt}{4.818pt}}
\put(625.0,113.0){\rule[-0.200pt]{0.400pt}{4.818pt}}
\put(625,68){\makebox(0,0){8}}
\put(625.0,857.0){\rule[-0.200pt]{0.400pt}{4.818pt}}
\put(828.0,113.0){\rule[-0.200pt]{0.400pt}{4.818pt}}
\put(828,68){\makebox(0,0){10}}
\put(828.0,857.0){\rule[-0.200pt]{0.400pt}{4.818pt}}
\put(1031.0,113.0){\rule[-0.200pt]{0.400pt}{4.818pt}}
\put(1031,68){\makebox(0,0){12}}
\put(1031.0,857.0){\rule[-0.200pt]{0.400pt}{4.818pt}}
\put(1233.0,113.0){\rule[-0.200pt]{0.400pt}{4.818pt}}
\put(1233,68){\makebox(0,0){14}}
\put(1233.0,857.0){\rule[-0.200pt]{0.400pt}{4.818pt}}
\put(1436.0,113.0){\rule[-0.200pt]{0.400pt}{4.818pt}}
\put(1436,68){\makebox(0,0){16}}
\put(1436.0,857.0){\rule[-0.200pt]{0.400pt}{4.818pt}}
\put(220.0,113.0){\rule[-0.200pt]{292.934pt}{0.400pt}}
\put(1436.0,113.0){\rule[-0.200pt]{0.400pt}{184.048pt}}
\put(220.0,877.0){\rule[-0.200pt]{292.934pt}{0.400pt}}
\put(45,495){\makebox(0,0){$M_a$[TeV]}}
\put(828,23){\makebox(0,0){$\log_{10}\mu_0$[GeV]}}
\put(1335,393){\makebox(0,0)[r]{$M_3$}}
\put(1335,202){\makebox(0,0)[r]{$M_2$}}
\put(1335,151){\makebox(0,0)[r]{$M_1$}}
\put(220.0,113.0){\rule[-0.200pt]{0.400pt}{184.048pt}}
\sbox{\plotpoint}{\rule[-0.400pt]{0.800pt}{0.800pt}}%
\put(220,786){\usebox{\plotpoint}}
\multiput(220.00,784.09)(1.375,-0.506){31}{\rule{2.347pt}{0.122pt}}
\multiput(220.00,784.34)(46.128,-19.000){2}{\rule{1.174pt}{0.800pt}}
\multiput(271.00,765.09)(1.427,-0.506){29}{\rule{2.422pt}{0.122pt}}
\multiput(271.00,765.34)(44.973,-18.000){2}{\rule{1.211pt}{0.800pt}}
\multiput(321.00,747.09)(1.303,-0.505){33}{\rule{2.240pt}{0.122pt}}
\multiput(321.00,747.34)(46.351,-20.000){2}{\rule{1.120pt}{0.800pt}}
\multiput(372.00,727.09)(1.375,-0.506){31}{\rule{2.347pt}{0.122pt}}
\multiput(372.00,727.34)(46.128,-19.000){2}{\rule{1.174pt}{0.800pt}}
\multiput(423.00,708.09)(1.348,-0.506){31}{\rule{2.305pt}{0.122pt}}
\multiput(423.00,708.34)(45.215,-19.000){2}{\rule{1.153pt}{0.800pt}}
\multiput(473.00,689.09)(1.456,-0.506){29}{\rule{2.467pt}{0.122pt}}
\multiput(473.00,689.34)(45.880,-18.000){2}{\rule{1.233pt}{0.800pt}}
\multiput(524.00,671.09)(1.375,-0.506){31}{\rule{2.347pt}{0.122pt}}
\multiput(524.00,671.34)(46.128,-19.000){2}{\rule{1.174pt}{0.800pt}}
\multiput(575.00,652.09)(1.515,-0.507){27}{\rule{2.553pt}{0.122pt}}
\multiput(575.00,652.34)(44.701,-17.000){2}{\rule{1.276pt}{0.800pt}}
\multiput(625.00,635.09)(1.546,-0.507){27}{\rule{2.600pt}{0.122pt}}
\multiput(625.00,635.34)(45.604,-17.000){2}{\rule{1.300pt}{0.800pt}}
\multiput(676.00,618.09)(1.456,-0.506){29}{\rule{2.467pt}{0.122pt}}
\multiput(676.00,618.34)(45.880,-18.000){2}{\rule{1.233pt}{0.800pt}}
\multiput(727.00,600.09)(1.515,-0.507){27}{\rule{2.553pt}{0.122pt}}
\multiput(727.00,600.34)(44.701,-17.000){2}{\rule{1.276pt}{0.800pt}}
\multiput(777.00,583.09)(1.649,-0.507){25}{\rule{2.750pt}{0.122pt}}
\multiput(777.00,583.34)(45.292,-16.000){2}{\rule{1.375pt}{0.800pt}}
\multiput(828.00,567.09)(1.649,-0.507){25}{\rule{2.750pt}{0.122pt}}
\multiput(828.00,567.34)(45.292,-16.000){2}{\rule{1.375pt}{0.800pt}}
\multiput(879.00,551.09)(1.616,-0.507){25}{\rule{2.700pt}{0.122pt}}
\multiput(879.00,551.34)(44.396,-16.000){2}{\rule{1.350pt}{0.800pt}}
\multiput(929.00,535.09)(1.649,-0.507){25}{\rule{2.750pt}{0.122pt}}
\multiput(929.00,535.34)(45.292,-16.000){2}{\rule{1.375pt}{0.800pt}}
\multiput(980.00,519.09)(1.767,-0.508){23}{\rule{2.920pt}{0.122pt}}
\multiput(980.00,519.34)(44.939,-15.000){2}{\rule{1.460pt}{0.800pt}}
\multiput(1031.00,504.09)(1.732,-0.508){23}{\rule{2.867pt}{0.122pt}}
\multiput(1031.00,504.34)(44.050,-15.000){2}{\rule{1.433pt}{0.800pt}}
\multiput(1081.00,489.09)(1.904,-0.509){21}{\rule{3.114pt}{0.123pt}}
\multiput(1081.00,489.34)(44.536,-14.000){2}{\rule{1.557pt}{0.800pt}}
\multiput(1132.00,475.09)(1.904,-0.509){21}{\rule{3.114pt}{0.123pt}}
\multiput(1132.00,475.34)(44.536,-14.000){2}{\rule{1.557pt}{0.800pt}}
\multiput(1183.00,461.08)(2.022,-0.509){19}{\rule{3.277pt}{0.123pt}}
\multiput(1183.00,461.34)(43.199,-13.000){2}{\rule{1.638pt}{0.800pt}}
\multiput(1233.00,448.08)(2.254,-0.511){17}{\rule{3.600pt}{0.123pt}}
\multiput(1233.00,448.34)(43.528,-12.000){2}{\rule{1.800pt}{0.800pt}}
\multiput(1284.00,436.08)(2.484,-0.512){15}{\rule{3.909pt}{0.123pt}}
\multiput(1284.00,436.34)(42.886,-11.000){2}{\rule{1.955pt}{0.800pt}}
\multiput(1335.00,425.08)(2.714,-0.514){13}{\rule{4.200pt}{0.124pt}}
\multiput(1335.00,425.34)(41.283,-10.000){2}{\rule{2.100pt}{0.800pt}}
\multiput(1385.00,415.08)(4.329,-0.526){7}{\rule{6.029pt}{0.127pt}}
\multiput(1385.00,415.34)(38.487,-7.000){2}{\rule{3.014pt}{0.800pt}}
\put(220,311){\usebox{\plotpoint}}
\put(220,307.34){\rule{10.400pt}{0.800pt}}
\multiput(220.00,309.34)(29.414,-4.000){2}{\rule{5.200pt}{0.800pt}}
\put(271,303.34){\rule{10.200pt}{0.800pt}}
\multiput(271.00,305.34)(28.829,-4.000){2}{\rule{5.100pt}{0.800pt}}
\put(321,299.34){\rule{10.400pt}{0.800pt}}
\multiput(321.00,301.34)(29.414,-4.000){2}{\rule{5.200pt}{0.800pt}}
\put(372,295.34){\rule{10.400pt}{0.800pt}}
\multiput(372.00,297.34)(29.414,-4.000){2}{\rule{5.200pt}{0.800pt}}
\put(423,291.34){\rule{10.200pt}{0.800pt}}
\multiput(423.00,293.34)(28.829,-4.000){2}{\rule{5.100pt}{0.800pt}}
\put(473,287.34){\rule{10.400pt}{0.800pt}}
\multiput(473.00,289.34)(29.414,-4.000){2}{\rule{5.200pt}{0.800pt}}
\put(524,283.34){\rule{10.400pt}{0.800pt}}
\multiput(524.00,285.34)(29.414,-4.000){2}{\rule{5.200pt}{0.800pt}}
\put(575,279.34){\rule{10.200pt}{0.800pt}}
\multiput(575.00,281.34)(28.829,-4.000){2}{\rule{5.100pt}{0.800pt}}
\put(625,275.34){\rule{10.400pt}{0.800pt}}
\multiput(625.00,277.34)(29.414,-4.000){2}{\rule{5.200pt}{0.800pt}}
\put(676,271.34){\rule{10.400pt}{0.800pt}}
\multiput(676.00,273.34)(29.414,-4.000){2}{\rule{5.200pt}{0.800pt}}
\put(727,267.34){\rule{10.200pt}{0.800pt}}
\multiput(727.00,269.34)(28.829,-4.000){2}{\rule{5.100pt}{0.800pt}}
\put(777,263.34){\rule{10.400pt}{0.800pt}}
\multiput(777.00,265.34)(29.414,-4.000){2}{\rule{5.200pt}{0.800pt}}
\put(828,259.34){\rule{10.400pt}{0.800pt}}
\multiput(828.00,261.34)(29.414,-4.000){2}{\rule{5.200pt}{0.800pt}}
\put(879,255.34){\rule{10.200pt}{0.800pt}}
\multiput(879.00,257.34)(28.829,-4.000){2}{\rule{5.100pt}{0.800pt}}
\put(929,251.34){\rule{10.400pt}{0.800pt}}
\multiput(929.00,253.34)(29.414,-4.000){2}{\rule{5.200pt}{0.800pt}}
\put(980,247.34){\rule{10.400pt}{0.800pt}}
\multiput(980.00,249.34)(29.414,-4.000){2}{\rule{5.200pt}{0.800pt}}
\put(1031,243.34){\rule{10.200pt}{0.800pt}}
\multiput(1031.00,245.34)(28.829,-4.000){2}{\rule{5.100pt}{0.800pt}}
\put(1081,239.34){\rule{10.400pt}{0.800pt}}
\multiput(1081.00,241.34)(29.414,-4.000){2}{\rule{5.200pt}{0.800pt}}
\put(1132,235.84){\rule{12.286pt}{0.800pt}}
\multiput(1132.00,237.34)(25.500,-3.000){2}{\rule{6.143pt}{0.800pt}}
\put(1183,232.34){\rule{10.200pt}{0.800pt}}
\multiput(1183.00,234.34)(28.829,-4.000){2}{\rule{5.100pt}{0.800pt}}
\put(1233,228.84){\rule{12.286pt}{0.800pt}}
\multiput(1233.00,230.34)(25.500,-3.000){2}{\rule{6.143pt}{0.800pt}}
\put(1284,225.34){\rule{10.400pt}{0.800pt}}
\multiput(1284.00,227.34)(29.414,-4.000){2}{\rule{5.200pt}{0.800pt}}
\put(1335,222.34){\rule{12.045pt}{0.800pt}}
\multiput(1335.00,223.34)(25.000,-2.000){2}{\rule{6.022pt}{0.800pt}}
\put(1385,219.84){\rule{12.286pt}{0.800pt}}
\multiput(1385.00,221.34)(25.500,-3.000){2}{\rule{6.143pt}{0.800pt}}
\put(220,219){\usebox{\plotpoint}}
\put(220,216.34){\rule{12.286pt}{0.800pt}}
\multiput(220.00,217.34)(25.500,-2.000){2}{\rule{6.143pt}{0.800pt}}
\put(271,213.84){\rule{12.045pt}{0.800pt}}
\multiput(271.00,215.34)(25.000,-3.000){2}{\rule{6.022pt}{0.800pt}}
\put(321,211.34){\rule{12.286pt}{0.800pt}}
\multiput(321.00,212.34)(25.500,-2.000){2}{\rule{6.143pt}{0.800pt}}
\put(372,209.34){\rule{12.286pt}{0.800pt}}
\multiput(372.00,210.34)(25.500,-2.000){2}{\rule{6.143pt}{0.800pt}}
\put(423,207.34){\rule{12.045pt}{0.800pt}}
\multiput(423.00,208.34)(25.000,-2.000){2}{\rule{6.022pt}{0.800pt}}
\put(473,205.34){\rule{12.286pt}{0.800pt}}
\multiput(473.00,206.34)(25.500,-2.000){2}{\rule{6.143pt}{0.800pt}}
\put(524,202.84){\rule{12.286pt}{0.800pt}}
\multiput(524.00,204.34)(25.500,-3.000){2}{\rule{6.143pt}{0.800pt}}
\put(575,200.34){\rule{12.045pt}{0.800pt}}
\multiput(575.00,201.34)(25.000,-2.000){2}{\rule{6.022pt}{0.800pt}}
\put(625,198.34){\rule{12.286pt}{0.800pt}}
\multiput(625.00,199.34)(25.500,-2.000){2}{\rule{6.143pt}{0.800pt}}
\put(676,196.34){\rule{12.286pt}{0.800pt}}
\multiput(676.00,197.34)(25.500,-2.000){2}{\rule{6.143pt}{0.800pt}}
\put(727,194.34){\rule{12.045pt}{0.800pt}}
\multiput(727.00,195.34)(25.000,-2.000){2}{\rule{6.022pt}{0.800pt}}
\put(777,191.84){\rule{12.286pt}{0.800pt}}
\multiput(777.00,193.34)(25.500,-3.000){2}{\rule{6.143pt}{0.800pt}}
\put(828,189.34){\rule{12.286pt}{0.800pt}}
\multiput(828.00,190.34)(25.500,-2.000){2}{\rule{6.143pt}{0.800pt}}
\put(879,187.34){\rule{12.045pt}{0.800pt}}
\multiput(879.00,188.34)(25.000,-2.000){2}{\rule{6.022pt}{0.800pt}}
\put(929,185.34){\rule{12.286pt}{0.800pt}}
\multiput(929.00,186.34)(25.500,-2.000){2}{\rule{6.143pt}{0.800pt}}
\put(980,183.34){\rule{12.286pt}{0.800pt}}
\multiput(980.00,184.34)(25.500,-2.000){2}{\rule{6.143pt}{0.800pt}}
\put(1031,181.34){\rule{12.045pt}{0.800pt}}
\multiput(1031.00,182.34)(25.000,-2.000){2}{\rule{6.022pt}{0.800pt}}
\put(1081,179.34){\rule{12.286pt}{0.800pt}}
\multiput(1081.00,180.34)(25.500,-2.000){2}{\rule{6.143pt}{0.800pt}}
\put(1132,177.34){\rule{12.286pt}{0.800pt}}
\multiput(1132.00,178.34)(25.500,-2.000){2}{\rule{6.143pt}{0.800pt}}
\put(1183,175.34){\rule{12.045pt}{0.800pt}}
\multiput(1183.00,176.34)(25.000,-2.000){2}{\rule{6.022pt}{0.800pt}}
\put(1233,173.34){\rule{12.286pt}{0.800pt}}
\multiput(1233.00,174.34)(25.500,-2.000){2}{\rule{6.143pt}{0.800pt}}
\put(1284,171.34){\rule{12.286pt}{0.800pt}}
\multiput(1284.00,172.34)(25.500,-2.000){2}{\rule{6.143pt}{0.800pt}}
\put(1335,169.84){\rule{12.045pt}{0.800pt}}
\multiput(1335.00,170.34)(25.000,-1.000){2}{\rule{6.022pt}{0.800pt}}
\put(1385,168.84){\rule{12.286pt}{0.800pt}}
\multiput(1385.00,169.34)(25.500,-1.000){2}{\rule{6.143pt}{0.800pt}}
\sbox{\plotpoint}{\rule[-0.200pt]{0.400pt}{0.400pt}}%
\put(220,240){\usebox{\plotpoint}}
\put(220.00,240.00){\usebox{\plotpoint}}
\put(240.76,240.00){\usebox{\plotpoint}}
\multiput(245,240)(20.756,0.000){0}{\usebox{\plotpoint}}
\put(261.51,240.00){\usebox{\plotpoint}}
\multiput(269,240)(20.756,0.000){0}{\usebox{\plotpoint}}
\put(282.27,240.00){\usebox{\plotpoint}}
\put(303.02,240.00){\usebox{\plotpoint}}
\multiput(306,240)(20.756,0.000){0}{\usebox{\plotpoint}}
\put(323.78,240.00){\usebox{\plotpoint}}
\multiput(331,240)(20.756,0.000){0}{\usebox{\plotpoint}}
\put(344.53,240.00){\usebox{\plotpoint}}
\put(365.29,240.00){\usebox{\plotpoint}}
\multiput(367,240)(20.756,0.000){0}{\usebox{\plotpoint}}
\put(386.04,240.00){\usebox{\plotpoint}}
\multiput(392,240)(20.756,0.000){0}{\usebox{\plotpoint}}
\put(406.80,240.00){\usebox{\plotpoint}}
\put(427.55,240.00){\usebox{\plotpoint}}
\multiput(429,240)(20.756,0.000){0}{\usebox{\plotpoint}}
\put(448.31,240.00){\usebox{\plotpoint}}
\multiput(453,240)(20.756,0.000){0}{\usebox{\plotpoint}}
\put(469.07,240.00){\usebox{\plotpoint}}
\put(489.82,240.00){\usebox{\plotpoint}}
\multiput(490,240)(20.756,0.000){0}{\usebox{\plotpoint}}
\put(510.58,240.00){\usebox{\plotpoint}}
\multiput(515,240)(20.756,0.000){0}{\usebox{\plotpoint}}
\put(531.33,240.00){\usebox{\plotpoint}}
\multiput(539,240)(20.756,0.000){0}{\usebox{\plotpoint}}
\put(552.09,240.00){\usebox{\plotpoint}}
\put(572.84,240.00){\usebox{\plotpoint}}
\multiput(576,240)(20.756,0.000){0}{\usebox{\plotpoint}}
\put(593.60,240.00){\usebox{\plotpoint}}
\multiput(601,240)(20.756,0.000){0}{\usebox{\plotpoint}}
\put(614.35,240.00){\usebox{\plotpoint}}
\put(635.11,240.00){\usebox{\plotpoint}}
\multiput(638,240)(20.756,0.000){0}{\usebox{\plotpoint}}
\put(655.87,240.00){\usebox{\plotpoint}}
\multiput(662,240)(20.756,0.000){0}{\usebox{\plotpoint}}
\put(676.62,240.00){\usebox{\plotpoint}}
\put(697.38,240.00){\usebox{\plotpoint}}
\multiput(699,240)(20.756,0.000){0}{\usebox{\plotpoint}}
\put(718.13,240.00){\usebox{\plotpoint}}
\multiput(724,240)(20.756,0.000){0}{\usebox{\plotpoint}}
\put(738.89,240.00){\usebox{\plotpoint}}
\put(759.64,240.00){\usebox{\plotpoint}}
\multiput(760,240)(20.756,0.000){0}{\usebox{\plotpoint}}
\put(780.40,240.00){\usebox{\plotpoint}}
\multiput(785,240)(20.756,0.000){0}{\usebox{\plotpoint}}
\put(801.15,240.00){\usebox{\plotpoint}}
\put(821.91,240.00){\usebox{\plotpoint}}
\multiput(822,240)(20.756,0.000){0}{\usebox{\plotpoint}}
\put(842.66,240.00){\usebox{\plotpoint}}
\multiput(846,240)(20.756,0.000){0}{\usebox{\plotpoint}}
\put(863.42,240.00){\usebox{\plotpoint}}
\multiput(871,240)(20.756,0.000){0}{\usebox{\plotpoint}}
\put(884.18,240.00){\usebox{\plotpoint}}
\put(904.93,240.00){\usebox{\plotpoint}}
\multiput(908,240)(20.756,0.000){0}{\usebox{\plotpoint}}
\put(925.69,240.00){\usebox{\plotpoint}}
\multiput(932,240)(20.756,0.000){0}{\usebox{\plotpoint}}
\put(946.44,240.00){\usebox{\plotpoint}}
\put(967.20,240.00){\usebox{\plotpoint}}
\multiput(969,240)(20.756,0.000){0}{\usebox{\plotpoint}}
\put(987.95,240.00){\usebox{\plotpoint}}
\multiput(994,240)(20.756,0.000){0}{\usebox{\plotpoint}}
\put(1008.71,240.00){\usebox{\plotpoint}}
\put(1029.46,240.00){\usebox{\plotpoint}}
\multiput(1031,240)(20.756,0.000){0}{\usebox{\plotpoint}}
\put(1050.22,240.00){\usebox{\plotpoint}}
\multiput(1055,240)(20.756,0.000){0}{\usebox{\plotpoint}}
\put(1070.98,240.00){\usebox{\plotpoint}}
\put(1091.73,240.00){\usebox{\plotpoint}}
\multiput(1092,240)(20.756,0.000){0}{\usebox{\plotpoint}}
\put(1112.49,240.00){\usebox{\plotpoint}}
\multiput(1117,240)(20.756,0.000){0}{\usebox{\plotpoint}}
\put(1133.24,240.00){\usebox{\plotpoint}}
\multiput(1141,240)(20.756,0.000){0}{\usebox{\plotpoint}}
\put(1154.00,240.00){\usebox{\plotpoint}}
\put(1174.75,240.00){\usebox{\plotpoint}}
\multiput(1178,240)(20.756,0.000){0}{\usebox{\plotpoint}}
\put(1195.51,240.00){\usebox{\plotpoint}}
\multiput(1203,240)(20.756,0.000){0}{\usebox{\plotpoint}}
\put(1216.26,240.00){\usebox{\plotpoint}}
\put(1237.02,240.00){\usebox{\plotpoint}}
\multiput(1239,240)(20.756,0.000){0}{\usebox{\plotpoint}}
\put(1257.77,240.00){\usebox{\plotpoint}}
\multiput(1264,240)(20.756,0.000){0}{\usebox{\plotpoint}}
\put(1278.53,240.00){\usebox{\plotpoint}}
\put(1299.29,240.00){\usebox{\plotpoint}}
\multiput(1301,240)(20.756,0.000){0}{\usebox{\plotpoint}}
\put(1320.04,240.00){\usebox{\plotpoint}}
\multiput(1325,240)(20.756,0.000){0}{\usebox{\plotpoint}}
\put(1340.80,240.00){\usebox{\plotpoint}}
\put(1361.55,240.00){\usebox{\plotpoint}}
\multiput(1362,240)(20.756,0.000){0}{\usebox{\plotpoint}}
\put(1382.31,240.00){\usebox{\plotpoint}}
\multiput(1387,240)(20.756,0.000){0}{\usebox{\plotpoint}}
\put(1403.06,240.00){\usebox{\plotpoint}}
\put(1423.82,240.00){\usebox{\plotpoint}}
\multiput(1424,240)(20.756,0.000){0}{\usebox{\plotpoint}}
\put(1436,240){\usebox{\plotpoint}}
\end{picture}

%% file: sfermi.tex
\setlength{\unitlength}{0.240900pt}
\ifx\plotpoint\undefined\newsavebox{\plotpoint}\fi
\sbox{\plotpoint}{\rule[-0.200pt]{0.400pt}{0.400pt}}%
\begin{picture}(1500,900)(0,0)
\font\gnuplot=cmr10 at 10pt
\gnuplot
\sbox{\plotpoint}{\rule[-0.200pt]{0.400pt}{0.400pt}}%
\put(220.0,113.0){\rule[-0.200pt]{292.934pt}{0.400pt}}
\put(220.0,113.0){\rule[-0.200pt]{4.818pt}{0.400pt}}
\put(198,113){\makebox(0,0)[r]{0}}
\put(1416.0,113.0){\rule[-0.200pt]{4.818pt}{0.400pt}}
\put(220.0,266.0){\rule[-0.200pt]{4.818pt}{0.400pt}}
\put(198,266){\makebox(0,0)[r]{1}}
\put(1416.0,266.0){\rule[-0.200pt]{4.818pt}{0.400pt}}
\put(220.0,419.0){\rule[-0.200pt]{4.818pt}{0.400pt}}
\put(198,419){\makebox(0,0)[r]{2}}
\put(1416.0,419.0){\rule[-0.200pt]{4.818pt}{0.400pt}}
\put(220.0,571.0){\rule[-0.200pt]{4.818pt}{0.400pt}}
\put(198,571){\makebox(0,0)[r]{3}}
\put(1416.0,571.0){\rule[-0.200pt]{4.818pt}{0.400pt}}
\put(220.0,724.0){\rule[-0.200pt]{4.818pt}{0.400pt}}
\put(198,724){\makebox(0,0)[r]{4}}
\put(1416.0,724.0){\rule[-0.200pt]{4.818pt}{0.400pt}}
\put(220.0,877.0){\rule[-0.200pt]{4.818pt}{0.400pt}}
\put(198,877){\makebox(0,0)[r]{5}}
\put(1416.0,877.0){\rule[-0.200pt]{4.818pt}{0.400pt}}
\put(220.0,113.0){\rule[-0.200pt]{0.400pt}{4.818pt}}
\put(220,68){\makebox(0,0){4}}
\put(220.0,857.0){\rule[-0.200pt]{0.400pt}{4.818pt}}
\put(423.0,113.0){\rule[-0.200pt]{0.400pt}{4.818pt}}
\put(423,68){\makebox(0,0){6}}
\put(423.0,857.0){\rule[-0.200pt]{0.400pt}{4.818pt}}
\put(625.0,113.0){\rule[-0.200pt]{0.400pt}{4.818pt}}
\put(625,68){\makebox(0,0){8}}
\put(625.0,857.0){\rule[-0.200pt]{0.400pt}{4.818pt}}
\put(828.0,113.0){\rule[-0.200pt]{0.400pt}{4.818pt}}
\put(828,68){\makebox(0,0){10}}
\put(828.0,857.0){\rule[-0.200pt]{0.400pt}{4.818pt}}
\put(1031.0,113.0){\rule[-0.200pt]{0.400pt}{4.818pt}}
\put(1031,68){\makebox(0,0){12}}
\put(1031.0,857.0){\rule[-0.200pt]{0.400pt}{4.818pt}}
\put(1233.0,113.0){\rule[-0.200pt]{0.400pt}{4.818pt}}
\put(1233,68){\makebox(0,0){14}}
\put(1233.0,857.0){\rule[-0.200pt]{0.400pt}{4.818pt}}
\put(1436.0,113.0){\rule[-0.200pt]{0.400pt}{4.818pt}}
\put(1436,68){\makebox(0,0){16}}
\put(1436.0,857.0){\rule[-0.200pt]{0.400pt}{4.818pt}}
\put(220.0,113.0){\rule[-0.200pt]{292.934pt}{0.400pt}}
\put(1436.0,113.0){\rule[-0.200pt]{0.400pt}{184.048pt}}
\put(220.0,877.0){\rule[-0.200pt]{292.934pt}{0.400pt}}
\put(45,495){\makebox(0,0){$m_i$[TeV]}}
\put(828,23){\makebox(0,0){$\log_{10}\mu_0$[GeV]}}
\put(1335,439){\makebox(0,0)[r]{$\widetilde Q$}}
\put(1335,312){\makebox(0,0)[r]{$\widetilde L$}}
\put(1335,220){\makebox(0,0)[r]{$\widetilde E$}}
\put(220.0,113.0){\rule[-0.200pt]{0.400pt}{184.048pt}}
\put(220,235){\usebox{\plotpoint}}
\put(220.00,235.00){\usebox{\plotpoint}}
\put(240.76,235.00){\usebox{\plotpoint}}
\multiput(245,235)(20.756,0.000){0}{\usebox{\plotpoint}}
\put(261.51,235.00){\usebox{\plotpoint}}
\multiput(269,235)(20.756,0.000){0}{\usebox{\plotpoint}}
\put(282.27,235.00){\usebox{\plotpoint}}
\put(303.02,235.00){\usebox{\plotpoint}}
\multiput(306,235)(20.756,0.000){0}{\usebox{\plotpoint}}
\put(323.78,235.00){\usebox{\plotpoint}}
\multiput(331,235)(20.756,0.000){0}{\usebox{\plotpoint}}
\put(344.53,235.00){\usebox{\plotpoint}}
\put(365.29,235.00){\usebox{\plotpoint}}
\multiput(367,235)(20.756,0.000){0}{\usebox{\plotpoint}}
\put(386.04,235.00){\usebox{\plotpoint}}
\multiput(392,235)(20.756,0.000){0}{\usebox{\plotpoint}}
\put(406.80,235.00){\usebox{\plotpoint}}
\put(427.55,235.00){\usebox{\plotpoint}}
\multiput(429,235)(20.756,0.000){0}{\usebox{\plotpoint}}
\put(448.31,235.00){\usebox{\plotpoint}}
\multiput(453,235)(20.756,0.000){0}{\usebox{\plotpoint}}
\put(469.07,235.00){\usebox{\plotpoint}}
\put(489.82,235.00){\usebox{\plotpoint}}
\multiput(490,235)(20.756,0.000){0}{\usebox{\plotpoint}}
\put(510.58,235.00){\usebox{\plotpoint}}
\multiput(515,235)(20.756,0.000){0}{\usebox{\plotpoint}}
\put(531.33,235.00){\usebox{\plotpoint}}
\multiput(539,235)(20.756,0.000){0}{\usebox{\plotpoint}}
\put(552.09,235.00){\usebox{\plotpoint}}
\put(572.84,235.00){\usebox{\plotpoint}}
\multiput(576,235)(20.756,0.000){0}{\usebox{\plotpoint}}
\put(593.60,235.00){\usebox{\plotpoint}}
\multiput(601,235)(20.756,0.000){0}{\usebox{\plotpoint}}
\put(614.35,235.00){\usebox{\plotpoint}}
\put(635.11,235.00){\usebox{\plotpoint}}
\multiput(638,235)(20.756,0.000){0}{\usebox{\plotpoint}}
\put(655.87,235.00){\usebox{\plotpoint}}
\multiput(662,235)(20.756,0.000){0}{\usebox{\plotpoint}}
\put(676.62,235.00){\usebox{\plotpoint}}
\put(697.38,235.00){\usebox{\plotpoint}}
\multiput(699,235)(20.756,0.000){0}{\usebox{\plotpoint}}
\put(718.13,235.00){\usebox{\plotpoint}}
\multiput(724,235)(20.756,0.000){0}{\usebox{\plotpoint}}
\put(738.89,235.00){\usebox{\plotpoint}}
\put(759.64,235.00){\usebox{\plotpoint}}
\multiput(760,235)(20.756,0.000){0}{\usebox{\plotpoint}}
\put(780.40,235.00){\usebox{\plotpoint}}
\multiput(785,235)(20.756,0.000){0}{\usebox{\plotpoint}}
\put(801.15,235.00){\usebox{\plotpoint}}
\put(821.91,235.00){\usebox{\plotpoint}}
\multiput(822,235)(20.756,0.000){0}{\usebox{\plotpoint}}
\put(842.66,235.00){\usebox{\plotpoint}}
\multiput(846,235)(20.756,0.000){0}{\usebox{\plotpoint}}
\put(863.42,235.00){\usebox{\plotpoint}}
\multiput(871,235)(20.756,0.000){0}{\usebox{\plotpoint}}
\put(884.18,235.00){\usebox{\plotpoint}}
\put(904.93,235.00){\usebox{\plotpoint}}
\multiput(908,235)(20.756,0.000){0}{\usebox{\plotpoint}}
\put(925.69,235.00){\usebox{\plotpoint}}
\multiput(932,235)(20.756,0.000){0}{\usebox{\plotpoint}}
\put(946.44,235.00){\usebox{\plotpoint}}
\put(967.20,235.00){\usebox{\plotpoint}}
\multiput(969,235)(20.756,0.000){0}{\usebox{\plotpoint}}
\put(987.95,235.00){\usebox{\plotpoint}}
\multiput(994,235)(20.756,0.000){0}{\usebox{\plotpoint}}
\put(1008.71,235.00){\usebox{\plotpoint}}
\put(1029.46,235.00){\usebox{\plotpoint}}
\multiput(1031,235)(20.756,0.000){0}{\usebox{\plotpoint}}
\put(1050.22,235.00){\usebox{\plotpoint}}
\multiput(1055,235)(20.756,0.000){0}{\usebox{\plotpoint}}
\put(1070.98,235.00){\usebox{\plotpoint}}
\put(1091.73,235.00){\usebox{\plotpoint}}
\multiput(1092,235)(20.756,0.000){0}{\usebox{\plotpoint}}
\put(1112.49,235.00){\usebox{\plotpoint}}
\multiput(1117,235)(20.756,0.000){0}{\usebox{\plotpoint}}
\put(1133.24,235.00){\usebox{\plotpoint}}
\multiput(1141,235)(20.756,0.000){0}{\usebox{\plotpoint}}
\put(1154.00,235.00){\usebox{\plotpoint}}
\put(1174.75,235.00){\usebox{\plotpoint}}
\multiput(1178,235)(20.756,0.000){0}{\usebox{\plotpoint}}
\put(1195.51,235.00){\usebox{\plotpoint}}
\multiput(1203,235)(20.756,0.000){0}{\usebox{\plotpoint}}
\put(1216.26,235.00){\usebox{\plotpoint}}
\put(1237.02,235.00){\usebox{\plotpoint}}
\multiput(1239,235)(20.756,0.000){0}{\usebox{\plotpoint}}
\put(1257.77,235.00){\usebox{\plotpoint}}
\multiput(1264,235)(20.756,0.000){0}{\usebox{\plotpoint}}
\put(1278.53,235.00){\usebox{\plotpoint}}
\put(1299.29,235.00){\usebox{\plotpoint}}
\multiput(1301,235)(20.756,0.000){0}{\usebox{\plotpoint}}
\put(1320.04,235.00){\usebox{\plotpoint}}
\multiput(1325,235)(20.756,0.000){0}{\usebox{\plotpoint}}
\put(1340.80,235.00){\usebox{\plotpoint}}
\put(1361.55,235.00){\usebox{\plotpoint}}
\multiput(1362,235)(20.756,0.000){0}{\usebox{\plotpoint}}
\put(1382.31,235.00){\usebox{\plotpoint}}
\multiput(1387,235)(20.756,0.000){0}{\usebox{\plotpoint}}
\put(1403.06,235.00){\usebox{\plotpoint}}
\put(1423.82,235.00){\usebox{\plotpoint}}
\multiput(1424,235)(20.756,0.000){0}{\usebox{\plotpoint}}
\put(1436,235){\usebox{\plotpoint}}
\sbox{\plotpoint}{\rule[-0.400pt]{0.800pt}{0.800pt}}%
\put(220,750){\usebox{\plotpoint}}
\put(220,746.84){\rule{12.286pt}{0.800pt}}
\multiput(220.00,748.34)(25.500,-3.000){2}{\rule{6.143pt}{0.800pt}}
\multiput(271.00,745.06)(7.980,-0.560){3}{\rule{8.200pt}{0.135pt}}
\multiput(271.00,745.34)(32.980,-5.000){2}{\rule{4.100pt}{0.800pt}}
\multiput(321.00,740.08)(3.135,-0.516){11}{\rule{4.733pt}{0.124pt}}
\multiput(321.00,740.34)(41.176,-9.000){2}{\rule{2.367pt}{0.800pt}}
\multiput(372.00,731.08)(2.769,-0.514){13}{\rule{4.280pt}{0.124pt}}
\multiput(372.00,731.34)(42.117,-10.000){2}{\rule{2.140pt}{0.800pt}}
\multiput(423.00,721.08)(2.434,-0.512){15}{\rule{3.836pt}{0.123pt}}
\multiput(423.00,721.34)(42.037,-11.000){2}{\rule{1.918pt}{0.800pt}}
\multiput(473.00,710.08)(2.063,-0.509){19}{\rule{3.338pt}{0.123pt}}
\multiput(473.00,710.34)(44.071,-13.000){2}{\rule{1.669pt}{0.800pt}}
\multiput(524.00,697.09)(1.904,-0.509){21}{\rule{3.114pt}{0.123pt}}
\multiput(524.00,697.34)(44.536,-14.000){2}{\rule{1.557pt}{0.800pt}}
\multiput(575.00,683.08)(2.022,-0.509){19}{\rule{3.277pt}{0.123pt}}
\multiput(575.00,683.34)(43.199,-13.000){2}{\rule{1.638pt}{0.800pt}}
\multiput(625.00,670.09)(1.904,-0.509){21}{\rule{3.114pt}{0.123pt}}
\multiput(625.00,670.34)(44.536,-14.000){2}{\rule{1.557pt}{0.800pt}}
\multiput(676.00,656.09)(1.767,-0.508){23}{\rule{2.920pt}{0.122pt}}
\multiput(676.00,656.34)(44.939,-15.000){2}{\rule{1.460pt}{0.800pt}}
\multiput(727.00,641.09)(1.732,-0.508){23}{\rule{2.867pt}{0.122pt}}
\multiput(727.00,641.34)(44.050,-15.000){2}{\rule{1.433pt}{0.800pt}}
\multiput(777.00,626.09)(1.904,-0.509){21}{\rule{3.114pt}{0.123pt}}
\multiput(777.00,626.34)(44.536,-14.000){2}{\rule{1.557pt}{0.800pt}}
\multiput(828.00,612.09)(1.767,-0.508){23}{\rule{2.920pt}{0.122pt}}
\multiput(828.00,612.34)(44.939,-15.000){2}{\rule{1.460pt}{0.800pt}}
\multiput(879.00,597.09)(1.732,-0.508){23}{\rule{2.867pt}{0.122pt}}
\multiput(879.00,597.34)(44.050,-15.000){2}{\rule{1.433pt}{0.800pt}}
\multiput(929.00,582.09)(1.767,-0.508){23}{\rule{2.920pt}{0.122pt}}
\multiput(929.00,582.34)(44.939,-15.000){2}{\rule{1.460pt}{0.800pt}}
\multiput(980.00,567.09)(1.767,-0.508){23}{\rule{2.920pt}{0.122pt}}
\multiput(980.00,567.34)(44.939,-15.000){2}{\rule{1.460pt}{0.800pt}}
\multiput(1031.00,552.09)(1.865,-0.509){21}{\rule{3.057pt}{0.123pt}}
\multiput(1031.00,552.34)(43.655,-14.000){2}{\rule{1.529pt}{0.800pt}}
\multiput(1081.00,538.09)(1.767,-0.508){23}{\rule{2.920pt}{0.122pt}}
\multiput(1081.00,538.34)(44.939,-15.000){2}{\rule{1.460pt}{0.800pt}}
\multiput(1132.00,523.08)(2.063,-0.509){19}{\rule{3.338pt}{0.123pt}}
\multiput(1132.00,523.34)(44.071,-13.000){2}{\rule{1.669pt}{0.800pt}}
\multiput(1183.00,510.09)(1.865,-0.509){21}{\rule{3.057pt}{0.123pt}}
\multiput(1183.00,510.34)(43.655,-14.000){2}{\rule{1.529pt}{0.800pt}}
\multiput(1233.00,496.08)(2.063,-0.509){19}{\rule{3.338pt}{0.123pt}}
\multiput(1233.00,496.34)(44.071,-13.000){2}{\rule{1.669pt}{0.800pt}}
\multiput(1284.00,483.08)(2.254,-0.511){17}{\rule{3.600pt}{0.123pt}}
\multiput(1284.00,483.34)(43.528,-12.000){2}{\rule{1.800pt}{0.800pt}}
\multiput(1335.00,471.08)(2.714,-0.514){13}{\rule{4.200pt}{0.124pt}}
\multiput(1335.00,471.34)(41.283,-10.000){2}{\rule{2.100pt}{0.800pt}}
\multiput(1385.00,461.08)(3.625,-0.520){9}{\rule{5.300pt}{0.125pt}}
\multiput(1385.00,461.34)(40.000,-8.000){2}{\rule{2.650pt}{0.800pt}}
\put(220,314){\usebox{\plotpoint}}
\put(321,311.84){\rule{12.286pt}{0.800pt}}
\multiput(321.00,312.34)(25.500,-1.000){2}{\rule{6.143pt}{0.800pt}}
\put(220.0,314.0){\rule[-0.400pt]{24.331pt}{0.800pt}}
\put(423,310.84){\rule{12.045pt}{0.800pt}}
\multiput(423.00,311.34)(25.000,-1.000){2}{\rule{6.022pt}{0.800pt}}
\put(473,309.84){\rule{12.286pt}{0.800pt}}
\multiput(473.00,310.34)(25.500,-1.000){2}{\rule{6.143pt}{0.800pt}}
\put(524,308.84){\rule{12.286pt}{0.800pt}}
\multiput(524.00,309.34)(25.500,-1.000){2}{\rule{6.143pt}{0.800pt}}
\put(575,307.84){\rule{12.045pt}{0.800pt}}
\multiput(575.00,308.34)(25.000,-1.000){2}{\rule{6.022pt}{0.800pt}}
\put(625,306.84){\rule{12.286pt}{0.800pt}}
\multiput(625.00,307.34)(25.500,-1.000){2}{\rule{6.143pt}{0.800pt}}
\put(676,305.84){\rule{12.286pt}{0.800pt}}
\multiput(676.00,306.34)(25.500,-1.000){2}{\rule{6.143pt}{0.800pt}}
\put(727,304.34){\rule{12.045pt}{0.800pt}}
\multiput(727.00,305.34)(25.000,-2.000){2}{\rule{6.022pt}{0.800pt}}
\put(777,302.34){\rule{12.286pt}{0.800pt}}
\multiput(777.00,303.34)(25.500,-2.000){2}{\rule{6.143pt}{0.800pt}}
\put(828,300.34){\rule{12.286pt}{0.800pt}}
\multiput(828.00,301.34)(25.500,-2.000){2}{\rule{6.143pt}{0.800pt}}
\put(879,298.34){\rule{12.045pt}{0.800pt}}
\multiput(879.00,299.34)(25.000,-2.000){2}{\rule{6.022pt}{0.800pt}}
\put(929,296.34){\rule{12.286pt}{0.800pt}}
\multiput(929.00,297.34)(25.500,-2.000){2}{\rule{6.143pt}{0.800pt}}
\put(980,294.34){\rule{12.286pt}{0.800pt}}
\multiput(980.00,295.34)(25.500,-2.000){2}{\rule{6.143pt}{0.800pt}}
\put(1031,291.84){\rule{12.045pt}{0.800pt}}
\multiput(1031.00,293.34)(25.000,-3.000){2}{\rule{6.022pt}{0.800pt}}
\put(1081,289.34){\rule{12.286pt}{0.800pt}}
\multiput(1081.00,290.34)(25.500,-2.000){2}{\rule{6.143pt}{0.800pt}}
\put(1132,286.84){\rule{12.286pt}{0.800pt}}
\multiput(1132.00,288.34)(25.500,-3.000){2}{\rule{6.143pt}{0.800pt}}
\put(1183,283.84){\rule{12.045pt}{0.800pt}}
\multiput(1183.00,285.34)(25.000,-3.000){2}{\rule{6.022pt}{0.800pt}}
\put(1233,281.34){\rule{12.286pt}{0.800pt}}
\multiput(1233.00,282.34)(25.500,-2.000){2}{\rule{6.143pt}{0.800pt}}
\put(1284,278.84){\rule{12.286pt}{0.800pt}}
\multiput(1284.00,280.34)(25.500,-3.000){2}{\rule{6.143pt}{0.800pt}}
\put(1335,276.34){\rule{12.045pt}{0.800pt}}
\multiput(1335.00,277.34)(25.000,-2.000){2}{\rule{6.022pt}{0.800pt}}
\put(1385,274.34){\rule{12.286pt}{0.800pt}}
\multiput(1385.00,275.34)(25.500,-2.000){2}{\rule{6.143pt}{0.800pt}}
\put(372.0,313.0){\rule[-0.400pt]{12.286pt}{0.800pt}}
\put(220,252){\usebox{\plotpoint}}
\put(220,250.84){\rule{12.286pt}{0.800pt}}
\multiput(220.00,250.34)(25.500,1.000){2}{\rule{6.143pt}{0.800pt}}
\put(473,251.84){\rule{12.286pt}{0.800pt}}
\multiput(473.00,251.34)(25.500,1.000){2}{\rule{6.143pt}{0.800pt}}
\put(271.0,253.0){\rule[-0.400pt]{48.662pt}{0.800pt}}
\put(929,251.84){\rule{12.286pt}{0.800pt}}
\multiput(929.00,252.34)(25.500,-1.000){2}{\rule{6.143pt}{0.800pt}}
\put(524.0,254.0){\rule[-0.400pt]{97.564pt}{0.800pt}}
\put(1081,250.84){\rule{12.286pt}{0.800pt}}
\multiput(1081.00,251.34)(25.500,-1.000){2}{\rule{6.143pt}{0.800pt}}
\put(980.0,253.0){\rule[-0.400pt]{24.331pt}{0.800pt}}
\put(1183,249.84){\rule{12.045pt}{0.800pt}}
\multiput(1183.00,250.34)(25.000,-1.000){2}{\rule{6.022pt}{0.800pt}}
\put(1132.0,252.0){\rule[-0.400pt]{12.286pt}{0.800pt}}
\put(1284,248.84){\rule{12.286pt}{0.800pt}}
\multiput(1284.00,249.34)(25.500,-1.000){2}{\rule{6.143pt}{0.800pt}}
\put(1335,247.84){\rule{12.045pt}{0.800pt}}
\multiput(1335.00,248.34)(25.000,-1.000){2}{\rule{6.022pt}{0.800pt}}
\put(1233.0,251.0){\rule[-0.400pt]{12.286pt}{0.800pt}}
\put(1385.0,249.0){\rule[-0.400pt]{12.286pt}{0.800pt}}
\end{picture}

%% file: sfermi32.tex
\setlength{\unitlength}{0.240900pt}
\ifx\plotpoint\undefined\newsavebox{\plotpoint}\fi
\sbox{\plotpoint}{\rule[-0.200pt]{0.400pt}{0.400pt}}%
\begin{picture}(1500,900)(0,0)
\font\gnuplot=cmr10 at 10pt
\gnuplot
\sbox{\plotpoint}{\rule[-0.200pt]{0.400pt}{0.400pt}}%
\put(220.0,113.0){\rule[-0.200pt]{292.934pt}{0.400pt}}
\put(220.0,113.0){\rule[-0.200pt]{4.818pt}{0.400pt}}
\put(198,113){\makebox(0,0)[r]{0}}
\put(1416.0,113.0){\rule[-0.200pt]{4.818pt}{0.400pt}}
\put(220.0,266.0){\rule[-0.200pt]{4.818pt}{0.400pt}}
\put(198,266){\makebox(0,0)[r]{1}}
\put(1416.0,266.0){\rule[-0.200pt]{4.818pt}{0.400pt}}
\put(220.0,419.0){\rule[-0.200pt]{4.818pt}{0.400pt}}
\put(198,419){\makebox(0,0)[r]{2}}
\put(1416.0,419.0){\rule[-0.200pt]{4.818pt}{0.400pt}}
\put(220.0,571.0){\rule[-0.200pt]{4.818pt}{0.400pt}}
\put(198,571){\makebox(0,0)[r]{3}}
\put(1416.0,571.0){\rule[-0.200pt]{4.818pt}{0.400pt}}
\put(220.0,724.0){\rule[-0.200pt]{4.818pt}{0.400pt}}
\put(198,724){\makebox(0,0)[r]{4}}
\put(1416.0,724.0){\rule[-0.200pt]{4.818pt}{0.400pt}}
\put(220.0,877.0){\rule[-0.200pt]{4.818pt}{0.400pt}}
\put(198,877){\makebox(0,0)[r]{5}}
\put(1416.0,877.0){\rule[-0.200pt]{4.818pt}{0.400pt}}
\put(220.0,113.0){\rule[-0.200pt]{0.400pt}{4.818pt}}
\put(220,68){\makebox(0,0){4}}
\put(220.0,857.0){\rule[-0.200pt]{0.400pt}{4.818pt}}
\put(423.0,113.0){\rule[-0.200pt]{0.400pt}{4.818pt}}
\put(423,68){\makebox(0,0){6}}
\put(423.0,857.0){\rule[-0.200pt]{0.400pt}{4.818pt}}
\put(625.0,113.0){\rule[-0.200pt]{0.400pt}{4.818pt}}
\put(625,68){\makebox(0,0){8}}
\put(625.0,857.0){\rule[-0.200pt]{0.400pt}{4.818pt}}
\put(828.0,113.0){\rule[-0.200pt]{0.400pt}{4.818pt}}
\put(828,68){\makebox(0,0){10}}
\put(828.0,857.0){\rule[-0.200pt]{0.400pt}{4.818pt}}
\put(1031.0,113.0){\rule[-0.200pt]{0.400pt}{4.818pt}}
\put(1031,68){\makebox(0,0){12}}
\put(1031.0,857.0){\rule[-0.200pt]{0.400pt}{4.818pt}}
\put(1233.0,113.0){\rule[-0.200pt]{0.400pt}{4.818pt}}
\put(1233,68){\makebox(0,0){14}}
\put(1233.0,857.0){\rule[-0.200pt]{0.400pt}{4.818pt}}
\put(1436.0,113.0){\rule[-0.200pt]{0.400pt}{4.818pt}}
\put(1436,68){\makebox(0,0){16}}
\put(1436.0,857.0){\rule[-0.200pt]{0.400pt}{4.818pt}}
\put(220.0,113.0){\rule[-0.200pt]{292.934pt}{0.400pt}}
\put(1436.0,113.0){\rule[-0.200pt]{0.400pt}{184.048pt}}
\put(220.0,877.0){\rule[-0.200pt]{292.934pt}{0.400pt}}
\put(45,495){\makebox(0,0){$m_i$[TeV]}}
\put(828,23){\makebox(0,0){$\log_{10}\mu_0$[GeV]}}
\put(1335,495){\makebox(0,0)[r]{$\widetilde b$}}
\put(1335,403){\makebox(0,0)[r]{$\widetilde t$}}
\put(1335,281){\makebox(0,0)[r]{$\widetilde \tau$}}
\put(1335,144){\makebox(0,0)[r]{$\chi^0$}}
\put(220.0,113.0){\rule[-0.200pt]{0.400pt}{184.048pt}}
\put(220,235){\usebox{\plotpoint}}
\put(220.00,235.00){\usebox{\plotpoint}}
\put(240.76,235.00){\usebox{\plotpoint}}
\multiput(245,235)(20.756,0.000){0}{\usebox{\plotpoint}}
\put(261.51,235.00){\usebox{\plotpoint}}
\multiput(269,235)(20.756,0.000){0}{\usebox{\plotpoint}}
\put(282.27,235.00){\usebox{\plotpoint}}
\put(303.02,235.00){\usebox{\plotpoint}}
\multiput(306,235)(20.756,0.000){0}{\usebox{\plotpoint}}
\put(323.78,235.00){\usebox{\plotpoint}}
\multiput(331,235)(20.756,0.000){0}{\usebox{\plotpoint}}
\put(344.53,235.00){\usebox{\plotpoint}}
\put(365.29,235.00){\usebox{\plotpoint}}
\multiput(367,235)(20.756,0.000){0}{\usebox{\plotpoint}}
\put(386.04,235.00){\usebox{\plotpoint}}
\multiput(392,235)(20.756,0.000){0}{\usebox{\plotpoint}}
\put(406.80,235.00){\usebox{\plotpoint}}
\put(427.55,235.00){\usebox{\plotpoint}}
\multiput(429,235)(20.756,0.000){0}{\usebox{\plotpoint}}
\put(448.31,235.00){\usebox{\plotpoint}}
\multiput(453,235)(20.756,0.000){0}{\usebox{\plotpoint}}
\put(469.07,235.00){\usebox{\plotpoint}}
\put(489.82,235.00){\usebox{\plotpoint}}
\multiput(490,235)(20.756,0.000){0}{\usebox{\plotpoint}}
\put(510.58,235.00){\usebox{\plotpoint}}
\multiput(515,235)(20.756,0.000){0}{\usebox{\plotpoint}}
\put(531.33,235.00){\usebox{\plotpoint}}
\multiput(539,235)(20.756,0.000){0}{\usebox{\plotpoint}}
\put(552.09,235.00){\usebox{\plotpoint}}
\put(572.84,235.00){\usebox{\plotpoint}}
\multiput(576,235)(20.756,0.000){0}{\usebox{\plotpoint}}
\put(593.60,235.00){\usebox{\plotpoint}}
\multiput(601,235)(20.756,0.000){0}{\usebox{\plotpoint}}
\put(614.35,235.00){\usebox{\plotpoint}}
\put(635.11,235.00){\usebox{\plotpoint}}
\multiput(638,235)(20.756,0.000){0}{\usebox{\plotpoint}}
\put(655.87,235.00){\usebox{\plotpoint}}
\multiput(662,235)(20.756,0.000){0}{\usebox{\plotpoint}}
\put(676.62,235.00){\usebox{\plotpoint}}
\put(697.38,235.00){\usebox{\plotpoint}}
\multiput(699,235)(20.756,0.000){0}{\usebox{\plotpoint}}
\put(718.13,235.00){\usebox{\plotpoint}}
\multiput(724,235)(20.756,0.000){0}{\usebox{\plotpoint}}
\put(738.89,235.00){\usebox{\plotpoint}}
\put(759.64,235.00){\usebox{\plotpoint}}
\multiput(760,235)(20.756,0.000){0}{\usebox{\plotpoint}}
\put(780.40,235.00){\usebox{\plotpoint}}
\multiput(785,235)(20.756,0.000){0}{\usebox{\plotpoint}}
\put(801.15,235.00){\usebox{\plotpoint}}
\put(821.91,235.00){\usebox{\plotpoint}}
\multiput(822,235)(20.756,0.000){0}{\usebox{\plotpoint}}
\put(842.66,235.00){\usebox{\plotpoint}}
\multiput(846,235)(20.756,0.000){0}{\usebox{\plotpoint}}
\put(863.42,235.00){\usebox{\plotpoint}}
\multiput(871,235)(20.756,0.000){0}{\usebox{\plotpoint}}
\put(884.18,235.00){\usebox{\plotpoint}}
\put(904.93,235.00){\usebox{\plotpoint}}
\multiput(908,235)(20.756,0.000){0}{\usebox{\plotpoint}}
\put(925.69,235.00){\usebox{\plotpoint}}
\multiput(932,235)(20.756,0.000){0}{\usebox{\plotpoint}}
\put(946.44,235.00){\usebox{\plotpoint}}
\put(967.20,235.00){\usebox{\plotpoint}}
\multiput(969,235)(20.756,0.000){0}{\usebox{\plotpoint}}
\put(987.95,235.00){\usebox{\plotpoint}}
\multiput(994,235)(20.756,0.000){0}{\usebox{\plotpoint}}
\put(1008.71,235.00){\usebox{\plotpoint}}
\put(1029.46,235.00){\usebox{\plotpoint}}
\multiput(1031,235)(20.756,0.000){0}{\usebox{\plotpoint}}
\put(1050.22,235.00){\usebox{\plotpoint}}
\multiput(1055,235)(20.756,0.000){0}{\usebox{\plotpoint}}
\put(1070.98,235.00){\usebox{\plotpoint}}
\put(1091.73,235.00){\usebox{\plotpoint}}
\multiput(1092,235)(20.756,0.000){0}{\usebox{\plotpoint}}
\put(1112.49,235.00){\usebox{\plotpoint}}
\multiput(1117,235)(20.756,0.000){0}{\usebox{\plotpoint}}
\put(1133.24,235.00){\usebox{\plotpoint}}
\multiput(1141,235)(20.756,0.000){0}{\usebox{\plotpoint}}
\put(1154.00,235.00){\usebox{\plotpoint}}
\put(1174.75,235.00){\usebox{\plotpoint}}
\multiput(1178,235)(20.756,0.000){0}{\usebox{\plotpoint}}
\put(1195.51,235.00){\usebox{\plotpoint}}
\multiput(1203,235)(20.756,0.000){0}{\usebox{\plotpoint}}
\put(1216.26,235.00){\usebox{\plotpoint}}
\put(1237.02,235.00){\usebox{\plotpoint}}
\multiput(1239,235)(20.756,0.000){0}{\usebox{\plotpoint}}
\put(1257.77,235.00){\usebox{\plotpoint}}
\multiput(1264,235)(20.756,0.000){0}{\usebox{\plotpoint}}
\put(1278.53,235.00){\usebox{\plotpoint}}
\put(1299.29,235.00){\usebox{\plotpoint}}
\multiput(1301,235)(20.756,0.000){0}{\usebox{\plotpoint}}
\put(1320.04,235.00){\usebox{\plotpoint}}
\multiput(1325,235)(20.756,0.000){0}{\usebox{\plotpoint}}
\put(1340.80,235.00){\usebox{\plotpoint}}
\put(1361.55,235.00){\usebox{\plotpoint}}
\multiput(1362,235)(20.756,0.000){0}{\usebox{\plotpoint}}
\put(1382.31,235.00){\usebox{\plotpoint}}
\multiput(1387,235)(20.756,0.000){0}{\usebox{\plotpoint}}
\put(1403.06,235.00){\usebox{\plotpoint}}
\put(1423.82,235.00){\usebox{\plotpoint}}
\multiput(1424,235)(20.756,0.000){0}{\usebox{\plotpoint}}
\put(1436,235){\usebox{\plotpoint}}
\sbox{\plotpoint}{\rule[-0.400pt]{0.800pt}{0.800pt}}%
\put(220,613){\usebox{\plotpoint}}
\put(220,609.34){\rule{10.400pt}{0.800pt}}
\multiput(220.00,611.34)(29.414,-4.000){2}{\rule{5.200pt}{0.800pt}}
\multiput(271.00,607.08)(4.242,-0.526){7}{\rule{5.914pt}{0.127pt}}
\multiput(271.00,607.34)(37.725,-7.000){2}{\rule{2.957pt}{0.800pt}}
\multiput(321.00,600.08)(2.769,-0.514){13}{\rule{4.280pt}{0.124pt}}
\multiput(321.00,600.34)(42.117,-10.000){2}{\rule{2.140pt}{0.800pt}}
\multiput(372.00,590.08)(2.484,-0.512){15}{\rule{3.909pt}{0.123pt}}
\multiput(372.00,590.34)(42.886,-11.000){2}{\rule{1.955pt}{0.800pt}}
\multiput(423.00,579.08)(2.434,-0.512){15}{\rule{3.836pt}{0.123pt}}
\multiput(423.00,579.34)(42.037,-11.000){2}{\rule{1.918pt}{0.800pt}}
\multiput(473.00,568.08)(2.063,-0.509){19}{\rule{3.338pt}{0.123pt}}
\multiput(473.00,568.34)(44.071,-13.000){2}{\rule{1.669pt}{0.800pt}}
\multiput(524.00,555.08)(2.063,-0.509){19}{\rule{3.338pt}{0.123pt}}
\multiput(524.00,555.34)(44.071,-13.000){2}{\rule{1.669pt}{0.800pt}}
\multiput(575.00,542.09)(1.865,-0.509){21}{\rule{3.057pt}{0.123pt}}
\multiput(575.00,542.34)(43.655,-14.000){2}{\rule{1.529pt}{0.800pt}}
\multiput(625.00,528.08)(2.063,-0.509){19}{\rule{3.338pt}{0.123pt}}
\multiput(625.00,528.34)(44.071,-13.000){2}{\rule{1.669pt}{0.800pt}}
\multiput(676.00,515.08)(2.063,-0.509){19}{\rule{3.338pt}{0.123pt}}
\multiput(676.00,515.34)(44.071,-13.000){2}{\rule{1.669pt}{0.800pt}}
\multiput(727.00,502.09)(1.865,-0.509){21}{\rule{3.057pt}{0.123pt}}
\multiput(727.00,502.34)(43.655,-14.000){2}{\rule{1.529pt}{0.800pt}}
\multiput(777.00,488.08)(2.063,-0.509){19}{\rule{3.338pt}{0.123pt}}
\multiput(777.00,488.34)(44.071,-13.000){2}{\rule{1.669pt}{0.800pt}}
\multiput(828.00,475.08)(2.063,-0.509){19}{\rule{3.338pt}{0.123pt}}
\multiput(828.00,475.34)(44.071,-13.000){2}{\rule{1.669pt}{0.800pt}}
\multiput(879.00,462.09)(1.865,-0.509){21}{\rule{3.057pt}{0.123pt}}
\multiput(879.00,462.34)(43.655,-14.000){2}{\rule{1.529pt}{0.800pt}}
\multiput(929.00,448.08)(2.254,-0.511){17}{\rule{3.600pt}{0.123pt}}
\multiput(929.00,448.34)(43.528,-12.000){2}{\rule{1.800pt}{0.800pt}}
\multiput(980.00,436.08)(2.063,-0.509){19}{\rule{3.338pt}{0.123pt}}
\multiput(980.00,436.34)(44.071,-13.000){2}{\rule{1.669pt}{0.800pt}}
\multiput(1031.00,423.08)(2.434,-0.512){15}{\rule{3.836pt}{0.123pt}}
\multiput(1031.00,423.34)(42.037,-11.000){2}{\rule{1.918pt}{0.800pt}}
\multiput(1081.00,412.08)(2.254,-0.511){17}{\rule{3.600pt}{0.123pt}}
\multiput(1081.00,412.34)(43.528,-12.000){2}{\rule{1.800pt}{0.800pt}}
\multiput(1132.00,400.08)(2.484,-0.512){15}{\rule{3.909pt}{0.123pt}}
\multiput(1132.00,400.34)(42.886,-11.000){2}{\rule{1.955pt}{0.800pt}}
\multiput(1183.00,389.08)(2.434,-0.512){15}{\rule{3.836pt}{0.123pt}}
\multiput(1183.00,389.34)(42.037,-11.000){2}{\rule{1.918pt}{0.800pt}}
\multiput(1233.00,378.08)(3.135,-0.516){11}{\rule{4.733pt}{0.124pt}}
\multiput(1233.00,378.34)(41.176,-9.000){2}{\rule{2.367pt}{0.800pt}}
\multiput(1284.00,369.08)(3.135,-0.516){11}{\rule{4.733pt}{0.124pt}}
\multiput(1284.00,369.34)(41.176,-9.000){2}{\rule{2.367pt}{0.800pt}}
\multiput(1335.00,360.08)(3.552,-0.520){9}{\rule{5.200pt}{0.125pt}}
\multiput(1335.00,360.34)(39.207,-8.000){2}{\rule{2.600pt}{0.800pt}}
\multiput(1385.00,352.06)(8.148,-0.560){3}{\rule{8.360pt}{0.135pt}}
\multiput(1385.00,352.34)(33.648,-5.000){2}{\rule{4.180pt}{0.800pt}}
\put(220,694){\usebox{\plotpoint}}
\put(220,690.34){\rule{10.400pt}{0.800pt}}
\multiput(220.00,692.34)(29.414,-4.000){2}{\rule{5.200pt}{0.800pt}}
\multiput(271.00,688.07)(5.374,-0.536){5}{\rule{6.867pt}{0.129pt}}
\multiput(271.00,688.34)(35.748,-6.000){2}{\rule{3.433pt}{0.800pt}}
\multiput(321.00,682.08)(3.135,-0.516){11}{\rule{4.733pt}{0.124pt}}
\multiput(321.00,682.34)(41.176,-9.000){2}{\rule{2.367pt}{0.800pt}}
\multiput(372.00,673.08)(2.769,-0.514){13}{\rule{4.280pt}{0.124pt}}
\multiput(372.00,673.34)(42.117,-10.000){2}{\rule{2.140pt}{0.800pt}}
\multiput(423.00,663.08)(2.434,-0.512){15}{\rule{3.836pt}{0.123pt}}
\multiput(423.00,663.34)(42.037,-11.000){2}{\rule{1.918pt}{0.800pt}}
\multiput(473.00,652.08)(2.063,-0.509){19}{\rule{3.338pt}{0.123pt}}
\multiput(473.00,652.34)(44.071,-13.000){2}{\rule{1.669pt}{0.800pt}}
\multiput(524.00,639.08)(2.063,-0.509){19}{\rule{3.338pt}{0.123pt}}
\multiput(524.00,639.34)(44.071,-13.000){2}{\rule{1.669pt}{0.800pt}}
\multiput(575.00,626.08)(2.022,-0.509){19}{\rule{3.277pt}{0.123pt}}
\multiput(575.00,626.34)(43.199,-13.000){2}{\rule{1.638pt}{0.800pt}}
\multiput(625.00,613.08)(2.063,-0.509){19}{\rule{3.338pt}{0.123pt}}
\multiput(625.00,613.34)(44.071,-13.000){2}{\rule{1.669pt}{0.800pt}}
\multiput(676.00,600.09)(1.767,-0.508){23}{\rule{2.920pt}{0.122pt}}
\multiput(676.00,600.34)(44.939,-15.000){2}{\rule{1.460pt}{0.800pt}}
\multiput(727.00,585.09)(1.865,-0.509){21}{\rule{3.057pt}{0.123pt}}
\multiput(727.00,585.34)(43.655,-14.000){2}{\rule{1.529pt}{0.800pt}}
\multiput(777.00,571.09)(1.904,-0.509){21}{\rule{3.114pt}{0.123pt}}
\multiput(777.00,571.34)(44.536,-14.000){2}{\rule{1.557pt}{0.800pt}}
\multiput(828.00,557.09)(1.904,-0.509){21}{\rule{3.114pt}{0.123pt}}
\multiput(828.00,557.34)(44.536,-14.000){2}{\rule{1.557pt}{0.800pt}}
\multiput(879.00,543.09)(1.865,-0.509){21}{\rule{3.057pt}{0.123pt}}
\multiput(879.00,543.34)(43.655,-14.000){2}{\rule{1.529pt}{0.800pt}}
\multiput(929.00,529.09)(1.904,-0.509){21}{\rule{3.114pt}{0.123pt}}
\multiput(929.00,529.34)(44.536,-14.000){2}{\rule{1.557pt}{0.800pt}}
\multiput(980.00,515.08)(2.063,-0.509){19}{\rule{3.338pt}{0.123pt}}
\multiput(980.00,515.34)(44.071,-13.000){2}{\rule{1.669pt}{0.800pt}}
\multiput(1031.00,502.08)(2.022,-0.509){19}{\rule{3.277pt}{0.123pt}}
\multiput(1031.00,502.34)(43.199,-13.000){2}{\rule{1.638pt}{0.800pt}}
\multiput(1081.00,489.09)(1.904,-0.509){21}{\rule{3.114pt}{0.123pt}}
\multiput(1081.00,489.34)(44.536,-14.000){2}{\rule{1.557pt}{0.800pt}}
\multiput(1132.00,475.08)(2.254,-0.511){17}{\rule{3.600pt}{0.123pt}}
\multiput(1132.00,475.34)(43.528,-12.000){2}{\rule{1.800pt}{0.800pt}}
\multiput(1183.00,463.08)(2.022,-0.509){19}{\rule{3.277pt}{0.123pt}}
\multiput(1183.00,463.34)(43.199,-13.000){2}{\rule{1.638pt}{0.800pt}}
\multiput(1233.00,450.08)(2.484,-0.512){15}{\rule{3.909pt}{0.123pt}}
\multiput(1233.00,450.34)(42.886,-11.000){2}{\rule{1.955pt}{0.800pt}}
\multiput(1284.00,439.08)(2.484,-0.512){15}{\rule{3.909pt}{0.123pt}}
\multiput(1284.00,439.34)(42.886,-11.000){2}{\rule{1.955pt}{0.800pt}}
\multiput(1335.00,428.08)(3.072,-0.516){11}{\rule{4.644pt}{0.124pt}}
\multiput(1335.00,428.34)(40.360,-9.000){2}{\rule{2.322pt}{0.800pt}}
\multiput(1385.00,419.08)(4.329,-0.526){7}{\rule{6.029pt}{0.127pt}}
\multiput(1385.00,419.34)(38.487,-7.000){2}{\rule{3.014pt}{0.800pt}}
\put(220,252){\usebox{\plotpoint}}
\put(271,250.84){\rule{12.045pt}{0.800pt}}
\multiput(271.00,250.34)(25.000,1.000){2}{\rule{6.022pt}{0.800pt}}
\put(220.0,252.0){\rule[-0.400pt]{12.286pt}{0.800pt}}
\put(473,251.84){\rule{12.286pt}{0.800pt}}
\multiput(473.00,251.34)(25.500,1.000){2}{\rule{6.143pt}{0.800pt}}
\put(321.0,253.0){\rule[-0.400pt]{36.617pt}{0.800pt}}
\put(929,251.84){\rule{12.286pt}{0.800pt}}
\multiput(929.00,252.34)(25.500,-1.000){2}{\rule{6.143pt}{0.800pt}}
\put(524.0,254.0){\rule[-0.400pt]{97.564pt}{0.800pt}}
\put(1081,250.84){\rule{12.286pt}{0.800pt}}
\multiput(1081.00,251.34)(25.500,-1.000){2}{\rule{6.143pt}{0.800pt}}
\put(980.0,253.0){\rule[-0.400pt]{24.331pt}{0.800pt}}
\put(1183,249.84){\rule{12.045pt}{0.800pt}}
\multiput(1183.00,250.34)(25.000,-1.000){2}{\rule{6.022pt}{0.800pt}}
\put(1233,248.84){\rule{12.286pt}{0.800pt}}
\multiput(1233.00,249.34)(25.500,-1.000){2}{\rule{6.143pt}{0.800pt}}
\put(1132.0,252.0){\rule[-0.400pt]{12.286pt}{0.800pt}}
\put(1335,247.84){\rule{12.045pt}{0.800pt}}
\multiput(1335.00,248.34)(25.000,-1.000){2}{\rule{6.022pt}{0.800pt}}
\put(1284.0,250.0){\rule[-0.400pt]{12.286pt}{0.800pt}}
\put(1385.0,249.0){\rule[-0.400pt]{12.286pt}{0.800pt}}
\put(220,240){\usebox{\plotpoint}}
\put(220,237.34){\rule{12.286pt}{0.800pt}}
\multiput(220.00,238.34)(25.500,-2.000){2}{\rule{6.143pt}{0.800pt}}
\put(271,234.84){\rule{12.045pt}{0.800pt}}
\multiput(271.00,236.34)(25.000,-3.000){2}{\rule{6.022pt}{0.800pt}}
\put(321,231.84){\rule{12.286pt}{0.800pt}}
\multiput(321.00,233.34)(25.500,-3.000){2}{\rule{6.143pt}{0.800pt}}
\put(372,228.84){\rule{12.286pt}{0.800pt}}
\multiput(372.00,230.34)(25.500,-3.000){2}{\rule{6.143pt}{0.800pt}}
\put(423,226.34){\rule{12.045pt}{0.800pt}}
\multiput(423.00,227.34)(25.000,-2.000){2}{\rule{6.022pt}{0.800pt}}
\put(473,223.84){\rule{12.286pt}{0.800pt}}
\multiput(473.00,225.34)(25.500,-3.000){2}{\rule{6.143pt}{0.800pt}}
\put(524,220.84){\rule{12.286pt}{0.800pt}}
\multiput(524.00,222.34)(25.500,-3.000){2}{\rule{6.143pt}{0.800pt}}
\put(575,218.34){\rule{12.045pt}{0.800pt}}
\multiput(575.00,219.34)(25.000,-2.000){2}{\rule{6.022pt}{0.800pt}}
\put(625,215.84){\rule{12.286pt}{0.800pt}}
\multiput(625.00,217.34)(25.500,-3.000){2}{\rule{6.143pt}{0.800pt}}
\put(676,213.34){\rule{12.286pt}{0.800pt}}
\multiput(676.00,214.34)(25.500,-2.000){2}{\rule{6.143pt}{0.800pt}}
\put(727,210.84){\rule{12.045pt}{0.800pt}}
\multiput(727.00,212.34)(25.000,-3.000){2}{\rule{6.022pt}{0.800pt}}
\put(777,208.34){\rule{12.286pt}{0.800pt}}
\multiput(777.00,209.34)(25.500,-2.000){2}{\rule{6.143pt}{0.800pt}}
\put(828,205.84){\rule{12.286pt}{0.800pt}}
\multiput(828.00,207.34)(25.500,-3.000){2}{\rule{6.143pt}{0.800pt}}
\put(879,202.84){\rule{12.045pt}{0.800pt}}
\multiput(879.00,204.34)(25.000,-3.000){2}{\rule{6.022pt}{0.800pt}}
\put(929,200.34){\rule{12.286pt}{0.800pt}}
\multiput(929.00,201.34)(25.500,-2.000){2}{\rule{6.143pt}{0.800pt}}
\put(980,197.84){\rule{12.286pt}{0.800pt}}
\multiput(980.00,199.34)(25.500,-3.000){2}{\rule{6.143pt}{0.800pt}}
\put(1031,195.34){\rule{12.045pt}{0.800pt}}
\multiput(1031.00,196.34)(25.000,-2.000){2}{\rule{6.022pt}{0.800pt}}
\put(1081,192.84){\rule{12.286pt}{0.800pt}}
\multiput(1081.00,194.34)(25.500,-3.000){2}{\rule{6.143pt}{0.800pt}}
\put(1132,190.34){\rule{12.286pt}{0.800pt}}
\multiput(1132.00,191.34)(25.500,-2.000){2}{\rule{6.143pt}{0.800pt}}
\put(1183,187.84){\rule{12.045pt}{0.800pt}}
\multiput(1183.00,189.34)(25.000,-3.000){2}{\rule{6.022pt}{0.800pt}}
\put(1233,185.34){\rule{12.286pt}{0.800pt}}
\multiput(1233.00,186.34)(25.500,-2.000){2}{\rule{6.143pt}{0.800pt}}
\put(1284,183.34){\rule{12.286pt}{0.800pt}}
\multiput(1284.00,184.34)(25.500,-2.000){2}{\rule{6.143pt}{0.800pt}}
\put(1335,181.34){\rule{12.045pt}{0.800pt}}
\multiput(1335.00,182.34)(25.000,-2.000){2}{\rule{6.022pt}{0.800pt}}
\put(1385,179.84){\rule{12.286pt}{0.800pt}}
\multiput(1385.00,180.34)(25.500,-1.000){2}{\rule{6.143pt}{0.800pt}}
\end{picture}

%% file: sferm350.tex
\setlength{\unitlength}{0.240900pt}
\ifx\plotpoint\undefined\newsavebox{\plotpoint}\fi
\sbox{\plotpoint}{\rule[-0.200pt]{0.400pt}{0.400pt}}%
\begin{picture}(1500,900)(0,0)
\font\gnuplot=cmr10 at 10pt
\gnuplot
\sbox{\plotpoint}{\rule[-0.200pt]{0.400pt}{0.400pt}}%
\put(220.0,113.0){\rule[-0.200pt]{292.934pt}{0.400pt}}
\put(220.0,113.0){\rule[-0.200pt]{4.818pt}{0.400pt}}
\put(198,113){\makebox(0,0)[r]{0}}
\put(1416.0,113.0){\rule[-0.200pt]{4.818pt}{0.400pt}}
\put(220.0,266.0){\rule[-0.200pt]{4.818pt}{0.400pt}}
\put(198,266){\makebox(0,0)[r]{1}}
\put(1416.0,266.0){\rule[-0.200pt]{4.818pt}{0.400pt}}
\put(220.0,419.0){\rule[-0.200pt]{4.818pt}{0.400pt}}
\put(198,419){\makebox(0,0)[r]{2}}
\put(1416.0,419.0){\rule[-0.200pt]{4.818pt}{0.400pt}}
\put(220.0,571.0){\rule[-0.200pt]{4.818pt}{0.400pt}}
\put(198,571){\makebox(0,0)[r]{3}}
\put(1416.0,571.0){\rule[-0.200pt]{4.818pt}{0.400pt}}
\put(220.0,724.0){\rule[-0.200pt]{4.818pt}{0.400pt}}
\put(198,724){\makebox(0,0)[r]{4}}
\put(1416.0,724.0){\rule[-0.200pt]{4.818pt}{0.400pt}}
\put(220.0,877.0){\rule[-0.200pt]{4.818pt}{0.400pt}}
\put(198,877){\makebox(0,0)[r]{5}}
\put(1416.0,877.0){\rule[-0.200pt]{4.818pt}{0.400pt}}
\put(220.0,113.0){\rule[-0.200pt]{0.400pt}{4.818pt}}
\put(220,68){\makebox(0,0){4}}
\put(220.0,857.0){\rule[-0.200pt]{0.400pt}{4.818pt}}
\put(423.0,113.0){\rule[-0.200pt]{0.400pt}{4.818pt}}
\put(423,68){\makebox(0,0){6}}
\put(423.0,857.0){\rule[-0.200pt]{0.400pt}{4.818pt}}
\put(625.0,113.0){\rule[-0.200pt]{0.400pt}{4.818pt}}
\put(625,68){\makebox(0,0){8}}
\put(625.0,857.0){\rule[-0.200pt]{0.400pt}{4.818pt}}
\put(828.0,113.0){\rule[-0.200pt]{0.400pt}{4.818pt}}
\put(828,68){\makebox(0,0){10}}
\put(828.0,857.0){\rule[-0.200pt]{0.400pt}{4.818pt}}
\put(1031.0,113.0){\rule[-0.200pt]{0.400pt}{4.818pt}}
\put(1031,68){\makebox(0,0){12}}
\put(1031.0,857.0){\rule[-0.200pt]{0.400pt}{4.818pt}}
\put(1233.0,113.0){\rule[-0.200pt]{0.400pt}{4.818pt}}
\put(1233,68){\makebox(0,0){14}}
\put(1233.0,857.0){\rule[-0.200pt]{0.400pt}{4.818pt}}
\put(1436.0,113.0){\rule[-0.200pt]{0.400pt}{4.818pt}}
\put(1436,68){\makebox(0,0){16}}
\put(1436.0,857.0){\rule[-0.200pt]{0.400pt}{4.818pt}}
\put(220.0,113.0){\rule[-0.200pt]{292.934pt}{0.400pt}}
\put(1436.0,113.0){\rule[-0.200pt]{0.400pt}{184.048pt}}
\put(220.0,877.0){\rule[-0.200pt]{292.934pt}{0.400pt}}
\put(45,495){\makebox(0,0){$m_i$[TeV]}}
\put(828,23){\makebox(0,0){$\log_{10}\mu_0$[GeV]}}
\put(1335,464){\makebox(0,0)[r]{$\widetilde b$}}
\put(1335,342){\makebox(0,0)[r]{$\widetilde t$}}
\put(625,144){\makebox(0,0)[r]{$\widetilde \tau$}}
\put(321,266){\makebox(0,0)[r]{$\chi^0$}}
\put(220.0,113.0){\rule[-0.200pt]{0.400pt}{184.048pt}}
\put(220,235){\usebox{\plotpoint}}
\put(220.00,235.00){\usebox{\plotpoint}}
\put(240.76,235.00){\usebox{\plotpoint}}
\multiput(245,235)(20.756,0.000){0}{\usebox{\plotpoint}}
\put(261.51,235.00){\usebox{\plotpoint}}
\multiput(269,235)(20.756,0.000){0}{\usebox{\plotpoint}}
\put(282.27,235.00){\usebox{\plotpoint}}
\put(303.02,235.00){\usebox{\plotpoint}}
\multiput(306,235)(20.756,0.000){0}{\usebox{\plotpoint}}
\put(323.78,235.00){\usebox{\plotpoint}}
\multiput(331,235)(20.756,0.000){0}{\usebox{\plotpoint}}
\put(344.53,235.00){\usebox{\plotpoint}}
\put(365.29,235.00){\usebox{\plotpoint}}
\multiput(367,235)(20.756,0.000){0}{\usebox{\plotpoint}}
\put(386.04,235.00){\usebox{\plotpoint}}
\multiput(392,235)(20.756,0.000){0}{\usebox{\plotpoint}}
\put(406.80,235.00){\usebox{\plotpoint}}
\put(427.55,235.00){\usebox{\plotpoint}}
\multiput(429,235)(20.756,0.000){0}{\usebox{\plotpoint}}
\put(448.31,235.00){\usebox{\plotpoint}}
\multiput(453,235)(20.756,0.000){0}{\usebox{\plotpoint}}
\put(469.07,235.00){\usebox{\plotpoint}}
\put(489.82,235.00){\usebox{\plotpoint}}
\multiput(490,235)(20.756,0.000){0}{\usebox{\plotpoint}}
\put(510.58,235.00){\usebox{\plotpoint}}
\multiput(515,235)(20.756,0.000){0}{\usebox{\plotpoint}}
\put(531.33,235.00){\usebox{\plotpoint}}
\multiput(539,235)(20.756,0.000){0}{\usebox{\plotpoint}}
\put(552.09,235.00){\usebox{\plotpoint}}
\put(572.84,235.00){\usebox{\plotpoint}}
\multiput(576,235)(20.756,0.000){0}{\usebox{\plotpoint}}
\put(593.60,235.00){\usebox{\plotpoint}}
\multiput(601,235)(20.756,0.000){0}{\usebox{\plotpoint}}
\put(614.35,235.00){\usebox{\plotpoint}}
\put(635.11,235.00){\usebox{\plotpoint}}
\multiput(638,235)(20.756,0.000){0}{\usebox{\plotpoint}}
\put(655.87,235.00){\usebox{\plotpoint}}
\multiput(662,235)(20.756,0.000){0}{\usebox{\plotpoint}}
\put(676.62,235.00){\usebox{\plotpoint}}
\put(697.38,235.00){\usebox{\plotpoint}}
\multiput(699,235)(20.756,0.000){0}{\usebox{\plotpoint}}
\put(718.13,235.00){\usebox{\plotpoint}}
\multiput(724,235)(20.756,0.000){0}{\usebox{\plotpoint}}
\put(738.89,235.00){\usebox{\plotpoint}}
\put(759.64,235.00){\usebox{\plotpoint}}
\multiput(760,235)(20.756,0.000){0}{\usebox{\plotpoint}}
\put(780.40,235.00){\usebox{\plotpoint}}
\multiput(785,235)(20.756,0.000){0}{\usebox{\plotpoint}}
\put(801.15,235.00){\usebox{\plotpoint}}
\put(821.91,235.00){\usebox{\plotpoint}}
\multiput(822,235)(20.756,0.000){0}{\usebox{\plotpoint}}
\put(842.66,235.00){\usebox{\plotpoint}}
\multiput(846,235)(20.756,0.000){0}{\usebox{\plotpoint}}
\put(863.42,235.00){\usebox{\plotpoint}}
\multiput(871,235)(20.756,0.000){0}{\usebox{\plotpoint}}
\put(884.18,235.00){\usebox{\plotpoint}}
\put(904.93,235.00){\usebox{\plotpoint}}
\multiput(908,235)(20.756,0.000){0}{\usebox{\plotpoint}}
\put(925.69,235.00){\usebox{\plotpoint}}
\multiput(932,235)(20.756,0.000){0}{\usebox{\plotpoint}}
\put(946.44,235.00){\usebox{\plotpoint}}
\put(967.20,235.00){\usebox{\plotpoint}}
\multiput(969,235)(20.756,0.000){0}{\usebox{\plotpoint}}
\put(987.95,235.00){\usebox{\plotpoint}}
\multiput(994,235)(20.756,0.000){0}{\usebox{\plotpoint}}
\put(1008.71,235.00){\usebox{\plotpoint}}
\put(1029.46,235.00){\usebox{\plotpoint}}
\multiput(1031,235)(20.756,0.000){0}{\usebox{\plotpoint}}
\put(1050.22,235.00){\usebox{\plotpoint}}
\multiput(1055,235)(20.756,0.000){0}{\usebox{\plotpoint}}
\put(1070.98,235.00){\usebox{\plotpoint}}
\put(1091.73,235.00){\usebox{\plotpoint}}
\multiput(1092,235)(20.756,0.000){0}{\usebox{\plotpoint}}
\put(1112.49,235.00){\usebox{\plotpoint}}
\multiput(1117,235)(20.756,0.000){0}{\usebox{\plotpoint}}
\put(1133.24,235.00){\usebox{\plotpoint}}
\multiput(1141,235)(20.756,0.000){0}{\usebox{\plotpoint}}
\put(1154.00,235.00){\usebox{\plotpoint}}
\put(1174.75,235.00){\usebox{\plotpoint}}
\multiput(1178,235)(20.756,0.000){0}{\usebox{\plotpoint}}
\put(1195.51,235.00){\usebox{\plotpoint}}
\multiput(1203,235)(20.756,0.000){0}{\usebox{\plotpoint}}
\put(1216.26,235.00){\usebox{\plotpoint}}
\put(1237.02,235.00){\usebox{\plotpoint}}
\multiput(1239,235)(20.756,0.000){0}{\usebox{\plotpoint}}
\put(1257.77,235.00){\usebox{\plotpoint}}
\multiput(1264,235)(20.756,0.000){0}{\usebox{\plotpoint}}
\put(1278.53,235.00){\usebox{\plotpoint}}
\put(1299.29,235.00){\usebox{\plotpoint}}
\multiput(1301,235)(20.756,0.000){0}{\usebox{\plotpoint}}
\put(1320.04,235.00){\usebox{\plotpoint}}
\multiput(1325,235)(20.756,0.000){0}{\usebox{\plotpoint}}
\put(1340.80,235.00){\usebox{\plotpoint}}
\put(1361.55,235.00){\usebox{\plotpoint}}
\multiput(1362,235)(20.756,0.000){0}{\usebox{\plotpoint}}
\put(1382.31,235.00){\usebox{\plotpoint}}
\multiput(1387,235)(20.756,0.000){0}{\usebox{\plotpoint}}
\put(1403.06,235.00){\usebox{\plotpoint}}
\put(1423.82,235.00){\usebox{\plotpoint}}
\multiput(1424,235)(20.756,0.000){0}{\usebox{\plotpoint}}
\put(1436,235){\usebox{\plotpoint}}
\sbox{\plotpoint}{\rule[-0.400pt]{0.800pt}{0.800pt}}%
\put(220,634){\usebox{\plotpoint}}
\put(220,630.34){\rule{10.400pt}{0.800pt}}
\multiput(220.00,632.34)(29.414,-4.000){2}{\rule{5.200pt}{0.800pt}}
\multiput(271.00,628.07)(5.374,-0.536){5}{\rule{6.867pt}{0.129pt}}
\multiput(271.00,628.34)(35.748,-6.000){2}{\rule{3.433pt}{0.800pt}}
\multiput(321.00,622.08)(2.484,-0.512){15}{\rule{3.909pt}{0.123pt}}
\multiput(321.00,622.34)(42.886,-11.000){2}{\rule{1.955pt}{0.800pt}}
\multiput(372.00,611.08)(2.484,-0.512){15}{\rule{3.909pt}{0.123pt}}
\multiput(372.00,611.34)(42.886,-11.000){2}{\rule{1.955pt}{0.800pt}}
\multiput(423.00,600.08)(2.208,-0.511){17}{\rule{3.533pt}{0.123pt}}
\multiput(423.00,600.34)(42.666,-12.000){2}{\rule{1.767pt}{0.800pt}}
\multiput(473.00,588.08)(2.063,-0.509){19}{\rule{3.338pt}{0.123pt}}
\multiput(473.00,588.34)(44.071,-13.000){2}{\rule{1.669pt}{0.800pt}}
\multiput(524.00,575.08)(2.063,-0.509){19}{\rule{3.338pt}{0.123pt}}
\multiput(524.00,575.34)(44.071,-13.000){2}{\rule{1.669pt}{0.800pt}}
\multiput(575.00,562.09)(1.865,-0.509){21}{\rule{3.057pt}{0.123pt}}
\multiput(575.00,562.34)(43.655,-14.000){2}{\rule{1.529pt}{0.800pt}}
\multiput(625.00,548.09)(1.904,-0.509){21}{\rule{3.114pt}{0.123pt}}
\multiput(625.00,548.34)(44.536,-14.000){2}{\rule{1.557pt}{0.800pt}}
\multiput(676.00,534.08)(2.063,-0.509){19}{\rule{3.338pt}{0.123pt}}
\multiput(676.00,534.34)(44.071,-13.000){2}{\rule{1.669pt}{0.800pt}}
\multiput(727.00,521.09)(1.732,-0.508){23}{\rule{2.867pt}{0.122pt}}
\multiput(727.00,521.34)(44.050,-15.000){2}{\rule{1.433pt}{0.800pt}}
\multiput(777.00,506.09)(1.904,-0.509){21}{\rule{3.114pt}{0.123pt}}
\multiput(777.00,506.34)(44.536,-14.000){2}{\rule{1.557pt}{0.800pt}}
\multiput(828.00,492.09)(1.904,-0.509){21}{\rule{3.114pt}{0.123pt}}
\multiput(828.00,492.34)(44.536,-14.000){2}{\rule{1.557pt}{0.800pt}}
\multiput(879.00,478.08)(2.022,-0.509){19}{\rule{3.277pt}{0.123pt}}
\multiput(879.00,478.34)(43.199,-13.000){2}{\rule{1.638pt}{0.800pt}}
\multiput(929.00,465.08)(2.063,-0.509){19}{\rule{3.338pt}{0.123pt}}
\multiput(929.00,465.34)(44.071,-13.000){2}{\rule{1.669pt}{0.800pt}}
\multiput(980.00,452.08)(2.063,-0.509){19}{\rule{3.338pt}{0.123pt}}
\multiput(980.00,452.34)(44.071,-13.000){2}{\rule{1.669pt}{0.800pt}}
\multiput(1031.00,439.08)(2.022,-0.509){19}{\rule{3.277pt}{0.123pt}}
\multiput(1031.00,439.34)(43.199,-13.000){2}{\rule{1.638pt}{0.800pt}}
\multiput(1081.00,426.08)(2.254,-0.511){17}{\rule{3.600pt}{0.123pt}}
\multiput(1081.00,426.34)(43.528,-12.000){2}{\rule{1.800pt}{0.800pt}}
\multiput(1132.00,414.08)(2.254,-0.511){17}{\rule{3.600pt}{0.123pt}}
\multiput(1132.00,414.34)(43.528,-12.000){2}{\rule{1.800pt}{0.800pt}}
\multiput(1183.00,402.08)(2.434,-0.512){15}{\rule{3.836pt}{0.123pt}}
\multiput(1183.00,402.34)(42.037,-11.000){2}{\rule{1.918pt}{0.800pt}}
\multiput(1233.00,391.08)(2.484,-0.512){15}{\rule{3.909pt}{0.123pt}}
\multiput(1233.00,391.34)(42.886,-11.000){2}{\rule{1.955pt}{0.800pt}}
\multiput(1284.00,380.08)(2.769,-0.514){13}{\rule{4.280pt}{0.124pt}}
\multiput(1284.00,380.34)(42.117,-10.000){2}{\rule{2.140pt}{0.800pt}}
\multiput(1335.00,370.08)(3.552,-0.520){9}{\rule{5.200pt}{0.125pt}}
\multiput(1335.00,370.34)(39.207,-8.000){2}{\rule{2.600pt}{0.800pt}}
\multiput(1385.00,362.07)(5.486,-0.536){5}{\rule{7.000pt}{0.129pt}}
\multiput(1385.00,362.34)(36.471,-6.000){2}{\rule{3.500pt}{0.800pt}}
\put(220,674){\usebox{\plotpoint}}
\multiput(220.00,672.07)(5.486,-0.536){5}{\rule{7.000pt}{0.129pt}}
\multiput(220.00,672.34)(36.471,-6.000){2}{\rule{3.500pt}{0.800pt}}
\multiput(271.00,666.06)(7.980,-0.560){3}{\rule{8.200pt}{0.135pt}}
\multiput(271.00,666.34)(32.980,-5.000){2}{\rule{4.100pt}{0.800pt}}
\multiput(321.00,661.08)(2.769,-0.514){13}{\rule{4.280pt}{0.124pt}}
\multiput(321.00,661.34)(42.117,-10.000){2}{\rule{2.140pt}{0.800pt}}
\multiput(372.00,651.08)(2.484,-0.512){15}{\rule{3.909pt}{0.123pt}}
\multiput(372.00,651.34)(42.886,-11.000){2}{\rule{1.955pt}{0.800pt}}
\multiput(423.00,640.08)(2.208,-0.511){17}{\rule{3.533pt}{0.123pt}}
\multiput(423.00,640.34)(42.666,-12.000){2}{\rule{1.767pt}{0.800pt}}
\multiput(473.00,628.09)(1.904,-0.509){21}{\rule{3.114pt}{0.123pt}}
\multiput(473.00,628.34)(44.536,-14.000){2}{\rule{1.557pt}{0.800pt}}
\multiput(524.00,614.08)(2.063,-0.509){19}{\rule{3.338pt}{0.123pt}}
\multiput(524.00,614.34)(44.071,-13.000){2}{\rule{1.669pt}{0.800pt}}
\multiput(575.00,601.09)(1.865,-0.509){21}{\rule{3.057pt}{0.123pt}}
\multiput(575.00,601.34)(43.655,-14.000){2}{\rule{1.529pt}{0.800pt}}
\multiput(625.00,587.09)(1.767,-0.508){23}{\rule{2.920pt}{0.122pt}}
\multiput(625.00,587.34)(44.939,-15.000){2}{\rule{1.460pt}{0.800pt}}
\multiput(676.00,572.09)(1.904,-0.509){21}{\rule{3.114pt}{0.123pt}}
\multiput(676.00,572.34)(44.536,-14.000){2}{\rule{1.557pt}{0.800pt}}
\multiput(727.00,558.09)(1.732,-0.508){23}{\rule{2.867pt}{0.122pt}}
\multiput(727.00,558.34)(44.050,-15.000){2}{\rule{1.433pt}{0.800pt}}
\multiput(777.00,543.09)(1.904,-0.509){21}{\rule{3.114pt}{0.123pt}}
\multiput(777.00,543.34)(44.536,-14.000){2}{\rule{1.557pt}{0.800pt}}
\multiput(828.00,529.09)(1.904,-0.509){21}{\rule{3.114pt}{0.123pt}}
\multiput(828.00,529.34)(44.536,-14.000){2}{\rule{1.557pt}{0.800pt}}
\multiput(879.00,515.09)(1.865,-0.509){21}{\rule{3.057pt}{0.123pt}}
\multiput(879.00,515.34)(43.655,-14.000){2}{\rule{1.529pt}{0.800pt}}
\multiput(929.00,501.09)(1.767,-0.508){23}{\rule{2.920pt}{0.122pt}}
\multiput(929.00,501.34)(44.939,-15.000){2}{\rule{1.460pt}{0.800pt}}
\multiput(980.00,486.08)(2.063,-0.509){19}{\rule{3.338pt}{0.123pt}}
\multiput(980.00,486.34)(44.071,-13.000){2}{\rule{1.669pt}{0.800pt}}
\multiput(1031.00,473.09)(1.865,-0.509){21}{\rule{3.057pt}{0.123pt}}
\multiput(1031.00,473.34)(43.655,-14.000){2}{\rule{1.529pt}{0.800pt}}
\multiput(1081.00,459.08)(2.063,-0.509){19}{\rule{3.338pt}{0.123pt}}
\multiput(1081.00,459.34)(44.071,-13.000){2}{\rule{1.669pt}{0.800pt}}
\multiput(1132.00,446.08)(2.254,-0.511){17}{\rule{3.600pt}{0.123pt}}
\multiput(1132.00,446.34)(43.528,-12.000){2}{\rule{1.800pt}{0.800pt}}
\multiput(1183.00,434.08)(2.022,-0.509){19}{\rule{3.277pt}{0.123pt}}
\multiput(1183.00,434.34)(43.199,-13.000){2}{\rule{1.638pt}{0.800pt}}
\multiput(1233.00,421.08)(2.484,-0.512){15}{\rule{3.909pt}{0.123pt}}
\multiput(1233.00,421.34)(42.886,-11.000){2}{\rule{1.955pt}{0.800pt}}
\multiput(1284.00,410.08)(2.484,-0.512){15}{\rule{3.909pt}{0.123pt}}
\multiput(1284.00,410.34)(42.886,-11.000){2}{\rule{1.955pt}{0.800pt}}
\multiput(1335.00,399.08)(3.072,-0.516){11}{\rule{4.644pt}{0.124pt}}
\multiput(1335.00,399.34)(40.360,-9.000){2}{\rule{2.322pt}{0.800pt}}
\multiput(1385.00,390.07)(5.486,-0.536){5}{\rule{7.000pt}{0.129pt}}
\multiput(1385.00,390.34)(36.471,-6.000){2}{\rule{3.500pt}{0.800pt}}
\put(220,158){\usebox{\plotpoint}}
\put(220,157.84){\rule{12.286pt}{0.800pt}}
\multiput(220.00,156.34)(25.500,3.000){2}{\rule{6.143pt}{0.800pt}}
\put(271,160.84){\rule{12.045pt}{0.800pt}}
\multiput(271.00,159.34)(25.000,3.000){2}{\rule{6.022pt}{0.800pt}}
\put(321,163.84){\rule{12.286pt}{0.800pt}}
\multiput(321.00,162.34)(25.500,3.000){2}{\rule{6.143pt}{0.800pt}}
\put(372,166.84){\rule{12.286pt}{0.800pt}}
\multiput(372.00,165.34)(25.500,3.000){2}{\rule{6.143pt}{0.800pt}}
\put(423,169.84){\rule{12.045pt}{0.800pt}}
\multiput(423.00,168.34)(25.000,3.000){2}{\rule{6.022pt}{0.800pt}}
\put(473,172.34){\rule{12.286pt}{0.800pt}}
\multiput(473.00,171.34)(25.500,2.000){2}{\rule{6.143pt}{0.800pt}}
\put(524,174.34){\rule{12.286pt}{0.800pt}}
\multiput(524.00,173.34)(25.500,2.000){2}{\rule{6.143pt}{0.800pt}}
\put(575,176.34){\rule{12.045pt}{0.800pt}}
\multiput(575.00,175.34)(25.000,2.000){2}{\rule{6.022pt}{0.800pt}}
\put(625,178.34){\rule{12.286pt}{0.800pt}}
\multiput(625.00,177.34)(25.500,2.000){2}{\rule{6.143pt}{0.800pt}}
\put(676,180.34){\rule{12.286pt}{0.800pt}}
\multiput(676.00,179.34)(25.500,2.000){2}{\rule{6.143pt}{0.800pt}}
\put(727,182.34){\rule{12.045pt}{0.800pt}}
\multiput(727.00,181.34)(25.000,2.000){2}{\rule{6.022pt}{0.800pt}}
\put(777,183.84){\rule{12.286pt}{0.800pt}}
\multiput(777.00,183.34)(25.500,1.000){2}{\rule{6.143pt}{0.800pt}}
\put(828,184.84){\rule{12.286pt}{0.800pt}}
\multiput(828.00,184.34)(25.500,1.000){2}{\rule{6.143pt}{0.800pt}}
\put(879,185.84){\rule{12.045pt}{0.800pt}}
\multiput(879.00,185.34)(25.000,1.000){2}{\rule{6.022pt}{0.800pt}}
\put(929,186.84){\rule{12.286pt}{0.800pt}}
\multiput(929.00,186.34)(25.500,1.000){2}{\rule{6.143pt}{0.800pt}}
\put(980,187.84){\rule{12.286pt}{0.800pt}}
\multiput(980.00,187.34)(25.500,1.000){2}{\rule{6.143pt}{0.800pt}}
\put(1031,188.84){\rule{12.045pt}{0.800pt}}
\multiput(1031.00,188.34)(25.000,1.000){2}{\rule{6.022pt}{0.800pt}}
\put(1132,189.84){\rule{12.286pt}{0.800pt}}
\multiput(1132.00,189.34)(25.500,1.000){2}{\rule{6.143pt}{0.800pt}}
\put(1081.0,191.0){\rule[-0.400pt]{12.286pt}{0.800pt}}
\put(1233,190.84){\rule{12.286pt}{0.800pt}}
\multiput(1233.00,190.34)(25.500,1.000){2}{\rule{6.143pt}{0.800pt}}
\put(1183.0,192.0){\rule[-0.400pt]{12.045pt}{0.800pt}}
\put(1284.0,193.0){\rule[-0.400pt]{36.617pt}{0.800pt}}
\put(220,240){\usebox{\plotpoint}}
\put(220,236.84){\rule{12.286pt}{0.800pt}}
\multiput(220.00,238.34)(25.500,-3.000){2}{\rule{6.143pt}{0.800pt}}
\put(271,234.34){\rule{12.045pt}{0.800pt}}
\multiput(271.00,235.34)(25.000,-2.000){2}{\rule{6.022pt}{0.800pt}}
\put(321,231.84){\rule{12.286pt}{0.800pt}}
\multiput(321.00,233.34)(25.500,-3.000){2}{\rule{6.143pt}{0.800pt}}
\put(372,228.84){\rule{12.286pt}{0.800pt}}
\multiput(372.00,230.34)(25.500,-3.000){2}{\rule{6.143pt}{0.800pt}}
\put(423,226.34){\rule{12.045pt}{0.800pt}}
\multiput(423.00,227.34)(25.000,-2.000){2}{\rule{6.022pt}{0.800pt}}
\put(473,223.84){\rule{12.286pt}{0.800pt}}
\multiput(473.00,225.34)(25.500,-3.000){2}{\rule{6.143pt}{0.800pt}}
\put(524,220.84){\rule{12.286pt}{0.800pt}}
\multiput(524.00,222.34)(25.500,-3.000){2}{\rule{6.143pt}{0.800pt}}
\put(575,218.34){\rule{12.045pt}{0.800pt}}
\multiput(575.00,219.34)(25.000,-2.000){2}{\rule{6.022pt}{0.800pt}}
\put(625,215.84){\rule{12.286pt}{0.800pt}}
\multiput(625.00,217.34)(25.500,-3.000){2}{\rule{6.143pt}{0.800pt}}
\put(676,213.34){\rule{12.286pt}{0.800pt}}
\multiput(676.00,214.34)(25.500,-2.000){2}{\rule{6.143pt}{0.800pt}}
\put(727,210.84){\rule{12.045pt}{0.800pt}}
\multiput(727.00,212.34)(25.000,-3.000){2}{\rule{6.022pt}{0.800pt}}
\put(777,207.84){\rule{12.286pt}{0.800pt}}
\multiput(777.00,209.34)(25.500,-3.000){2}{\rule{6.143pt}{0.800pt}}
\put(828,205.34){\rule{12.286pt}{0.800pt}}
\multiput(828.00,206.34)(25.500,-2.000){2}{\rule{6.143pt}{0.800pt}}
\put(879,202.84){\rule{12.045pt}{0.800pt}}
\multiput(879.00,204.34)(25.000,-3.000){2}{\rule{6.022pt}{0.800pt}}
\put(929,200.34){\rule{12.286pt}{0.800pt}}
\multiput(929.00,201.34)(25.500,-2.000){2}{\rule{6.143pt}{0.800pt}}
\put(980,197.84){\rule{12.286pt}{0.800pt}}
\multiput(980.00,199.34)(25.500,-3.000){2}{\rule{6.143pt}{0.800pt}}
\put(1031,195.34){\rule{12.045pt}{0.800pt}}
\multiput(1031.00,196.34)(25.000,-2.000){2}{\rule{6.022pt}{0.800pt}}
\put(1081,192.84){\rule{12.286pt}{0.800pt}}
\multiput(1081.00,194.34)(25.500,-3.000){2}{\rule{6.143pt}{0.800pt}}
\put(1132,190.34){\rule{12.286pt}{0.800pt}}
\multiput(1132.00,191.34)(25.500,-2.000){2}{\rule{6.143pt}{0.800pt}}
\put(1183,187.84){\rule{12.045pt}{0.800pt}}
\multiput(1183.00,189.34)(25.000,-3.000){2}{\rule{6.022pt}{0.800pt}}
\put(1233,185.34){\rule{12.286pt}{0.800pt}}
\multiput(1233.00,186.34)(25.500,-2.000){2}{\rule{6.143pt}{0.800pt}}
\put(1284,183.34){\rule{12.286pt}{0.800pt}}
\multiput(1284.00,184.34)(25.500,-2.000){2}{\rule{6.143pt}{0.800pt}}
\put(1335,181.34){\rule{12.045pt}{0.800pt}}
\multiput(1335.00,182.34)(25.000,-2.000){2}{\rule{6.022pt}{0.800pt}}
\put(1385,179.84){\rule{12.286pt}{0.800pt}}
\multiput(1385.00,180.34)(25.500,-1.000){2}{\rule{6.143pt}{0.800pt}}
\end{picture}

%% file: lsp13.tex
\setlength{\unitlength}{0.240900pt}
\ifx\plotpoint\undefined\newsavebox{\plotpoint}\fi
\sbox{\plotpoint}{\rule[-0.200pt]{0.400pt}{0.400pt}}%
\begin{picture}(1500,900)(0,0)
\font\gnuplot=cmr10 at 10pt
\gnuplot
\sbox{\plotpoint}{\rule[-0.200pt]{0.400pt}{0.400pt}}%
\put(220.0,113.0){\rule[-0.200pt]{292.934pt}{0.400pt}}
\put(220.0,113.0){\rule[-0.200pt]{4.818pt}{0.400pt}}
\put(198,113){\makebox(0,0)[r]{0}}
\put(1416.0,113.0){\rule[-0.200pt]{4.818pt}{0.400pt}}
\put(220.0,189.0){\rule[-0.200pt]{4.818pt}{0.400pt}}
\put(198,189){\makebox(0,0)[r]{0.2}}
\put(1416.0,189.0){\rule[-0.200pt]{4.818pt}{0.400pt}}
\put(220.0,266.0){\rule[-0.200pt]{4.818pt}{0.400pt}}
\put(198,266){\makebox(0,0)[r]{0.4}}
\put(1416.0,266.0){\rule[-0.200pt]{4.818pt}{0.400pt}}
\put(220.0,342.0){\rule[-0.200pt]{4.818pt}{0.400pt}}
\put(198,342){\makebox(0,0)[r]{0.6}}
\put(1416.0,342.0){\rule[-0.200pt]{4.818pt}{0.400pt}}
\put(220.0,419.0){\rule[-0.200pt]{4.818pt}{0.400pt}}
\put(198,419){\makebox(0,0)[r]{0.8}}
\put(1416.0,419.0){\rule[-0.200pt]{4.818pt}{0.400pt}}
\put(220.0,495.0){\rule[-0.200pt]{4.818pt}{0.400pt}}
\put(198,495){\makebox(0,0)[r]{1}}
\put(1416.0,495.0){\rule[-0.200pt]{4.818pt}{0.400pt}}
\put(220.0,571.0){\rule[-0.200pt]{4.818pt}{0.400pt}}
\put(198,571){\makebox(0,0)[r]{1.2}}
\put(1416.0,571.0){\rule[-0.200pt]{4.818pt}{0.400pt}}
\put(220.0,648.0){\rule[-0.200pt]{4.818pt}{0.400pt}}
\put(198,648){\makebox(0,0)[r]{1.4}}
\put(1416.0,648.0){\rule[-0.200pt]{4.818pt}{0.400pt}}
\put(220.0,724.0){\rule[-0.200pt]{4.818pt}{0.400pt}}
\put(198,724){\makebox(0,0)[r]{1.6}}
\put(1416.0,724.0){\rule[-0.200pt]{4.818pt}{0.400pt}}
\put(220.0,801.0){\rule[-0.200pt]{4.818pt}{0.400pt}}
\put(198,801){\makebox(0,0)[r]{1.8}}
\put(1416.0,801.0){\rule[-0.200pt]{4.818pt}{0.400pt}}
\put(220.0,877.0){\rule[-0.200pt]{4.818pt}{0.400pt}}
\put(198,877){\makebox(0,0)[r]{2}}
\put(1416.0,877.0){\rule[-0.200pt]{4.818pt}{0.400pt}}
\put(296.0,113.0){\rule[-0.200pt]{0.400pt}{4.818pt}}
\put(296,68){\makebox(0,0){5}}
\put(296.0,857.0){\rule[-0.200pt]{0.400pt}{4.818pt}}
\put(423.0,113.0){\rule[-0.200pt]{0.400pt}{4.818pt}}
\put(423,68){\makebox(0,0){10}}
\put(423.0,857.0){\rule[-0.200pt]{0.400pt}{4.818pt}}
\put(549.0,113.0){\rule[-0.200pt]{0.400pt}{4.818pt}}
\put(549,68){\makebox(0,0){15}}
\put(549.0,857.0){\rule[-0.200pt]{0.400pt}{4.818pt}}
\put(676.0,113.0){\rule[-0.200pt]{0.400pt}{4.818pt}}
\put(676,68){\makebox(0,0){20}}
\put(676.0,857.0){\rule[-0.200pt]{0.400pt}{4.818pt}}
\put(803.0,113.0){\rule[-0.200pt]{0.400pt}{4.818pt}}
\put(803,68){\makebox(0,0){25}}
\put(803.0,857.0){\rule[-0.200pt]{0.400pt}{4.818pt}}
\put(929.0,113.0){\rule[-0.200pt]{0.400pt}{4.818pt}}
\put(929,68){\makebox(0,0){30}}
\put(929.0,857.0){\rule[-0.200pt]{0.400pt}{4.818pt}}
\put(1056.0,113.0){\rule[-0.200pt]{0.400pt}{4.818pt}}
\put(1056,68){\makebox(0,0){35}}
\put(1056.0,857.0){\rule[-0.200pt]{0.400pt}{4.818pt}}
\put(1183.0,113.0){\rule[-0.200pt]{0.400pt}{4.818pt}}
\put(1183,68){\makebox(0,0){40}}
\put(1183.0,857.0){\rule[-0.200pt]{0.400pt}{4.818pt}}
\put(1309.0,113.0){\rule[-0.200pt]{0.400pt}{4.818pt}}
\put(1309,68){\makebox(0,0){45}}
\put(1309.0,857.0){\rule[-0.200pt]{0.400pt}{4.818pt}}
\put(1436.0,113.0){\rule[-0.200pt]{0.400pt}{4.818pt}}
\put(1436,68){\makebox(0,0){50}}
\put(1436.0,857.0){\rule[-0.200pt]{0.400pt}{4.818pt}}
\put(220.0,113.0){\rule[-0.200pt]{292.934pt}{0.400pt}}
\put(1436.0,113.0){\rule[-0.200pt]{0.400pt}{184.048pt}}
\put(220.0,877.0){\rule[-0.200pt]{292.934pt}{0.400pt}}
\put(45,495){\makebox(0,0){$m_0$[TeV]}}
\put(828,23){\makebox(0,0){$\tan \beta$}}
\put(220.0,113.0){\rule[-0.200pt]{0.400pt}{184.048pt}}
\put(1056,113){\makebox(0,0){$\star$}}
\put(1183,113){\makebox(0,0){$\star$}}
\put(1309,113){\makebox(0,0){$\star$}}
\put(1436,113){\makebox(0,0){$\star$}}
\put(1056,151){\makebox(0,0){$\star$}}
\put(1183,151){\makebox(0,0){$\star$}}
\put(1309,151){\makebox(0,0){$\star$}}
\put(1436,151){\makebox(0,0){$\star$}}
\put(1183,189){\makebox(0,0){$\star$}}
\put(1309,189){\makebox(0,0){$\star$}}
\put(1436,189){\makebox(0,0){$\star$}}
\put(1309,228){\makebox(0,0){$\star$}}
\put(1436,228){\makebox(0,0){$\star$}}
\put(220,113){\raisebox{-.8pt}{\makebox(0,0){$\Box$}}}
\put(220,151){\raisebox{-.8pt}{\makebox(0,0){$\Box$}}}
\put(220,189){\raisebox{-.8pt}{\makebox(0,0){$\Box$}}}
\put(296,113){\raisebox{-.8pt}{\makebox(0,0){$\Box$}}}
\put(423,113){\raisebox{-.8pt}{\makebox(0,0){$\Box$}}}
\put(549,113){\raisebox{-.8pt}{\makebox(0,0){$\Box$}}}
\put(676,113){\raisebox{-.8pt}{\makebox(0,0){$\Box$}}}
\put(803,113){\raisebox{-.8pt}{\makebox(0,0){$\Box$}}}
\put(929,113){\raisebox{-.8pt}{\makebox(0,0){$\Box$}}}
\put(1056,113){\raisebox{-.8pt}{\makebox(0,0){$\Box$}}}
\put(1183,113){\raisebox{-.8pt}{\makebox(0,0){$\Box$}}}
\put(1309,113){\raisebox{-.8pt}{\makebox(0,0){$\Box$}}}
\put(1436,113){\raisebox{-.8pt}{\makebox(0,0){$\Box$}}}
\put(296,151){\raisebox{-.8pt}{\makebox(0,0){$\Box$}}}
\put(423,151){\raisebox{-.8pt}{\makebox(0,0){$\Box$}}}
\put(549,151){\raisebox{-.8pt}{\makebox(0,0){$\Box$}}}
\put(676,151){\raisebox{-.8pt}{\makebox(0,0){$\Box$}}}
\put(803,151){\raisebox{-.8pt}{\makebox(0,0){$\Box$}}}
\put(929,151){\raisebox{-.8pt}{\makebox(0,0){$\Box$}}}
\put(1056,151){\raisebox{-.8pt}{\makebox(0,0){$\Box$}}}
\put(1183,151){\raisebox{-.8pt}{\makebox(0,0){$\Box$}}}
\put(1309,151){\raisebox{-.8pt}{\makebox(0,0){$\Box$}}}
\put(1436,151){\raisebox{-.8pt}{\makebox(0,0){$\Box$}}}
\put(296,189){\raisebox{-.8pt}{\makebox(0,0){$\Box$}}}
\put(423,189){\raisebox{-.8pt}{\makebox(0,0){$\Box$}}}
\put(549,189){\raisebox{-.8pt}{\makebox(0,0){$\Box$}}}
\put(676,189){\raisebox{-.8pt}{\makebox(0,0){$\Box$}}}
\put(803,189){\raisebox{-.8pt}{\makebox(0,0){$\Box$}}}
\put(929,189){\raisebox{-.8pt}{\makebox(0,0){$\Box$}}}
\put(1056,189){\raisebox{-.8pt}{\makebox(0,0){$\Box$}}}
\put(1183,189){\raisebox{-.8pt}{\makebox(0,0){$\Box$}}}
\put(1309,189){\raisebox{-.8pt}{\makebox(0,0){$\Box$}}}
\put(1436,189){\raisebox{-.8pt}{\makebox(0,0){$\Box$}}}
\put(676,228){\raisebox{-.8pt}{\makebox(0,0){$\Box$}}}
\put(803,228){\raisebox{-.8pt}{\makebox(0,0){$\Box$}}}
\put(929,228){\raisebox{-.8pt}{\makebox(0,0){$\Box$}}}
\put(1056,228){\raisebox{-.8pt}{\makebox(0,0){$\Box$}}}
\put(1183,228){\raisebox{-.8pt}{\makebox(0,0){$\Box$}}}
\put(1309,228){\raisebox{-.8pt}{\makebox(0,0){$\Box$}}}
\put(1436,228){\raisebox{-.8pt}{\makebox(0,0){$\Box$}}}
\put(929,266){\raisebox{-.8pt}{\makebox(0,0){$\Box$}}}
\put(1056,266){\raisebox{-.8pt}{\makebox(0,0){$\Box$}}}
\put(1183,266){\raisebox{-.8pt}{\makebox(0,0){$\Box$}}}
\put(1309,266){\raisebox{-.8pt}{\makebox(0,0){$\Box$}}}
\put(1436,266){\raisebox{-.8pt}{\makebox(0,0){$\Box$}}}
\put(1056,304){\raisebox{-.8pt}{\makebox(0,0){$\Box$}}}
\put(1183,304){\raisebox{-.8pt}{\makebox(0,0){$\Box$}}}
\put(1309,304){\raisebox{-.8pt}{\makebox(0,0){$\Box$}}}
\put(1436,304){\raisebox{-.8pt}{\makebox(0,0){$\Box$}}}
\put(1309,342){\raisebox{-.8pt}{\makebox(0,0){$\Box$}}}
\put(1436,342){\raisebox{-.8pt}{\makebox(0,0){$\Box$}}}
\put(1436,380){\raisebox{-.8pt}{\makebox(0,0){$\Box$}}}
\end{picture}

%% file: lsp8.tex
\setlength{\unitlength}{0.240900pt}
\ifx\plotpoint\undefined\newsavebox{\plotpoint}\fi
\sbox{\plotpoint}{\rule[-0.200pt]{0.400pt}{0.400pt}}%
\begin{picture}(1500,900)(0,0)
\font\gnuplot=cmr10 at 10pt
\gnuplot
\sbox{\plotpoint}{\rule[-0.200pt]{0.400pt}{0.400pt}}%
\put(220.0,113.0){\rule[-0.200pt]{292.934pt}{0.400pt}}
\put(220.0,113.0){\rule[-0.200pt]{4.818pt}{0.400pt}}
\put(198,113){\makebox(0,0)[r]{0}}
\put(1416.0,113.0){\rule[-0.200pt]{4.818pt}{0.400pt}}
\put(220.0,189.0){\rule[-0.200pt]{4.818pt}{0.400pt}}
\put(198,189){\makebox(0,0)[r]{0.2}}
\put(1416.0,189.0){\rule[-0.200pt]{4.818pt}{0.400pt}}
\put(220.0,266.0){\rule[-0.200pt]{4.818pt}{0.400pt}}
\put(198,266){\makebox(0,0)[r]{0.4}}
\put(1416.0,266.0){\rule[-0.200pt]{4.818pt}{0.400pt}}
\put(220.0,342.0){\rule[-0.200pt]{4.818pt}{0.400pt}}
\put(198,342){\makebox(0,0)[r]{0.6}}
\put(1416.0,342.0){\rule[-0.200pt]{4.818pt}{0.400pt}}
\put(220.0,419.0){\rule[-0.200pt]{4.818pt}{0.400pt}}
\put(198,419){\makebox(0,0)[r]{0.8}}
\put(1416.0,419.0){\rule[-0.200pt]{4.818pt}{0.400pt}}
\put(220.0,495.0){\rule[-0.200pt]{4.818pt}{0.400pt}}
\put(198,495){\makebox(0,0)[r]{1}}
\put(1416.0,495.0){\rule[-0.200pt]{4.818pt}{0.400pt}}
\put(220.0,571.0){\rule[-0.200pt]{4.818pt}{0.400pt}}
\put(198,571){\makebox(0,0)[r]{1.2}}
\put(1416.0,571.0){\rule[-0.200pt]{4.818pt}{0.400pt}}
\put(220.0,648.0){\rule[-0.200pt]{4.818pt}{0.400pt}}
\put(198,648){\makebox(0,0)[r]{1.4}}
\put(1416.0,648.0){\rule[-0.200pt]{4.818pt}{0.400pt}}
\put(220.0,724.0){\rule[-0.200pt]{4.818pt}{0.400pt}}
\put(198,724){\makebox(0,0)[r]{1.6}}
\put(1416.0,724.0){\rule[-0.200pt]{4.818pt}{0.400pt}}
\put(220.0,801.0){\rule[-0.200pt]{4.818pt}{0.400pt}}
\put(198,801){\makebox(0,0)[r]{1.8}}
\put(1416.0,801.0){\rule[-0.200pt]{4.818pt}{0.400pt}}
\put(220.0,877.0){\rule[-0.200pt]{4.818pt}{0.400pt}}
\put(198,877){\makebox(0,0)[r]{2}}
\put(1416.0,877.0){\rule[-0.200pt]{4.818pt}{0.400pt}}
\put(296.0,113.0){\rule[-0.200pt]{0.400pt}{4.818pt}}
\put(296,68){\makebox(0,0){5}}
\put(296.0,857.0){\rule[-0.200pt]{0.400pt}{4.818pt}}
\put(423.0,113.0){\rule[-0.200pt]{0.400pt}{4.818pt}}
\put(423,68){\makebox(0,0){10}}
\put(423.0,857.0){\rule[-0.200pt]{0.400pt}{4.818pt}}
\put(549.0,113.0){\rule[-0.200pt]{0.400pt}{4.818pt}}
\put(549,68){\makebox(0,0){15}}
\put(549.0,857.0){\rule[-0.200pt]{0.400pt}{4.818pt}}
\put(676.0,113.0){\rule[-0.200pt]{0.400pt}{4.818pt}}
\put(676,68){\makebox(0,0){20}}
\put(676.0,857.0){\rule[-0.200pt]{0.400pt}{4.818pt}}
\put(803.0,113.0){\rule[-0.200pt]{0.400pt}{4.818pt}}
\put(803,68){\makebox(0,0){25}}
\put(803.0,857.0){\rule[-0.200pt]{0.400pt}{4.818pt}}
\put(929.0,113.0){\rule[-0.200pt]{0.400pt}{4.818pt}}
\put(929,68){\makebox(0,0){30}}
\put(929.0,857.0){\rule[-0.200pt]{0.400pt}{4.818pt}}
\put(1056.0,113.0){\rule[-0.200pt]{0.400pt}{4.818pt}}
\put(1056,68){\makebox(0,0){35}}
\put(1056.0,857.0){\rule[-0.200pt]{0.400pt}{4.818pt}}
\put(1183.0,113.0){\rule[-0.200pt]{0.400pt}{4.818pt}}
\put(1183,68){\makebox(0,0){40}}
\put(1183.0,857.0){\rule[-0.200pt]{0.400pt}{4.818pt}}
\put(1309.0,113.0){\rule[-0.200pt]{0.400pt}{4.818pt}}
\put(1309,68){\makebox(0,0){45}}
\put(1309.0,857.0){\rule[-0.200pt]{0.400pt}{4.818pt}}
\put(1436.0,113.0){\rule[-0.200pt]{0.400pt}{4.818pt}}
\put(1436,68){\makebox(0,0){50}}
\put(1436.0,857.0){\rule[-0.200pt]{0.400pt}{4.818pt}}
\put(220.0,113.0){\rule[-0.200pt]{292.934pt}{0.400pt}}
\put(1436.0,113.0){\rule[-0.200pt]{0.400pt}{184.048pt}}
\put(220.0,877.0){\rule[-0.200pt]{292.934pt}{0.400pt}}
\put(45,495){\makebox(0,0){$m_0$[TeV]}}
\put(828,23){\makebox(0,0){$\tan \beta$}}
\put(220.0,113.0){\rule[-0.200pt]{0.400pt}{184.048pt}}
\put(1309,113){\makebox(0,0){$\star$}}
\put(1436,113){\makebox(0,0){$\star$}}
\put(1309,151){\makebox(0,0){$\star$}}
\put(1436,151){\makebox(0,0){$\star$}}
\put(1436,189){\makebox(0,0){$\star$}}
\put(1436,228){\makebox(0,0){$\star$}}
\put(220,113){\raisebox{-.8pt}{\makebox(0,0){$\Box$}}}
\put(220,151){\raisebox{-.8pt}{\makebox(0,0){$\Box$}}}
\put(220,189){\raisebox{-.8pt}{\makebox(0,0){$\Box$}}}
\put(220,228){\raisebox{-.8pt}{\makebox(0,0){$\Box$}}}
\put(296,113){\raisebox{-.8pt}{\makebox(0,0){$\Box$}}}
\put(423,113){\raisebox{-.8pt}{\makebox(0,0){$\Box$}}}
\put(549,113){\raisebox{-.8pt}{\makebox(0,0){$\Box$}}}
\put(676,113){\raisebox{-.8pt}{\makebox(0,0){$\Box$}}}
\put(803,113){\raisebox{-.8pt}{\makebox(0,0){$\Box$}}}
\put(929,113){\raisebox{-.8pt}{\makebox(0,0){$\Box$}}}
\put(1056,113){\raisebox{-.8pt}{\makebox(0,0){$\Box$}}}
\put(1183,113){\raisebox{-.8pt}{\makebox(0,0){$\Box$}}}
\put(1309,113){\raisebox{-.8pt}{\makebox(0,0){$\Box$}}}
\put(1436,113){\raisebox{-.8pt}{\makebox(0,0){$\Box$}}}
\put(296,151){\raisebox{-.8pt}{\makebox(0,0){$\Box$}}}
\put(423,151){\raisebox{-.8pt}{\makebox(0,0){$\Box$}}}
\put(549,151){\raisebox{-.8pt}{\makebox(0,0){$\Box$}}}
\put(676,151){\raisebox{-.8pt}{\makebox(0,0){$\Box$}}}
\put(803,151){\raisebox{-.8pt}{\makebox(0,0){$\Box$}}}
\put(929,151){\raisebox{-.8pt}{\makebox(0,0){$\Box$}}}
\put(1056,151){\raisebox{-.8pt}{\makebox(0,0){$\Box$}}}
\put(1183,151){\raisebox{-.8pt}{\makebox(0,0){$\Box$}}}
\put(1309,151){\raisebox{-.8pt}{\makebox(0,0){$\Box$}}}
\put(1436,151){\raisebox{-.8pt}{\makebox(0,0){$\Box$}}}
\put(296,189){\raisebox{-.8pt}{\makebox(0,0){$\Box$}}}
\put(423,189){\raisebox{-.8pt}{\makebox(0,0){$\Box$}}}
\put(549,189){\raisebox{-.8pt}{\makebox(0,0){$\Box$}}}
\put(676,189){\raisebox{-.8pt}{\makebox(0,0){$\Box$}}}
\put(803,189){\raisebox{-.8pt}{\makebox(0,0){$\Box$}}}
\put(929,189){\raisebox{-.8pt}{\makebox(0,0){$\Box$}}}
\put(1056,189){\raisebox{-.8pt}{\makebox(0,0){$\Box$}}}
\put(1183,189){\raisebox{-.8pt}{\makebox(0,0){$\Box$}}}
\put(1309,189){\raisebox{-.8pt}{\makebox(0,0){$\Box$}}}
\put(1436,189){\raisebox{-.8pt}{\makebox(0,0){$\Box$}}}
\put(296,228){\raisebox{-.8pt}{\makebox(0,0){$\Box$}}}
\put(423,228){\raisebox{-.8pt}{\makebox(0,0){$\Box$}}}
\put(549,228){\raisebox{-.8pt}{\makebox(0,0){$\Box$}}}
\put(676,228){\raisebox{-.8pt}{\makebox(0,0){$\Box$}}}
\put(803,228){\raisebox{-.8pt}{\makebox(0,0){$\Box$}}}
\put(929,228){\raisebox{-.8pt}{\makebox(0,0){$\Box$}}}
\put(1056,228){\raisebox{-.8pt}{\makebox(0,0){$\Box$}}}
\put(1183,228){\raisebox{-.8pt}{\makebox(0,0){$\Box$}}}
\put(1309,228){\raisebox{-.8pt}{\makebox(0,0){$\Box$}}}
\put(1436,228){\raisebox{-.8pt}{\makebox(0,0){$\Box$}}}
\put(549,266){\raisebox{-.8pt}{\makebox(0,0){$\Box$}}}
\put(676,266){\raisebox{-.8pt}{\makebox(0,0){$\Box$}}}
\put(803,266){\raisebox{-.8pt}{\makebox(0,0){$\Box$}}}
\put(929,266){\raisebox{-.8pt}{\makebox(0,0){$\Box$}}}
\put(1056,266){\raisebox{-.8pt}{\makebox(0,0){$\Box$}}}
\put(1183,266){\raisebox{-.8pt}{\makebox(0,0){$\Box$}}}
\put(1309,266){\raisebox{-.8pt}{\makebox(0,0){$\Box$}}}
\put(1436,266){\raisebox{-.8pt}{\makebox(0,0){$\Box$}}}
\put(803,304){\raisebox{-.8pt}{\makebox(0,0){$\Box$}}}
\put(929,304){\raisebox{-.8pt}{\makebox(0,0){$\Box$}}}
\put(1056,304){\raisebox{-.8pt}{\makebox(0,0){$\Box$}}}
\put(1183,304){\raisebox{-.8pt}{\makebox(0,0){$\Box$}}}
\put(1309,304){\raisebox{-.8pt}{\makebox(0,0){$\Box$}}}
\put(1436,304){\raisebox{-.8pt}{\makebox(0,0){$\Box$}}}
\put(1056,342){\raisebox{-.8pt}{\makebox(0,0){$\Box$}}}
\put(1183,342){\raisebox{-.8pt}{\makebox(0,0){$\Box$}}}
\put(1309,342){\raisebox{-.8pt}{\makebox(0,0){$\Box$}}}
\put(1436,342){\raisebox{-.8pt}{\makebox(0,0){$\Box$}}}
\put(1183,380){\raisebox{-.8pt}{\makebox(0,0){$\Box$}}}
\put(1309,380){\raisebox{-.8pt}{\makebox(0,0){$\Box$}}}
\put(1436,380){\raisebox{-.8pt}{\makebox(0,0){$\Box$}}}
\put(1309,419){\raisebox{-.8pt}{\makebox(0,0){$\Box$}}}
\put(1436,419){\raisebox{-.8pt}{\makebox(0,0){$\Box$}}}
\put(1436,457){\raisebox{-.8pt}{\makebox(0,0){$\Box$}}}
\end{picture}

%% file: lsp2.tex
\setlength{\unitlength}{0.240900pt}
\ifx\plotpoint\undefined\newsavebox{\plotpoint}\fi
\sbox{\plotpoint}{\rule[-0.200pt]{0.400pt}{0.400pt}}%
\begin{picture}(1500,900)(0,0)
\font\gnuplot=cmr10 at 10pt
\gnuplot
\sbox{\plotpoint}{\rule[-0.200pt]{0.400pt}{0.400pt}}%
\put(220.0,113.0){\rule[-0.200pt]{292.934pt}{0.400pt}}
\put(220.0,113.0){\rule[-0.200pt]{4.818pt}{0.400pt}}
\put(198,113){\makebox(0,0)[r]{0}}
\put(1416.0,113.0){\rule[-0.200pt]{4.818pt}{0.400pt}}
\put(220.0,189.0){\rule[-0.200pt]{4.818pt}{0.400pt}}
\put(198,189){\makebox(0,0)[r]{0.2}}
\put(1416.0,189.0){\rule[-0.200pt]{4.818pt}{0.400pt}}
\put(220.0,266.0){\rule[-0.200pt]{4.818pt}{0.400pt}}
\put(198,266){\makebox(0,0)[r]{0.4}}
\put(1416.0,266.0){\rule[-0.200pt]{4.818pt}{0.400pt}}
\put(220.0,342.0){\rule[-0.200pt]{4.818pt}{0.400pt}}
\put(198,342){\makebox(0,0)[r]{0.6}}
\put(1416.0,342.0){\rule[-0.200pt]{4.818pt}{0.400pt}}
\put(220.0,419.0){\rule[-0.200pt]{4.818pt}{0.400pt}}
\put(198,419){\makebox(0,0)[r]{0.8}}
\put(1416.0,419.0){\rule[-0.200pt]{4.818pt}{0.400pt}}
\put(220.0,495.0){\rule[-0.200pt]{4.818pt}{0.400pt}}
\put(198,495){\makebox(0,0)[r]{1}}
\put(1416.0,495.0){\rule[-0.200pt]{4.818pt}{0.400pt}}
\put(220.0,571.0){\rule[-0.200pt]{4.818pt}{0.400pt}}
\put(198,571){\makebox(0,0)[r]{1.2}}
\put(1416.0,571.0){\rule[-0.200pt]{4.818pt}{0.400pt}}
\put(220.0,648.0){\rule[-0.200pt]{4.818pt}{0.400pt}}
\put(198,648){\makebox(0,0)[r]{1.4}}
\put(1416.0,648.0){\rule[-0.200pt]{4.818pt}{0.400pt}}
\put(220.0,724.0){\rule[-0.200pt]{4.818pt}{0.400pt}}
\put(198,724){\makebox(0,0)[r]{1.6}}
\put(1416.0,724.0){\rule[-0.200pt]{4.818pt}{0.400pt}}
\put(220.0,801.0){\rule[-0.200pt]{4.818pt}{0.400pt}}
\put(198,801){\makebox(0,0)[r]{1.8}}
\put(1416.0,801.0){\rule[-0.200pt]{4.818pt}{0.400pt}}
\put(220.0,877.0){\rule[-0.200pt]{4.818pt}{0.400pt}}
\put(198,877){\makebox(0,0)[r]{2}}
\put(1416.0,877.0){\rule[-0.200pt]{4.818pt}{0.400pt}}
\put(296.0,113.0){\rule[-0.200pt]{0.400pt}{4.818pt}}
\put(296,68){\makebox(0,0){5}}
\put(296.0,857.0){\rule[-0.200pt]{0.400pt}{4.818pt}}
\put(423.0,113.0){\rule[-0.200pt]{0.400pt}{4.818pt}}
\put(423,68){\makebox(0,0){10}}
\put(423.0,857.0){\rule[-0.200pt]{0.400pt}{4.818pt}}
\put(549.0,113.0){\rule[-0.200pt]{0.400pt}{4.818pt}}
\put(549,68){\makebox(0,0){15}}
\put(549.0,857.0){\rule[-0.200pt]{0.400pt}{4.818pt}}
\put(676.0,113.0){\rule[-0.200pt]{0.400pt}{4.818pt}}
\put(676,68){\makebox(0,0){20}}
\put(676.0,857.0){\rule[-0.200pt]{0.400pt}{4.818pt}}
\put(803.0,113.0){\rule[-0.200pt]{0.400pt}{4.818pt}}
\put(803,68){\makebox(0,0){25}}
\put(803.0,857.0){\rule[-0.200pt]{0.400pt}{4.818pt}}
\put(929.0,113.0){\rule[-0.200pt]{0.400pt}{4.818pt}}
\put(929,68){\makebox(0,0){30}}
\put(929.0,857.0){\rule[-0.200pt]{0.400pt}{4.818pt}}
\put(1056.0,113.0){\rule[-0.200pt]{0.400pt}{4.818pt}}
\put(1056,68){\makebox(0,0){35}}
\put(1056.0,857.0){\rule[-0.200pt]{0.400pt}{4.818pt}}
\put(1183.0,113.0){\rule[-0.200pt]{0.400pt}{4.818pt}}
\put(1183,68){\makebox(0,0){40}}
\put(1183.0,857.0){\rule[-0.200pt]{0.400pt}{4.818pt}}
\put(1309.0,113.0){\rule[-0.200pt]{0.400pt}{4.818pt}}
\put(1309,68){\makebox(0,0){45}}
\put(1309.0,857.0){\rule[-0.200pt]{0.400pt}{4.818pt}}
\put(1436.0,113.0){\rule[-0.200pt]{0.400pt}{4.818pt}}
\put(1436,68){\makebox(0,0){50}}
\put(1436.0,857.0){\rule[-0.200pt]{0.400pt}{4.818pt}}
\put(220.0,113.0){\rule[-0.200pt]{292.934pt}{0.400pt}}
\put(1436.0,113.0){\rule[-0.200pt]{0.400pt}{184.048pt}}
\put(220.0,877.0){\rule[-0.200pt]{292.934pt}{0.400pt}}
\put(45,495){\makebox(0,0){$m_0$[TeV]}}
\put(828,23){\makebox(0,0){$\tan \beta$}}
\put(220.0,113.0){\rule[-0.200pt]{0.400pt}{184.048pt}}
\put(1183,113){\makebox(0,0){$\star$}}
\put(1309,113){\makebox(0,0){$\star$}}
\put(1436,113){\makebox(0,0){$\star$}}
\put(1183,151){\makebox(0,0){$\star$}}
\put(1309,151){\makebox(0,0){$\star$}}
\put(1436,151){\makebox(0,0){$\star$}}
\put(1309,189){\makebox(0,0){$\star$}}
\put(1436,189){\makebox(0,0){$\star$}}
\put(1309,228){\makebox(0,0){$\star$}}
\put(1436,228){\makebox(0,0){$\star$}}
\put(1436,266){\makebox(0,0){$\star$}}
\put(1436,304){\makebox(0,0){$\star$}}
\put(220,113){\raisebox{-.8pt}{\makebox(0,0){$\Box$}}}
\put(220,151){\raisebox{-.8pt}{\makebox(0,0){$\Box$}}}
\put(220,189){\raisebox{-.8pt}{\makebox(0,0){$\Box$}}}
\put(220,228){\raisebox{-.8pt}{\makebox(0,0){$\Box$}}}
\put(220,266){\raisebox{-.8pt}{\makebox(0,0){$\Box$}}}
\put(220,304){\raisebox{-.8pt}{\makebox(0,0){$\Box$}}}
\put(220,342){\raisebox{-.8pt}{\makebox(0,0){$\Box$}}}
\put(296,113){\raisebox{-.8pt}{\makebox(0,0){$\Box$}}}
\put(423,113){\raisebox{-.8pt}{\makebox(0,0){$\Box$}}}
\put(549,113){\raisebox{-.8pt}{\makebox(0,0){$\Box$}}}
\put(676,113){\raisebox{-.8pt}{\makebox(0,0){$\Box$}}}
\put(803,113){\raisebox{-.8pt}{\makebox(0,0){$\Box$}}}
\put(929,113){\raisebox{-.8pt}{\makebox(0,0){$\Box$}}}
\put(1056,113){\raisebox{-.8pt}{\makebox(0,0){$\Box$}}}
\put(1183,113){\raisebox{-.8pt}{\makebox(0,0){$\Box$}}}
\put(1309,113){\raisebox{-.8pt}{\makebox(0,0){$\Box$}}}
\put(1436,113){\raisebox{-.8pt}{\makebox(0,0){$\Box$}}}
\put(296,151){\raisebox{-.8pt}{\makebox(0,0){$\Box$}}}
\put(423,151){\raisebox{-.8pt}{\makebox(0,0){$\Box$}}}
\put(549,151){\raisebox{-.8pt}{\makebox(0,0){$\Box$}}}
\put(676,151){\raisebox{-.8pt}{\makebox(0,0){$\Box$}}}
\put(803,151){\raisebox{-.8pt}{\makebox(0,0){$\Box$}}}
\put(929,151){\raisebox{-.8pt}{\makebox(0,0){$\Box$}}}
\put(1056,151){\raisebox{-.8pt}{\makebox(0,0){$\Box$}}}
\put(1183,151){\raisebox{-.8pt}{\makebox(0,0){$\Box$}}}
\put(1309,151){\raisebox{-.8pt}{\makebox(0,0){$\Box$}}}
\put(1436,151){\raisebox{-.8pt}{\makebox(0,0){$\Box$}}}
\put(296,189){\raisebox{-.8pt}{\makebox(0,0){$\Box$}}}
\put(423,189){\raisebox{-.8pt}{\makebox(0,0){$\Box$}}}
\put(549,189){\raisebox{-.8pt}{\makebox(0,0){$\Box$}}}
\put(676,189){\raisebox{-.8pt}{\makebox(0,0){$\Box$}}}
\put(803,189){\raisebox{-.8pt}{\makebox(0,0){$\Box$}}}
\put(929,189){\raisebox{-.8pt}{\makebox(0,0){$\Box$}}}
\put(1056,189){\raisebox{-.8pt}{\makebox(0,0){$\Box$}}}
\put(1183,189){\raisebox{-.8pt}{\makebox(0,0){$\Box$}}}
\put(1309,189){\raisebox{-.8pt}{\makebox(0,0){$\Box$}}}
\put(1436,189){\raisebox{-.8pt}{\makebox(0,0){$\Box$}}}
\put(296,228){\raisebox{-.8pt}{\makebox(0,0){$\Box$}}}
\put(423,228){\raisebox{-.8pt}{\makebox(0,0){$\Box$}}}
\put(549,228){\raisebox{-.8pt}{\makebox(0,0){$\Box$}}}
\put(676,228){\raisebox{-.8pt}{\makebox(0,0){$\Box$}}}
\put(803,228){\raisebox{-.8pt}{\makebox(0,0){$\Box$}}}
\put(929,228){\raisebox{-.8pt}{\makebox(0,0){$\Box$}}}
\put(1056,228){\raisebox{-.8pt}{\makebox(0,0){$\Box$}}}
\put(1183,228){\raisebox{-.8pt}{\makebox(0,0){$\Box$}}}
\put(1309,228){\raisebox{-.8pt}{\makebox(0,0){$\Box$}}}
\put(1436,228){\raisebox{-.8pt}{\makebox(0,0){$\Box$}}}
\put(296,266){\raisebox{-.8pt}{\makebox(0,0){$\Box$}}}
\put(423,266){\raisebox{-.8pt}{\makebox(0,0){$\Box$}}}
\put(549,266){\raisebox{-.8pt}{\makebox(0,0){$\Box$}}}
\put(676,266){\raisebox{-.8pt}{\makebox(0,0){$\Box$}}}
\put(803,266){\raisebox{-.8pt}{\makebox(0,0){$\Box$}}}
\put(929,266){\raisebox{-.8pt}{\makebox(0,0){$\Box$}}}
\put(1056,266){\raisebox{-.8pt}{\makebox(0,0){$\Box$}}}
\put(1183,266){\raisebox{-.8pt}{\makebox(0,0){$\Box$}}}
\put(1309,266){\raisebox{-.8pt}{\makebox(0,0){$\Box$}}}
\put(1436,266){\raisebox{-.8pt}{\makebox(0,0){$\Box$}}}
\put(296,304){\raisebox{-.8pt}{\makebox(0,0){$\Box$}}}
\put(423,304){\raisebox{-.8pt}{\makebox(0,0){$\Box$}}}
\put(549,304){\raisebox{-.8pt}{\makebox(0,0){$\Box$}}}
\put(676,304){\raisebox{-.8pt}{\makebox(0,0){$\Box$}}}
\put(803,304){\raisebox{-.8pt}{\makebox(0,0){$\Box$}}}
\put(929,304){\raisebox{-.8pt}{\makebox(0,0){$\Box$}}}
\put(1056,304){\raisebox{-.8pt}{\makebox(0,0){$\Box$}}}
\put(1183,304){\raisebox{-.8pt}{\makebox(0,0){$\Box$}}}
\put(1309,304){\raisebox{-.8pt}{\makebox(0,0){$\Box$}}}
\put(1436,304){\raisebox{-.8pt}{\makebox(0,0){$\Box$}}}
\put(296,342){\raisebox{-.8pt}{\makebox(0,0){$\Box$}}}
\put(423,342){\raisebox{-.8pt}{\makebox(0,0){$\Box$}}}
\put(549,342){\raisebox{-.8pt}{\makebox(0,0){$\Box$}}}
\put(676,342){\raisebox{-.8pt}{\makebox(0,0){$\Box$}}}
\put(803,342){\raisebox{-.8pt}{\makebox(0,0){$\Box$}}}
\put(929,342){\raisebox{-.8pt}{\makebox(0,0){$\Box$}}}
\put(1056,342){\raisebox{-.8pt}{\makebox(0,0){$\Box$}}}
\put(1183,342){\raisebox{-.8pt}{\makebox(0,0){$\Box$}}}
\put(1309,342){\raisebox{-.8pt}{\makebox(0,0){$\Box$}}}
\put(1436,342){\raisebox{-.8pt}{\makebox(0,0){$\Box$}}}
\put(676,380){\raisebox{-.8pt}{\makebox(0,0){$\Box$}}}
\put(803,380){\raisebox{-.8pt}{\makebox(0,0){$\Box$}}}
\put(929,380){\raisebox{-.8pt}{\makebox(0,0){$\Box$}}}
\put(1056,380){\raisebox{-.8pt}{\makebox(0,0){$\Box$}}}
\put(1183,380){\raisebox{-.8pt}{\makebox(0,0){$\Box$}}}
\put(1309,380){\raisebox{-.8pt}{\makebox(0,0){$\Box$}}}
\put(1436,380){\raisebox{-.8pt}{\makebox(0,0){$\Box$}}}
\put(929,419){\raisebox{-.8pt}{\makebox(0,0){$\Box$}}}
\put(1056,419){\raisebox{-.8pt}{\makebox(0,0){$\Box$}}}
\put(1183,419){\raisebox{-.8pt}{\makebox(0,0){$\Box$}}}
\put(1309,419){\raisebox{-.8pt}{\makebox(0,0){$\Box$}}}
\put(1436,419){\raisebox{-.8pt}{\makebox(0,0){$\Box$}}}
\put(1056,457){\raisebox{-.8pt}{\makebox(0,0){$\Box$}}}
\put(1183,457){\raisebox{-.8pt}{\makebox(0,0){$\Box$}}}
\put(1309,457){\raisebox{-.8pt}{\makebox(0,0){$\Box$}}}
\put(1436,457){\raisebox{-.8pt}{\makebox(0,0){$\Box$}}}
\put(1183,495){\raisebox{-.8pt}{\makebox(0,0){$\Box$}}}
\put(1309,495){\raisebox{-.8pt}{\makebox(0,0){$\Box$}}}
\put(1436,495){\raisebox{-.8pt}{\makebox(0,0){$\Box$}}}
\put(1309,533){\raisebox{-.8pt}{\makebox(0,0){$\Box$}}}
\put(1436,533){\raisebox{-.8pt}{\makebox(0,0){$\Box$}}}
\put(1309,571){\raisebox{-.8pt}{\makebox(0,0){$\Box$}}}
\put(1436,571){\raisebox{-.8pt}{\makebox(0,0){$\Box$}}}
\put(1436,610){\raisebox{-.8pt}{\makebox(0,0){$\Box$}}}
\put(1436,648){\raisebox{-.8pt}{\makebox(0,0){$\Box$}}}
\end{picture}

%% file: draft6.bbl
\newpage
\begin{thebibliography}{99}

\bi{witten1}
E. Witten, Nucl. Phys. {\bf B471} (1996) 135;
P. Horava and E. Witten, Nucl. Phys. {\bf B475} (1996) 94.

\bi{witten2}
E. Witten, Nucl. Phys. {\bf B443} (1995) 85;
A. Dabholkar, Phys. Lett. {\bf B357} (1995) 307;
C.M. Hull, Phys. Lett. {\bf B357} (1995) 545;
J. Polchinski and E. Witten, Nucl. Phys. {\bf B460} (1996) 525.

\bi{antoniadis1}
I. Antoniadis and M. Quiros,
Phys. Lett. B392 (1997) 61.

\bi{antoniadis2}
I. Antoniadis, N. Arkani-Hamed, S. Dimopoulos and G. Dvali,
``{\em New Dimensions at a Millimeter to a 
Fermi and Superstring at TeV}'',
hep-ph/9804398.

\bi{arkani1}
N. Arkani-Hamed, S. Dimopoulos and G. Dvali,
``{\em The Hierarchy Problem and New Dimensions at a
Millimeter}'',
hep-ph/9803315;
``{\em Phenomenology, Astrophysics and Cosmology
of Theories with Sub-Millimeter 
Dimensions and TeV Scale Quantum Gravity}'',
hep-ph/9807344.

\bi{arkani2}
N. Arkani-Hamed, S. Dimopoulos and J. March-Russell,
``{\em Stabilization of Sub-Millimeter Dimensions: 
The New Guise of the Hierarchy Problem}'',
hep-th/9809124.

\bi{antoniadis3}
I. Antoniadis, S. Dimopoulos, A. Pomarol and M. Quir\'os,
``{\em Soft Masses in Theories with Supersymmetry Breaking 
by TeV-Compactification}'',
hep-ph/9810410.

\bi{arkani3}
N. Arkani-Hamed and S. Dimopoulos,
``{\em New origin for approximate symmetries from distant breaking 
in extra dimensions}'', 
hep-ph/9811353.

\bi{berezhiani}
Z. Berezhiani and G. Dvali,
``{\em Flavor Violation in Theories with TeV Scale Quantum Gravity}'', 
hep-ph/9811378.

\bi{arkani4}
N. Arkani-Hamed, S. Dimopoulos, G. Dvali and J. March-Russell,
``{\em  Neutrino Masses from Large Extra Dimensions}'', 
hep-ph/9811448.




\bi{sundrum}
R. Sundrum, ``{\em Effective Field Theory for a Three-Brane Universe}'', 
hep-ph/9805471;
``{\em Compactification for a Three-Brane Universe}'', 
hep-ph/9807348.

\bi{shiu}
G. Shiu and S.-H.H. Tye,
``{\em TeV Scale Superstring and Extra Dimensions}'', 
hep-th/9805157.

\bi{pomarol}
A. Pomarol and M. Quir\'os,
``{\em The Standard Model from extra dimensions}'', 
hep-ph/9806263.

\bi{lykken11}
J. Lykken, E. Poppitz and S. Trivedi,
``{\em Branes with GUTs and Supersymmetry Breaking}'',
hep-th/9806080.

\bi{poppitz11}
E. Poppitz,
``{\em Chirality, D-Branes and Gauge/String Unification}'',
hep-ph/9806309.


\bi{dienes1}
K. Dienes, E. Dudas and T. Gherghetta,
``{\em Grand Unification at Intermediate 
Mass Scales through Extra Dimensions}'', hep-ph/9806292.


\bi{bachas1}
C. Bachas,
``{\em Unification with Low String Scale}'', hep-ph/9807415.



\bi{kakushadze1}
Z. Kakushadze and S.-H. Tye,
``{\em Three Generation in Type I
Compactifications}'', hep-th/9806143.
``{\em Brane World}'', hep-th/9809147.

\bi{benakli1}
K. Benakli, 
``{\em Phenomenology of Low Quantum Gravity Scale Models}'', 
hep-ph/9809582.

\bi{dienes2}
K. Dienes, E. Dudas, T. Gherghetta and A. Riotto,
``{\em Cosmological Phase Transitions and 
Radius Stabilization in Higher Dimensions}'', 
hep-ph/9809406.

\bi{randall}
L. Randall and R. Sundrum,
``{\em Out of This World Supersymmetry Breaking}'', 
hep-th/9810155.

\bi{ibanez}
C.P. Burgess, L.E. Ib\'a\~nez and F. Quevedo,
``{\em Strings at the Intermediate Scale or is the Fermi Scale Dual 
to the Planck Scale ?}'', hep-ph/9810535.


\bi{han}
T. Han, J.D. Lykken and R.J. Zhang,
``{\em  On Kaluza-Klein States from Large Extra Dimensions}'', 
hep-ph/9811350.

\bi{dienes2}
K.R. Dienes, E. Dudas and T. Gherghetta,
``{\em Neutrino Oscillattions without Neutrino Masses or 
Heavy Mass Scales: A Higher-Dimensional Seesaw Mechanism }'', 
hep-ph/9811428.

\bi{kakushadze2}
Z. Kakushadze, 
``{\em  Novel Extension of MSSM and "TeV Scale" Coupling Unification}'', 
hep-th/9811193.




\bi{abel1}
S. Abel and S. King,
``{\em On fixed points and fermion mass structure
from large extra dimensions}'', hep-ph/9809467.

\bi{ross1}
D. Ghilencea and G.G. Ross,
``{\em Unification and extra space-time
dimensions}'', hep-ph/9809217.

\bi{lim}
H. Hatanaka, T. Inami and C.S. Lim,
``{\em The Gauge Hierarchy Problem and Higher Dimensional Gauge Theories}'', 
hep-th/9805067.



\bi{giudice}
G.F. Giudice, R. Rattazzi and J.D. Wells,
``{\em Quantum Gravity and Extra Dimensions at High-Energy Colliders}'', 
hep-ph/9811291.



\bi{nussinov}
S. Nussinov and R. Shrock,
``{\em Some Remarks on Theories with Large Compact Dimensions 
and TeV-Scale Quantum Gravity}'', 
hep-ph/9811323.

\bi{mirabelli}
E.A. Mirabelli, M. Perelstein and M.E. Peskin,
``{\em Collider Signatures of New Large Space Dimensions}'', 
hep-ph/9811337.

\bi{hewett}
J. L. Hewett,
``{\em  Indirect Collider Signals for Extra Dimensions}'', 
hep-ph/9811356.



\bi{wilson1}
K. Wilson and J. Kogut,
Phys. Rep. {\bf 12C} (1974) 75.

\bi{wegner}
F.J. Wegner and A. Houghton,
Phys. ReV. {\bf A8} (1973) 401.



\bi{polchinski1} 
J. Polchinski, Nucl. Phys. {\bf B231} (1984) 269;
C. Wetterich, Phys. Lett.
{\bf B301} (1993) 90;
N. Teradis and C. Wetterich, Nucl. Phys.
{\bf B442} [FS] (1994) 541.


\bibitem{yamada1} Y. Yamada, Phys.Rev. {\bf D50} (1994) 3537.

\bibitem{hisano1} J. Hisano and M. Shifman, 
Phys. Rev. {\bf D56} (1997) 5475.

\bibitem{jack3} I. Jack and D.R.T. Jones,
Phys. Lett. {\bf B415} (1997) 383; 
I. Jack, D.R.T. Jones and
 A. Pickering, Phys. Lett. {\bf B426} (1998) 73. 
 

\bibitem{avdeev1} L.V. Avdeev, 
D.I. Kazakov and I.N. Kondrashuk,
Nucl. Phys. {\bf B510} (1998) 289. 


\bi{morris1} T.R. Morris, 
Int. J. Mod. Phys. {\bf A9} (1994) 2411.


\bi{nicoll}
J.F. Nicoll, T.S. Chang and H.E. Stanly,,
Phys. ReV. {\bf A13} (1976) 1251.

\bi{hasen1}
A. Hasenfratz and P. Hasenfratz,
Nucl. Phys. {\bf B270} (1986) 687.

\bi{morris2}
T.R. Morris, Phys. Lett. {\bf 334} (1994) 355;
K.-I. Aoki {\em et al},
Prog. Theor. Phys. {\bf 95} (1996) 409;
T.R. Morris, 
{\em Elements of the Continuous Renormalization
Group}, hep-th/9802039 and
references therein.

\bi{jack6}
I. Jack, D.R.T. Jones, S.P. Martin, M.T. Vaughn  
and Y. Yamada, Phys. Rev. {\bf D50} (1994) R5481.

\bi{kkz1}
T. Kobayashi, J. Kubo and G. Zoupanos,
Phys. Lett. {\bf B427} (1998) 291.

\bibitem{jack4} I. Jack, D.R.T. Jones and
 A. Pickering, Phys. Lett. {\bf B432} (1998) 114. 

\bibitem{novikov1} V. Novikov, M. Shifman, 
A. Vainstein and V. Zakharov,  Nucl. Phys.
{\bf B229} (1983) 381; Phys. Lett. {\bf B166}
 (1986) 329;  M. Shifman, Int. J. Mod.
      Phys.{\bf  A11} (1996) 5761 and references therein. 



\bi{pdg}Particle Data Group, C. Caso {\em et al}.,
Eur. Phys. J. {\bf C3} (1998) 1.

\bi{kmoz}
J.~Kubo, M.~Mondrag\'on, M. Olechowski and G.~Zoupanos, 
Nucl. Phys. {\bf B479} (1996) 25.


\bi{level}
L.E. Ib\'a\~nez, Phys. Lett. {\bf B318} (1993) 73; 
H. Kawabe, T. Kobayashi and N. Ohtsubo, 
Phys. Lett. {\bf B325} (1994) 77; 
Nucl. Phys. {\bf B434} (1995) 210.

\bibitem{bmass1}
S.~Mart\'i i Garc\'ia, J.~Fuster and S.~Cabrera, 
hep-ex/9708030.

\bibitem{bmass2}
M.~Jamin and A.~Pich, Nucl. Phys. {\bf B507} (1997) 334.

\bibitem{bmass3}
V.~Gim\'enez, G.~Martinelli and C.T.~Sachrajda,
Phys. Lett. {\bf B393} (1997) 124.

\bibitem{bmass4}
G.~Rodrigo, A.~Santamaria and M.~Bilenky, 
Phys. Rev. Lett. {\bf 79} (1997) 193.



\bibitem{hall}
L.J.~Hall, R.~Rattazzi and U.~Sarid, 
Phys. Rev. {\bf D50} (1994) 7048; 
R.~Hempfling, Phys. Rev. {\bf D49} (1994) 6168.

\bibitem{COPW}
M.~Carena, M.~Olechowski, S.~Pokorski and C.E.M.~Wagner, 
Nucl. Phys. {\bf B426} (1994) 269.


\bi{BIM}
A.~Brignole, L.E.~Ib\'a\~nez and C.~Mu\~noz,
 Nucl. Phys. {\bf B422} (1994) 125; 
A.~Brignole, L.E.~Ib\'a\~nez, C.~Mu\~noz and C.~Scheich,
Z. Phys. {\bf C74} (1997) 157.

\bibitem{JJ}
D.R.T.~Jones, L.~Mezincescu and Y.P.~Yau, 
Phys. Lett. {\bf B148} (1984) 317; 
I.~Jack and D.R.T.~Jones, Phys. Lett. {\bf B349} (1995) 294; 
D.I.~Kazakov, M.Yu.~Kalmykov, I.N.~Kondrashuk and A.V.~Gladyshev, 
Nucl. Phys. {\bf B471} (1996) 387.

\bi{sumrule}
Y.~Kawamura, T.~Kobayashi and J.~Kubo, 
Phys. Lett. {\bf B405} (1997) 64.

\bi{kkmz}
T.~Kobayashi, J.~Kubo, M.~Mondrag\'on and G.~Zoupanos, 
Nucl. Phys. {\bf B511} (1998) 45.

\bi{kks}
Y. Kawamura, T. Kobayashi and H. Shimabukuro, 
Phys. Lett. {\bf B436} (1998) 108.

\end{thebibliography}
